\documentclass[a4paper,12pt]{article}
\usepackage{amsfonts,amsmath,amssymb,color,comment,listings}
\usepackage[paper=letterpaper,margin=1.0in]{geometry}
\usepackage{graphicx,ytableau}
\usepackage{cite}
\usepackage{hyperref}
\usepackage{young} 
\usepackage{tikz,pgfplots}
\usetikzlibrary{angles,quotes,arrows.meta,chains,matrix,scopes,calc,intersections,positioning,shapes.misc,decorations.markings,shapes.geometric}
\tikzset{cross/.style={cross out, draw=black,thick, minimum size=3*(#1-\pgflinewidth), inner sep=0pt, outer sep=0pt},
cross/.default={3pt}}

\newcommand{\be}{\begin{equation}}
\newcommand{\ee}{\end{equation}}
\newcommand{\bea}{\begin{eqnarray}\displaystyle}
\newcommand{\eea}{\end{eqnarray}}

\newcommand{\ba}{\begin{array}}
\newcommand{\ea}{\end{array}}
\newcommand{\ben}{\begin{enumerate}}
\newcommand{\een}{\end{enumerate}}
\newcommand{\bi}{\begin{itemize}}
\newcommand{\ei}{\end{itemize}}
\newcommand{\bc}{\begin{center}}
\newcommand{\ec}{\end{center}}
\newcommand{\bfig}{\begin{figure}}
\newcommand{\efig}{\end{figure}}
\newcommand{\bq}{\begin{quotation}}
\newcommand{\eq}{\end{quotation}}
\newcommand{\bt}{\begin{table}}
\newcommand{\et}{\end{table}}
\newcommand{\btab}{\begin{tabular}}
\newcommand{\etab}{\end{tabular}}
\newcommand{\bmi}{\begin{minipage}}
\newcommand{\emi}{\end{minipage}}
\newcommand{\bs}{\begin{slide}}
\newcommand{\es}{\end{slide}}

\newcommand{\la}{\langle}
\newcommand{\ra}{\rangle}

\def\tr{ {\rm tr } }

\def\cA{{\cal A}}  \def\cC{{\cal C}}
  
\def\cG{{\cal G}}  
 \def\cK{{\cal K}} 
 \def\cN{{\cal N}} \def\cO{{\cal O}}

\def\cV{{\cal V}}  
 \def\cZ{{\cal Z}}

\newcommand{\mC}{{\mathbb C}}

\newcommand{\mcal}{\mathcal} 
\def\cC{ \mathcal{C} }

\newcommand{\wchi}{\widehat{\chi}} 
 
\def\c2{ \widehat \chi_2^R }

\begin{document}

\rightline{QMUL-PH-19-28}
\vspace{0.8truecm}

\vspace{15pt}


{\LARGE{
\centerline{ \bf BPS states, conserved charges and  }
\centerline{\bf  centres of symmetric group algebras  }
}}

\vskip.5cm

\thispagestyle{empty} \centerline{
   {\large \bf  Garreth Kemp${}^{a, }$\footnote{\tt gkemp@uj.ac.za},  Sanjaye Ramgoolam${}^{b,c,}$\footnote{ {\tt s.ramgoolam@qmul.ac.uk}}
       }}

\vspace{.2cm}
\centerline{ {\it ${}^a$ Department of Mathematics and Applied Mathematics, 
}}
\centerline{ {\it University of Johannesburg,
Auckland Park, 2006  }}

\vspace{.4cm}
\centerline{{\it ${}^b$ Centre for Research in String Theory,  },}
\centerline{ {\it School of Physics and Astronomy, Queen Mary University of London} } 
\centerline{ {\it    Mile End Road, London E1 4NS, UK}}

\vspace{.2cm}
\centerline{{\it ${}^c$ National Institute for Theoretical Physics,}}
\centerline{{\it School of Physics and Mandelstam Institute for Theoretical Physics,}}
\centerline{{\it University of the Witwatersrand, Wits, 2050, South Africa } }

\vspace{1.4truecm}

\thispagestyle{empty}

\centerline{\bf ABSTRACT}

\vskip.4cm

In ${\mathcal{N}} =4$ SYM with $U(N)$ gauge symmetry,  the multiplicity of half-BPS states with fixed dimension can be labelled by Young diagrams  and can be  distinguished using 
conserved charges corresponding to Casimirs of  $U(N)$. The information theoretic  study of LLM geometries and superstars in the dual $AdS_5 \times S^5$  background has raised a number of questions about the distinguishability of Young diagrams when a finite set of Casimirs are known. Using Schur-Weyl duality relations between unitary groups and symmetric groups, these questions translate into structural questions about the centres of symmetric group algebras. We obtain analytic and computational results about these structural properties and related Shannon entropies, and generate associated number sequences. A characterization of Young diagrams in terms of  content distribution functions  relates these number sequences to diophantine equations. These content distribution functions can be visualized as connected, segmented,  open strings in content space.

\setcounter{page}{0}
\setcounter{tocdepth}{2}

\tableofcontents

\section{Introduction}

One of the best understood instances of the AdS/CFT correspondence \cite{malda} 
is the duality between $ \cN =4$ SYM with $U(N)$ gauge group and type IIB superstring theory
on $ AdS_5 \times S^5$ with $N$ units of five-form flux. In particular, in the half-BPS sector, giant gravitons \cite{mst} were identified as important non-perturbative objects in the string theory which demonstrate remarkable sensitivity to finite $N$ effects, notably the stringy exclusion principle \cite{malstrom}, in their classical properties. Sub-determinant operators in the CFT were identified as duals for an interesting class of giant gravitons \cite{BBNS}.  The construction of CFT duals of general giant gravitons was obtained by using Young diagrams to organize a finite $N$ orthogonal basis of CFT operators\cite{CJR2001}. An underlying free-fermion description of this sector was identified \cite{CJR2001,Berenstein}.  Recent computations have been able to reproduce giant graviton correlators from the CFT using calculations in space-time \cite{BKYZ1103,CDZ1208,Lin:2012ey,KSY1507} 

Large numbers of giant gravitons back-react on the space-time to produce new geometries with $ AdS_5 \times S^5$ asymptotics. The general smooth half-BPS solutions with these asymptotics were foundby Lin,Lunin and Maldacena (LLM) \cite{LLM}. The LLM picture can also be used to describe singular geometries called superstars and their relation to giant gravitons \cite{MT0109,BJJS0505,BJS0505}.  
 An important feature of the half-BPS sector is the existence of conserved charges, related to the Casimirs of $u(N)$ \cite{CJR2001}. The interpretation of these charges in the space-time picture of LLM geometries and superstars was discussed in \cite{IntInfLoss}. They were found to correspond to multi-pole moments of the  gravitational solutions. It was also argued that measuring all the $N$ independent Casimir eigenvalues needed to determine a geometry requires measurement of the variations of the metric over distances shorter than the Planck distance. This was proposed as an important mechanism of information loss, where the semiclassical observer fails to identify the exact quantum state corresponding to a given geometry. A precise and general mathematical understanding of the limitations of information accessible to the semi-classical observer in gravity can be expected to have deep implications in black hole physics. More recent discussions on this theme include \cite{BerMill1605,BC1102,Simon1810,BBLMPR1810}.

The generalization from half-BPS to sectors with lower symmetry is a problem of multi-matrix quantum field theory combinatorics. Constructions of multi-matrix gauge invariant operators with multiple Young diagrams were given in \cite{BBFH05,BHLN02}. Bases for multi-matrix gauge invariant operators, which diagonalise 
the free field theory inner products, were found in \cite{KR1,BHR1,BHR2,BCD0801,BDS0806}, by using symmetric groups to organize the gauge invariants. The key reason for symmetric groups entering these questions about $U(N)$ gauge invariants is 
Schur-Weyl duality, which relates the action of $U(N)$ on the $n$-fold tensor product of the fundamental $V^{ \otimes n }$ to the action of $ S_n$ - the symmetric group of all permutations of $n$ distinct objects. 
These multi-matrix orthogonal  bases included bases covariant under the global symmetry group, e.g. $U(2)$ for the quarter-BPS and $U(3)$ for the eighth-BPS sector \cite{BHR1,BHR2}, as well as restricted Schur bases with quantum numbers associated with subgroups of $S_n$ \cite{BCD0801,BDS0806}. It was found that the different bases amounted to diagonalizing different sets of ``generalized Casimir operators'' associated with different kinds of 
enhanced symmetries in the free field limit \cite{EHS}. 
The goal of further understanding  the structure of the diagonal bases for multi-matrix systems, and associated expressions for correlators, prompted a systematic study of the structure of permutation algebras related to these bases.  The Wedderburn-Artin decomposition of semi-simple algebras into matrix blocks \cite{PCA,YK1701} was found to illuminate the properties of the permutation algebras underlying multi-matrix orthogonal bases and correlators. These semi-simple algebras are also related to the formulation of the combinatorics of multi-matrix gauge invariants under classical Lie group gauge symmetries in terms of two dimensional topological field theories \cite{QuivCalc,YK1403}.  
 This matrix block structure of permutation algebras underlying the construction of gauge invariants is also applicable to gauge invariants in tensor models \cite{BR1,BR2}. Closely related discussions of 
tensor model correlators are in  \cite{DR1706,DGT1707,DGHM1910,IMM1909,ABD1907}. 
In particular,  Casimir operators have also been studied in the guise of cut-and-join operators  in matrix and tensor models, see for example \cite{IMM1909} and references therein.

In this paper, motivated by the discussion in \cite{IntInfLoss} and the subsequent developments in   the mathematics of the space of gauge invariant operators - particularly the relevance of the structure theory of permutation algebras  -   we initiate a systematic study of the quantitative characterisation of the uncertainty 
in the determination of Young diagram operators in the half-BPS sector, when a finite set of Casimirs is specified. In section \ref{sec:CasimirsChargesMatInvts} we review some key elements of the connections between BPS operators of dimension $n$,  Casimir operators of $U(N)$, and the symmetric group $S_n$ of all permutations of $n$ distinct objects. The group algebra $ \mC ( S_n)$ of formal linear combinations of 
$S_n$ group elements with complex coefficients plays an important role, along with the subspace of this group algebra which commutes with all $ \mC ( S_n)$.  This subspace is a commutative sub-algebra called the centre of $ \mC ( S_n)$, or the central algebra,  and denoted $ \cZ ( \mC ( S_n)) $.   The eigenvalues of Casimirs of $ U(N)$ are related to the normalized characters of central elements in $ \cZ ( \mC ( S_n)) $. 

In section \ref{sec:centreSn}, we consider two linear bases for  $ \cZ ( \mC ( S_n)) $: one corresponds to conjugacy classes of $ S_n$ and another to irreducible representations. As is well-known 
the conjugacy classes correspond to cycle structures of permutations. Thus $ \cZ ( \mC ( S_n ) $ is a vector space of dimension equal to the number of partitions of $n$, denoted $p(n)$, with a commutative  and associative product. A distinguished set of conjugacy classes correspond to permutations have a single cycle of length $k$: the corresponding central element is denoted $T_k$. We prove that for any $n$, the set $\cG_n = \{ T_2 , T_3 , \cdots , T_n \}$ form a generating subspace of the central algebra. This means that by taking linear combinations of  these elements and their products, we can get any element of $ \cZ ( \mC ( S_n) ) $. In fact, for a fixed $n$, we generically only need a subset of $ \cG_n $ to 
generate the central algebra.  The connection between cycle structures and irreps, which may be viewed as a Fourier transform,  leads to a formulation of the distinguishability of Young diagrams in terms of minimal generating  subspaces  $ \cZ ( \mC ( S_n ) ) $. A simple inspection of normalized characters of the cycle operators in irreps shows, for example, that for $n$ up to $5$ and $n=7$, but not $n=6$, $T_2$ alone suffices to generate the centre: in other words, $T_2$ and its powers form a linear basis for $ \cZ ( \mC ( S_n) ) $. This is demonstrated directly by writing out the powers of $T_2$ in terms of 
linear combinations of central elements corresponding to the conjugacy classes.

 In section \ref{sec:exptsdata}, we investigate the dimensions of the subspaces of $ \cZ ( \mC ( S_n ) ) $
generated by $ T_2$, by $T_3$ and by the pair $ \{ T_2 , T_3 \}$. These dimensions, as shown in section \ref{sec:centreSn}, are given respectively by the number of distinct normalized characters $ { \chi_R ( T_2 ) \over d_R } $, the number of distinct ${ \chi_R ( T_3 ) \over d_R } $, and the number of distinct pairs $ \{ {\chi_R ( T_2 )  \over d_R }  , { \chi_R ( T_3 ) \over d_R }  \}$, as $R$ runs over the set of Young diagrams.  In each case, for small enough $n$, there are no degeneracies as $ R $ runs over all the Young diagrams. However as $n$ increases, one or more $R$ give the same normalized character, or list of normalized characters. The distribution of degeneracies can be used to define a probability distribution over the space of possible normalized characters. For each fixed value or list of values,  the Shannon entropy - which is the logarithm of the multiplicity -  gives a measure of the uncertainty associated with having knowledge of the value or value sets but not the exact identity of the Young diagram. Depending on a choice of probability distribution over the spectrum of values we can get an expectation value for this uncertainty associated with multiplicities.  We study two natural ways of averaging this entropy and discuss the data measuring these entropy averages. This gives involves developing an interesting AdS/CFT-based information theoretic physical perspective on mathematical data of fundamental interest, namely normalized characters of specified sets of conjugacy classes in $S_n$.

 In section \ref{sec:CDFs} we define and study  a number sequence $n_{*} ( k)$ : for a given $k$, $n_{*} (k)$ is the smallest $n$, where the normalised characters of $\{ T_2 , \cdots , T_k \} $ or equivalently the Casimirs $C_2 , \cdots , C_k$ fail to distinguish all the Young diagrams. The mathematics literature \cite{Lassalle2008,CGS2004} contains elegant formulae for the normalised characters in terms of {\it content polynomials} which are explained in this section. The transformations between Casimirs, normalized characters, and content polynomials have a useful triangularity property which allows us to compute $n_{*} (k)$ in terms of content polynomials which are efficiently programmable in Mathematica. For $ k = 6$, we find that $n_{*} =80$. 
Our present computational approach becomes prohibitively inefficient  beyond $n = 80$, so the determination of $n_{*} (7)$ is an interesting computational challenge. As a first step in the direction of developing 
analytic approaches to the determination of $n_{*} ( k )$, for large $k$, we introduce a notion of {\it content distribution functions}  which are shown to uniquely characterise Young diagrams: the content polynomials are moments of the content distribution functions.  We express our earlier result about $ \cG_n $ forming a generating set in terms of the content distribution functions and observe that at $ n = n_{*} ( k )$, a set of vanishing moment equations are satisfied by the differences of content distribution functions. These content distribution functions can be visualized as segmented, connected, open strings in content space - which may be useful in the future as a tool to develop new techniques to determine the properties of $ n_{*} ( k )$.  

We conclude with a summary and discussion of future research directions.

\section{ Casimirs, charges and Matrix Invariants  } \label{sec:CasimirsChargesMatInvts}

In this section we recall the definition of the Schur polynomial basis for the half-BPS sector \cite{CJR2001}, where half-BPS operators  are labelled by Young diagrams of $U(N)$ and are linear combinations of multi-traces of one complex matrix $Z$.  Multi-traces with scaling dimension $n$, where each $ Z$ has dimension $1$, are parametrized by permutations which control the contraction of $U(N) $ indices. 
We review how the action of the $U\left(N\right)$ Casimirs on the multi traces can be expressed in terms of central elements of $ \cZ ( \mC ( S_n )) $ acting on the permutations labels \cite{CJR2001,EHS}.  We explain a diagrammatic algorithm for finding the map between the  Casimirs and the central elements. 
We then show that knowledge of the Casimirs $ \{ C_{2}, C_{3}, \cdots, C_{k} \} $ is equivalent to 
knowing the normalized characters for $ \{ T_{2}, T_{3}, \cdots T_{k} \} $.

\subsection{The map from Casimirs to central elements for 1-matrix problem } 

The half-BPS operators in $N=4$ SYM with $U(N)$ gauge group 
are gauge invariant functions of one complex matrix $ Z$ 
transforming in the adjoint of the gauge group
\bea 
Z \rightarrow U Z U^{ \dagger}.
\eea
$Z$ is a quantum field with scaling dimension one. 
The gauge invariant functions are traces and products of traces. By the operator-state correspondence 
of CFT, these correspond to quantum states in CFT and hence quantum states in the AdS. 
For scaling dimension $n \le N$, the linearly independent gauge invariants 
correspond to partitions of $n$. The scaling dimension corresponds to the energy operator for 
translation along the time direction of global coordinates in AdS \cite{Witten}. 
 For example at $n=3$, we have the following basis for gauge invariants 
\bea
\tr Z^3 , \tr Z^2 \tr Z , ( \tr Z )^3
\eea
General multi-trace operators of degree $n$ can be parametrized by permutations $ \sigma $ in $S_n$, 
the symmetric group of all permutations of $ \{ 1, 2, \cdots , n \}$. 
\bea\label{Opsig}  
\cO_{ \sigma } ( Z )   = \sum_{ i_1 , \cdots , i_n } Z^{ i_1}_{ i_{ \sigma (1) } } Z^{ i_2}_{ i_{ \sigma (2)} } \cdots Z^{ i_n }_{ i_{ \sigma (n) } } =  \tr \left (  Z^{ \otimes n } \sigma \right )  
\eea
 In the second expression $ Z^{ \otimes n } $ and $ \sigma $ are both being 
viewed as linear operators on the $n$-fold tensor product $ V_N^{ \otimes n }$  of the fundamental representation $V_N$ of $U(N)$. 

The matrix differential operators $ E^i_{ m} $
\bea 
E^i_m = \sum_{  j=1 }^N  Z^{i}_j { \partial \over \partial Z_j^m} 
\eea
obey the commutation relations of the $ gl(N)$ Lie algebra. 
\bea 
[ E^{i}_j , E^k_l ]  = \delta_j^k E^{i}_l - \delta^i_l E^{k}_j 
\eea
Appropriate anti-hermitian linear combinations generate the $u(N)$ Lie algebra. 
The Casimirs $\tilde C_k$ generate the centre of the enveloping algebra $ U ( u (N) )$. 
\bea\label{CasCk}  
\tilde C_{k } = \sum_{ i_1 , i_2 , \cdots , i_k = 1 }^N E^{i_1}_{ i_2} E^{ i_2}_{ i_3} \cdots E^{i_k }_{ i_1} 
\eea
The commutators of $ E^i_m$ with $Z^{p}_q$ are 
\bea 
[ E^{ i}_m , Z^p_q ] = \delta_m^p E^i_q 
\eea
The lower $q$ index is left invariant, while the upper index transforms as the fundamental representation 
$V_N$. 
The commutator of $ E^i_k$ with a product is 
\bea 
[ E^i_m , Z^{p_1}_{q_1} Z^{p_2}_{q_2} \cdots Z^{p_n}_{q_n} ] 
= \sum_{ j =1}^n \delta_{m}^{p_j} Z^i_{q_j} \prod_{ l \ne j } Z^{p_l}_{q_l} 
\eea
The upper indices $\{ p_1  , p_2 , \cdots , p_n \}$ transform as $ V_{ N}^{ \otimes n }$. 

These equations can be used to show that the Casimirs (\ref{CasCk}) act on the 
operators (\ref{Opsig}) through left multiplication by elements $\hat C_k$ in the central sub-algebra 
$ \cZ ( \mC ( S_n ) )$ 
\bea 
\tilde C_k \cO_{ \sigma } = \cO_{  C_k \sigma } 
\eea
This is explained in more detail in section 3.1  and Appendix A.1 of the paper \cite{EHS}, where the Casimirs are related to Noether charges for an enhanced $U(N) \times U(N)$ symmetry in the free field limit of $ \cN =4 $ SYM. 
 It is shown how the quadratic Casimir of $U(N)$, expressed as a second order matrix differential operator, relates to $T_2$, the  central element of 
$ \cZ ( \mC ( S_n ) ) $  which is related to permutations having  one non-trivial cycle of length $2$. 

This map between $ C_k$, viewed as operators on $ V_{ N}^{ \otimes n }$
 and central elements of the group algebra $ \mC ( S_n )$ has been studied
systematically  in the context of 2d Yang Mills theory \cite{CMR,GSY}.
For example, we have the results 
\bea\label{CtoT}  
&& C_2 =  \sum_{ r \ne s } ( r s ) + N n  =  2 T_2 + N n  \cr 
&& C_3 = 3 T_3 + 4 N T_2 + N^2 n + n ( n-1) \cr 
&& C_4 = 4 T_4 + 9 N T_3 +  ( 6 N^2 + ( 6n -10 ) ) T_2 + 3 N n ( n-1) + N^3 n \cr
&& C_5 = 5 T_{5} + 16 T_{(2,2)} + 16N T_4 + (18N^2+12(n-3) + 15)T_3 + (4N^3 + 24N(n-2))T_2 \cr
&& \hspace{30pt}+ (nN^4 + 3N^2 n(n-1) + 2n(n-1)(n-2))
\eea
Note the appearance of $T_{(2,2)}$ in $C_{5}$. For $C_6$, the central elements $T_{(2,2)}$ and $T_{(3,2)}$ will appear.

An orthogonal basis in the free field inner product is parametrised by Young diagrams \cite{CJR2001}
\bea 
\cO_{ R } && = { 1 \over n!  } \sum_{ \sigma \in S_n } \chi_R ( \sigma ) \cO_{ \sigma }  ( Z )  \cr  
&& = \sum_{ p \vdash n } { 1 \over Sym ( p ) } \prod_{ i } (  \tr Z^i )^{ p_i } 
\eea
These are referred to as Schur Polynomial operators of the half-BPS sector.
The commutation relations  of the Casimirs with the gauge invariant operators can be read off from the 
Casimir to central elements transformations. 
For example 
\bea 
[ C_2 , \cO_{ R } ]  && =   \left ( 2{  \chi_R  \over d_R } ( T_2 ) + Nn   \right ) \cO_R  \cr 
[ C_3 , \cO_{ R  } ] && =     
\left (  3 { \chi_R ( T_3 ) \over d_R } + 4 N {  \chi_R  \over d_R } ( T_2 ) +  N^2 n + n ( n-1) \right ) 
\cO_R 
\eea
In  the above examples, we see that knowing the degree $n$ and the normalized character of $T_2$
is equivalent to knowing $C_2$. Knowing $ n$ along with the normalized characters of 
$ T_2 , T_3 $ is equivalent to knowing $ C_2 , C_3 $. We will prove the following theorem.

{ \bf Theorem 2.1.1} The Casimir operators  $ C_2  , \cdots , C_k $ in $V_N^{ \otimes n}$ can be expressed in terms of  $T_2 , T_3, \cdots , T_k $.  This relies on the form of the transformations between the Casimirs 
and the central elements.\\
To prove this theorem express the Casimir $C_{k}$, acting on $ V_N^{ \otimes n }$  in the form 
\bea
	C_{k} = \sum\limits^{n}_{r_{1},r_{2},\cdots r_{k} = 1} \rho_{r_{1}}\left( E^{i_{1}}_{i_{2}} \right)\rho_{r_{2}}\left( E^{i_{2}}_{i_{3}} \right) \cdots \rho_{r_{k}}\left( E^{i_{k}}_{i_{1}} \right).
\label{eq:Ckcasimir}
\eea
where the $r_{i}$ label the different factors of $ V_N^{ \otimes n }$, and $ \rho_{ r_i } ( E^{j_1}_{j_2}  ) $ is the linear operator of $E^{j_1}_{j_2} $ acting on the $r_i $'th factor. 
There is a diagrammatic algorithm for converting the generating Casimirs to 
central class operators \cite{CMR,GSY}.   We will review this algorithm and use it to prove the Theorem.  
An immediate corollary given the above discussion is that the eigenvalues of the Casimir operators
$ C_2 (R)  , \cdots , C_k(R) $  
on the Young diagram operators  $\cO_R $ is determined in terms of  the normalised characters $ { \chi_R  ( T_2)\over d_R } , \cdots , { \chi_{R} ( T_k ) \over d_R } $.  


We now describe the diagrammatic algorithm for $C_{k}$. Draw a circle with $k$ crosses, labeled 1 to $k$ with orientation as shown in figure \ref{fig:Ck}. The sum over the $r$ indices can be separated into different coincidences between the $ \{ r_{1} , r_{2} , r_{3} , \cdots, r_{k} \}$. This is a sum over set partitions: partitions of the set $\{ 1,2,3, \cdots ,k \}$ into collections of subsets \cite{Wiki-set-partitions}. Thus, a particular contribution to $C_{k}$ will be a partition of  the set $\{ 1,2,3, \cdots ,k \}$ into $p$ subsets, where $p\leq k$. The total number of set partitions of $k$ elements into $p$ subsets is given by Stirling's number of the second kind \cite{Stirling-second}. All the crosses labeled by the elements in a particular subset are joined with a chord or a line. To each of the lines, apply the following procedure.  Thicken the line, let the cross with the smaller label disappear, and let the cross with the largest number slide along  the graph in the direction of the orientation to join the edge of the  thickening - this will be illustrated in examples shortly. It is a diagrammatic translation of the multiplication of $E$ operators acting on the same $V_N$ : two crosses correspond to two $E$ operators, their multiplication produces a single $E$ and a $ \delta $ function which results in a reconnection of index lines.    After this operation is applied to all the chords, the effect of this is to separate the graph into a set of  loops. In general a graph will separate into a loop with $k'_{1}$ crosses, a loop with $k'_{2}$ crosses and so on, which we denote by   $D_{k'_{1}, k'_{2},\cdots}$. The central element is obtained from $D_{k'_{1}, k'_{2},\cdots}$ by retaining all the $k'_{i}>1$. Relabel the $k'_{i}>1$ with $k_{i}$ and drop the $k'_{i}$ which are equal to $  0$ or $1$. We thus obtain $T_{(k_{1},k_{2},\cdots)}$. Let $n_{0}$ be the number of loops with zero crosses and let $n_{1}$ be the number of loops with one cross. There is a factor of $N^{n_{0}}$ and a multiplicative factor of
\bea
	\prod^{n_{1}-1}_{i=0}\left( n - \left( k_{1}+k_{2}+\cdots  \right) - i \right).
\label{eq:n1counting}
\eea
There is one last symmetry factor generated. Consider $T_{(k_{1},k_{2},\cdots)}$. If $m_{\mu}\, (\mu>1)$ is the number of $k_{i}$ values equal to $\mu$, we obtain a multiplicative factor of
\bea
{ \rm{ Symm} }  ( \vec k ) = 	\prod\limits_{\mu > 1}\mu^{m_{\mu}}m_{\mu}!
\label{eq:Symmfactor}
\eea
which accounts for the cyclic variations and permutations of all the $r_{i}$ leaving the overall contribution to $C_{k}$ invariant. Thus the formula for $C_k$ takes the form 
\bea 
C_k = \sum_{ \rm { set ~ partitions ~ of ~ \{ 1, 2, \cdots , k \} } }
N^{ n_0 } \left ( \prod^{n_{1}-1}_{i=0}\left( n - \left( k_{1}+k_{2}+\cdots   \right ) - i \right)  \right ) ~ { \rm{ Symm} }  ( \vec k )  ~  T_{ k_1 , k_2 , \cdots } 
\eea

\begin{center}
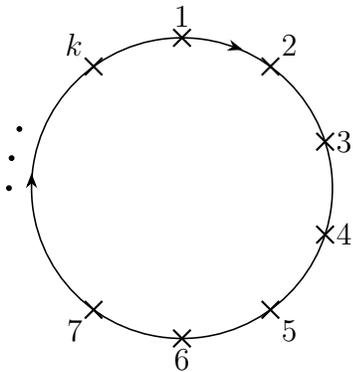
\begin{figure}[h!]
\begin{tikzpicture}[semithick, >=Stealth]
\coordinate (A0) at (-7.5,0);
\filldraw[black] (A0) circle (0pt) ;
\coordinate (A) at (0,0);
\path (A) + (90:2) coordinate (A1); 
\path (A) + (54:2) coordinate (A2); 
\path (A) + (18:2) coordinate (A3);
\path (A) + (-18:2) coordinate (A4);
\path (A) + (-54:2) coordinate (A5);
\path (A) + (-90:2) coordinate (A6);   
\path (A) + (-126:2) coordinate (A7);
\path (A) + (-162:2) coordinate (A8);  
\path (A) + (-198:2) coordinate (A9);   
\path (A) + (-234:2) coordinate (A10);
\path (A) + (-180:2.3) coordinate (A11);
\path (A) + (-190:2.3) coordinate (A12);
\path (A) + (-200:2.3) coordinate (A13);
\draw[
        decoration={markings, mark=at position 0.2with {\arrow{<}}},
        decoration={markings, mark=at position 0.5with {\arrow{<}}},
        postaction={decorate}
        ]
	 (A) circle [radius = 2]; 
\draw (A1) node[cross] {};
\draw (A2) node[cross] {};
\draw (A3) node[cross] {};
\draw (A4) node[cross] {};
\draw (A5) node[cross] {};
\draw (A6) node[cross] {};
\draw (A7) node[cross] {};
\draw (A10) node[cross] {};
\filldraw[black] (A11) circle (0.8pt) ;
\filldraw[black] (A12) circle (0.8pt) ;
\filldraw[black] (A13) circle (0.8pt) ;
\node[above] at (A1) {$1$};
\node[above right] at (A2) {$2$};
\node[right] at (A3) {$3$};
\node[right] at (A4) {$4$};
\node[below right] at (A5) {$5$};
\node[below] at (A6) {$6$};
\node[below left] at (A7) {$7$};
\node[above left] at (A10) {$k$};
\end{tikzpicture}
\caption{To compute $C_{k}$, draw a circle with $k$ crosses labeled 1 to $k$. Each cross represents a $E$ in equation (\ref{eq:Ckcasimir}). }
\label{fig:Ck}
\end{figure}
\end{center}
We study some examples now. For $k=2$, we are computing $C_{2}$. There are only two partitions of $\{1,2\}$ to sum over. They are $\{1,2\}$ and $\{1|2\}$. The first is a partition of the set into one subset and the second is a partition of the set into two subsets. The first corresponds to the case where $r_{1} = r_{2}$. Here crosses 1 and 2 are joined by a single chord. The second partition corresponds to the case where $r_{1} \neq r_{2}$ and the crosses are not joined. Thicken the chord in the first graph, erase the cross labelled $1$ and slide the cross labelled 2 along the graph in the direction of the orientation. The result is two separate loops - one with $k'_{1} = 1$ and the other with $k'_{2} = 0$. Thus, we get a $D_{1,0}$. Furthermore, $n_{0} = 1$ and $n_{1} = 1$ for this graph. If none of the $k'$ are bigger than 1, which is the case here, we write the central element labeled by the identity $T_{1}$. Since $n_{0}=1$, there is a factor of $N$. Since $n_{1} = 1$ and $k_{i} = 0$, we obtain the factor $n$. The overall result of this graph is $nNT_{1}$. This is shown in figure \ref{fig:C2}. The second graph gives a $D_{2}$, where $k'_{1} = 2$. Both $n_{0} = n_{1} = 0$ meaning that we have a single loop with two crosses and no factors of $N$ or $n$. Since $k'_{1} = 2$, we relabel to $k_{1} = 2$ and we finally obtain $2T_{2}$. The factor of $2$ comes from applying formula (\ref{eq:Symmfactor}).
\begin{center}
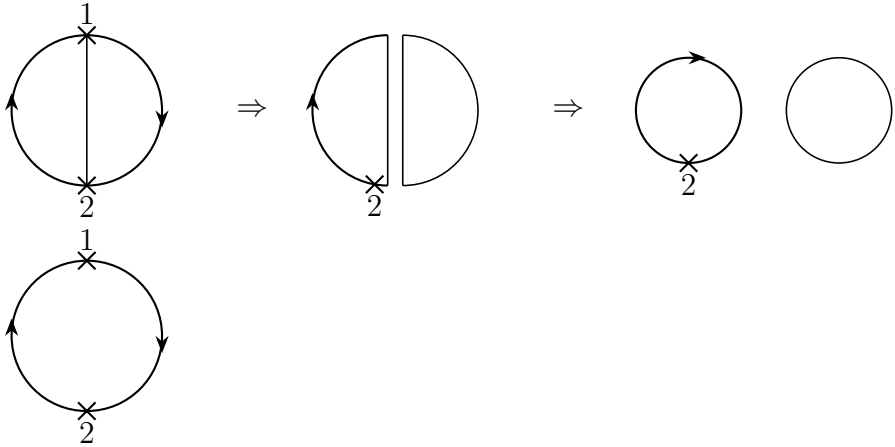
\begin{figure}
\begin{tikzpicture}[semithick, >=Stealth]
\coordinate (E) at (0,0);
\path (E) + (90:1) coordinate (E1); 
\path (E) + (270:1) coordinate (E2);
\path (E) + (0:1) coordinate (E3); 
\draw[thick,
        decoration={markings, mark=at position 0with {\arrow{<}}},
        decoration={markings, mark=at position 0.5with {\arrow{<}}},
        postaction={decorate}
        ]
        (E) circle [radius = 1]; 
\draw (E1) node[cross] {};
\draw (E2) node[cross] {};
\draw (E1)--(E2);
\node[above] at (E1) {$1$};
\node[below] at (E2) {$2$};
\path (E3) + (0:1.2) coordinate (E4); 
\node[] at (E4) {$\Rightarrow$};
\coordinate (G) at (4,0);
\coordinate (H) at (4.2,0);
\path (G) + (90:1) coordinate (G1); 
\path (G) + (270:1) coordinate (G2);
\path (G) + (260:1) coordinate (G3);
\path (H) + (90:1) coordinate (H1); 
\path (H) + (270:1) coordinate (H2);
\path (H) + (0:1) coordinate (H3);
\draw (H1) arc (90:-90:1);
\draw[thick,
        decoration={markings, mark=at position 0.5with {\arrow{<}}},
        postaction={decorate}
        ]
(G1) arc (90:270:1);
\draw (G3) node[cross] {};
\node[below] at (G3) {$2$};
\draw (G1) -- (G2);
\draw (H1) -- (H2); 
 \path (H3) + (0:1.2) coordinate (H4); 
\node[] at (H4) {$\Rightarrow$};
 \coordinate (I) at (8,0);
\path (I) + (270:0.7) coordinate (I1); 
\draw[thick,
        decoration={markings, mark=at position 0.25with {\arrow{<}}},
        postaction={decorate}
        ]
(I) circle [radius = 0.7]; 
\draw (I1) node[cross] {};
\node[below] at (I1) {$2$};
\coordinate (H) at (10,0);
\draw (H) circle [radius = 0.7];
 \coordinate (F) at (0,-3);
\path (F) + (90:1) coordinate (F1); 
\path (F) + (270:1) coordinate (F2);
\draw (F1) node[cross] {};
\draw (F2) node[cross] {};
\draw[thick,
        decoration={markings, mark=at position 0with {\arrow{<}}},
        decoration={markings, mark=at position 0.5with {\arrow{<}}},
        postaction={decorate}
        ]
        (F) circle [radius = 1]; 
\node[above] at (F1) {$1$};
\node[below] at (F2) {$2$};
\end{tikzpicture}
\caption{Computing $C_{2}$. The above graphs correspond to the two possible set partitions for $\{1,2\}$. The first graph depicts the case when $r_{1}$ and $r_{2}$ coincide. Thus, these two crosses are joined. After thickening the chord and erasing the cross with the smaller label, the graph splits into a loop with one cross labeled 2 and a loop with no crosses. The second graph depicts the case when $r_{1} \neq r_{2}$. Here, there is no joining of crosses and the graph just remains as is.} 
\label{fig:C2}
\end{figure}
\end{center}
For $k=5$, there are 52 set partitions in total to sum over, given by the Bell number $B_{5}$. One example is $\{1|234|5\}$. This corresponds to $r_{1}\neq r_{2} = r_{3} = r_{4} \neq r_{5}$. Crosses 2,3 and 4 are joined by a single chord. After thickening the chord, erasing crosses 2 and 3 and sliding 4 along the graph toward 5, the graph splits into 3 loops where $k'_{1} = 3, k'_{2} = 0$ and $k'_{3} = 0$. Thus, we get $D_{3,0,0}$, which leads to $T_{(3)}$. Here, $n_{0} = 2$ and $n_{1} = 0$. This contributes a factor of $N^{2}$. We finally obtain a contribution of $3NT_{3}$ to $C_{5}$ from this partition. The factor of $3$ is again obtained from (\ref{eq:Symmfactor}). This example is illustrated in figure \ref{fig:C5example}.
 \begin{center}
\begin{figure}
\begin{tikzpicture}[semithick, >=Stealth]
\coordinate (A) at (0,0);
\path (A) + (90:2) coordinate (A1); 
\path (A) + (18:2) coordinate (A2); 
\path (A) + (-54:2) coordinate (A3);
\path (A) + (-126:2) coordinate (A4);
\path (A) + (-198:2) coordinate (A5);
\path (A) + (-180:2.3) coordinate (A11);
\path (A) + (-190:2.3) coordinate (A12);
\path (A) + (-200:2.3) coordinate (A13);
\draw[
        decoration={markings, mark=at position 0.18with {\arrow{<}}},
        decoration={markings, mark=at position 0.5with {\arrow{<}}},
        postaction={decorate}
        ]
	 (A) circle [radius = 2]; 
\draw (A1) node[cross] {};
\draw (A2) node[cross] {};
\draw (A3) node[cross] {};
\draw (A4) node[cross] {};
\draw (A5) node[cross] {};
\node[above] at (A1) {$1$};
\node[above right] at (A2) {$2$};
\node[below right] at (A3) {$3$};
\node[below left] at (A4) {$4$};
\node[above left] at (A5) {$5$};
\draw [thick] (A2) to [out=200,in=115] (A3);
\draw [thick] (A3) to [out=120,in=60] (A4);
\path (A2) + (0:1.25) coordinate (A6); 
\node[] at (A6) {$\Rightarrow$};
\coordinate (B) at (6,0);
\path (B) + (90:2) coordinate (B1); 
\path (B) + (18:2) coordinate (B2); 
\path (B) + (-54:2) coordinate (B3);
\path (B) + (-136:2) coordinate (B4);
\path (B) + (-198:2) coordinate (B5);
\coordinate (B6) at ([shift=(15:2cm)]B);
\draw[thick,
       decoration={markings, mark=at position 0.18with {\arrow{<}}},
        postaction={decorate}
        ]
(B6) arc (15:235:2cm) coordinate (C);
\draw (B1) node[cross] {};
\draw (B4) node[cross] {};
\draw (B5) node[cross] {};
\node[above] at (B1) {$1$};
\node[below left] at (B4) {$4$};
\node[above left] at (B5) {$5$};
\draw [thick] (C) to [out=60,in=115] (B3);
\draw [thick] (B6) to [out=190,in=120] (B3);
\coordinate (B7) at ([shift=(270:0.2cm)]B6);
\coordinate (B8) at ([shift=(0:0.2cm)]B3);
\draw [thick] (B7) to [out=190,in=120] (B8);
\draw [thick] (B7) to [out=300,in=30] (B8);
\coordinate (B9) at ([shift=(-15:0.2cm)]C);
\coordinate (B10) at ([shift=(220:0.2cm)]B3);
\draw [thick] (B9) to [out=45,in=130] (B10);
\draw [thick] (B9) to [out=310,in=240] (B10);
\path (B2) + (0:1.25) coordinate (B11); 
\node[] at (B11) {$\Rightarrow$};
\coordinate (E) at (11,0);
\path (E) + (90:1) coordinate (E1); 
\path (E) + (-30:1) coordinate (E2);
\path (E) + (210:1) coordinate (E3);
\draw[thick,
       decoration={markings, mark=at position 0.18with {\arrow{<}}},
        postaction={decorate}
        ]
(E) circle [radius = 1]; 
\draw (E1) node[cross] {};
\draw (E2) node[cross] {};
\draw (E3) node[cross] {};
\node[above] at (E1) {$1$};
\node[below right] at (E2) {$4$};
\node[below left] at (E3) {$5$};
\draw[thick] ([shift=(0:2.35cm)]E) circle [radius =0.8];
\draw[thick] ([shift=(0:4.25cm)]E) circle [radius =0.8];
\end{tikzpicture}
\caption{{\small{Computing the contribution of $r_{1}\neq r_{2} = r_{3} = r_{4} \neq r_{5}$ to $C_{5}$. Crosses 2,3 and 4 are joined as shown on the left. After thickening the chord joining these crosses, erasing the smaller labels 2 and 3 and sliding 4 along toward 5, this diagram splits into three disconnected pieces. Following the recipe converts this term in the Casimir sum into $3N^{2}T_{(3)}$.} }}
\label{fig:C5example}
\end{figure}
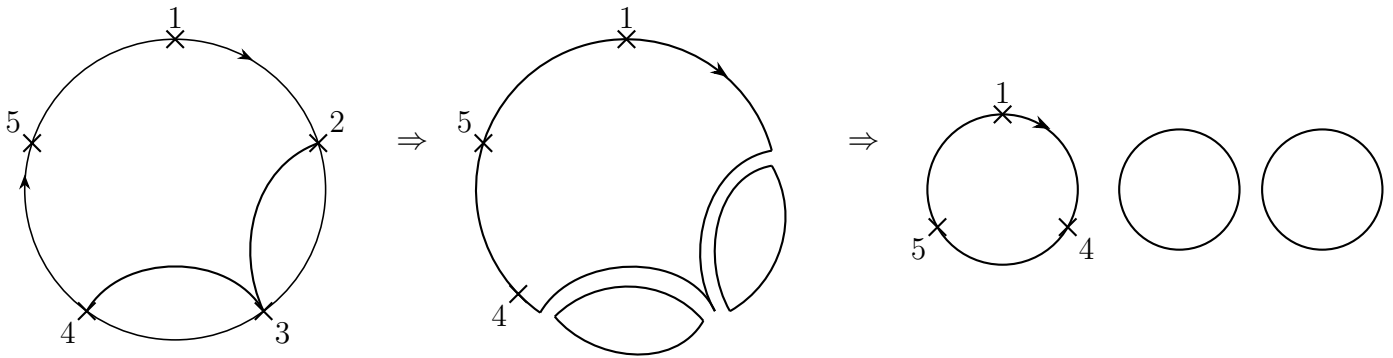
\end{center}
The key fact that we make use of is the following. When we sum over the different set partitions of $ \{ 1, 2, \cdots ,k \}$, the case 
$ r_1 \ne r_2 \cdots \ne r_k$ leads to $ kT_k$ as described by the diagrammatic rules. This has branching number $k -1$, where branching number is defined in equations (\ref{eq:branchingnum1}) and (\ref{eq:branchingnum2}). Any other set partition produces something of lower branching number. Suppose the set  $\{ r_1 ,  \cdots , r_k \}  $ is divided into $p$ disjoint subsets. Within each subset we have $r$'s which coincide. Setting aside the case where we get $ k T_k$, we have $ p < k$. After we multiply out the $E$'s within a subset, we get a single $E$. The total number of $E$'s left is $ p$ which is also equal to the sum of all the $k'_{i}$ values. Some of the $k'_{i}$ could equal 1 which corresponds to an $E^{i}_{i}$, which is the identity. These are not included when writing the $T$ operators. The conjugacy class we get is generally of the form $ T_{ k_1 , k_2 , \cdots , k_l  }$ where $k_1, \cdots k_l $ are positive integers larger than one. The remaining copies of $E$ are collected into $l$ cyclic collections. We thus have
\bea 
p \geq k_1 + k_2 + \cdots k_l.
\eea
The branching number is $ k_1 + k_2 + \cdots + k_l - l  \leq p - l < k -1 $. But $ T_2 , \cdots , T_{k-1}$ generate all the $T$ operators for branching number less than $k-1$, a result (Theorem 3.2.1) we prove in section \ref{sec:T1toTngens}. Thus $C_k$ can be expressed in terms of $T_k$ along with products involving $T_{ i }$ for $ i \le k-1$.  This means that knowing  the normalized characters of $ T_2 , \cdots , T_k$ for any Young diagram $R$ 
is equivalent to knowing the Casimir eigenvalue of $ C_2 , \cdots , C_k$  for the  Young diagram $R$.

\section{ $ \cZ ( \mC ( S_n ) $ : The centre of $ \mC ( S_n ) $. } \label{sec:centreSn} 
In this section, we consider the centre of the group algebra $\mC ( S_n )$,denoted as  $\cZ ( \mC ( S_n ) )$. First, we identify a basis for $\cZ(\mC ( S_n ) )$ from conjugacy classes labeled by partitions of $n$. Next, we show that a certain subset $\mathcal{G}_{n}$ of these basis elements is capable of generating $\cZ ( \mC ( S_n ) )$. However, at a given $n$ not all of the elements of $ \mathcal{G}_{n}$  are needed to generate $\cZ ( \mC ( S_n ) )$. Another useful basis for the centre comes from projectors associated with irreducible representations (irreps) of $S_{n}$. In the irrep basis we develop criteria for when elements of $\mathcal{G}_{n}$ generate $\cZ ( \mC ( S_n ) )$. Lastly, we present  an explicit  non-trivial example where a single element of $\mathcal{G}_{n}$ generates the centre.

\subsection{A linear basis for $ \cZ ( \mC ( S_n ) $  from conjugacy classes.}

It is well-known that the partitions of $n$ label the conjugacy classes in $S_{n}$. In particular $\lambda \vdash n$ labels the conjugacy class of permutations with cycle structure $\lambda$. Identify $T_{\lambda}$ with the formal sum over all elements of the conjugacy class $\lambda$ with equal coefficient. The elements $T_{\lambda}$ form a basis for the centre $\cZ ( \mC ( S_n ) ) $.
%
Consider cycle structures of the form $ [ k , 1^{n-k}] $, with one cycle of length $k$ and remaining cycles of length $1$.  
Denote the sum of permutations, in the group algebra $ \mC ( S_n ) $, with this cycle structure as $T_k$. 
So, for example 
\bea 
T_{ 2} &=&  \sum_{ i < j =1}^n ( i j ) \cr 
T_3 &=&  \sum_{ i < j < k } ( i j k ) + ( i kj ).
\eea
These cycle operators play an important role in this paper. 

\subsection{ Proving that $T_2 , T_3 , \cdots T_n  $ generate the centre  of $ \cZ ( \mC ( S_n )$ } \label{sec:T1toTngens}

In this section, we prove that the central elements $\{T_{2}, \cdots T_{n} \}$ generates the centre of the group algebra $ \mC (S_{n})$.

It is convenient to first define the branching number for the permutation $\sigma$ corresponding to modified cycle type $\lambda$. Say $\sigma \in S_{n}$ has a cycle type $\rho = \left( \rho_{1}, \rho_{2}, \cdots , \rho_{k} \right)$, i.e. $k$ cycles of length $ \rho_1 , \rho_2 \cdots \rho_k $ such that 
\bea 
\sum_{ i =1}^k \rho_i = n 
\eea
which is denoted as $\rho \vdash n$. The modified cycle type of $\rho$ is defined as  $\lambda = \left( \rho_{1} - 1, \rho_{2}-1, \cdots , \rho_{k}-1 \right)$ \cite{Mac}. Then the branching number is defined by
\bea
\label{eq:branchingnum1}
B(\sigma) &=&  \sum_{  \rm  { cycles } } ( ( {\rm { Cycle ~~ Length} } ) -1  ) = n - k  \cr 
\label{eq:branchingnum2}
	&=& \sum_{ i  } \lambda_i = | \lambda  | 
\eea
Define $C_{\lambda}$ to be the set of all permutations $\omega$ whose modified cycle type is the partition $\lambda$. For each partition $\lambda$, let $\mathcal{C}_{\lambda}$ denote the sum of all $\omega \in S_{n}$ whose modified cycle type is $\lambda$. For example, take $\mathcal{C}_{\left(2\right)}$. We have $\mathcal{C}_{\left(2\right)} = \sum\limits_{\omega \in S_{n} \bigcap C_{\left(2\right)}} \omega$. So, if $n = 10$, 
\bea
	\mathcal{C}_{\left(2\right)} = \left(1,2,3\right) + \left(1,3,2\right) + \cdots + (8,9,10).
\eea
Thus $ \mathcal{C}_{\left(2\right)} $ is equal to $T_3$; but it is convenient here to work with a notation that uses the reduced cycle type.  The set $\{T_{2}, \cdots , T_{n}\}$ has branching numbers $B = 1, \cdots , n-1$. Thus, the branching number of the $T_{i}$ can be read off from the labels of the corresponding $\mathcal{C}_{i-1}$.\\

{\bf Theorem 3.2.1 } Given the set of central elements in $ \cZ ( \mC ( S_n ))$,   $\mathcal{G}_{k} = \left\{ \mathcal{C}_{\left(1\right)}, \mathcal{C}_{\left(2\right)}, \cdots \mathcal{C}_{\left(k\right)} \right\}$, any $\mathcal{C}_{\lambda}$, where $\lambda$ is any partition such that $\left|\lambda \right| \leq k$ can be written in terms of linear combinations of products of elements in the set. 

The statement that $\{ T_2  , \cdots , T_n \} $ generate $ \cZ ( \mC ( S_n ) )$ is an immediate corollary.  

 We make use of the following result from \cite{Mac}  about the product of  central elements 
\bea
	\mathcal{C}_{\lambda}\cdot\mathcal{C}_{\mu} = \sum\limits_{\nu}a^{\nu}_{\lambda \mu}\mathcal{C}_{\nu},
	\label{eq:MacDMult}
\eea
where the coefficients $a^{\nu}_{\lambda \mu} = 0$ unless $\left| \nu \right| \leq \left| \lambda \right| + \left| \mu \right| $ \footnote{Note that the interpretation of permutations in terms of branched covers which plays an important role in the string theory of 2D Yang Mills \cite{grta} allows a physical interpretation of this inequality. $ \mu , \lambda $ describe the branching over two branch points. 
If we let the two branch points collide to have a single branching described by $ \nu$, the change in branching number due to the collision $ |\lambda| + |\mu| - |\nu| $ must be non-negative since this is accounted by the formation of a  number, positive or zero,  of collapsed handle singularities as a result of the collision. In the process of collision the Euler characteristic of the covering surface does not change, but contributions to the Riemann Hurwitz formula from branch points are traded for contributions form collapsed handles}. In what follows, we frequently consider the case, $\mathcal{C}_{\lambda} \cdot \mathcal{C}_{\left(r\right)}$, $r \geq 1$ and $\left| \lambda \right| + r = \left| \nu \right|$. We have the conditions \cite{Mac}
  \bea
  	a^{\nu}_{\lambda \left(r\right)} &=& 0 \hspace{20pt}  \mathrm{unless}\; \nu \geq \lambda \bigcup \left(r\right),\\
	\label{eq:coeffofC1}
  	a^{ \lambda \bigcup \left(r\right)}_{\lambda \left(r\right)} &>& 0.
	\label{eq:coeffofC2}
 \eea
In the first point, the so-called natural ordering of partitions is being used. A partition $ \lambda $ can be described by the sequence of its parts, listed in weakly decreasing order
$ ( \lambda_1 , \lambda_2 , \cdots , \lambda_r )$: 
\bea 
\lambda_1 \ge \lambda_2 \ge \cdots \ge \lambda_r.
\eea
Given two partitions $ \lambda = ( \lambda_1 , \lambda_2 , \cdots , \lambda_r )$ and $ \mu = ( \mu_1 , \mu_2 , \cdots , \mu_s ) $, the natural ordering is defined by saying that 
$ \mu \ge \lambda $ if
\bea 
\mu_1 + \cdots + \mu_k \ge \lambda_1 + \cdots + \lambda_k \hbox{ for all } k \ge 1.
\eea
In this definition partitions are extended by zero parts if necessary. This is also called dominance order. Taking the transpose of partitions reverses the dominance order. In other words, if $ \mu \ge \lambda $ then $\lambda^T \ge \mu^T$. We now present examples for small values of $k$:
\begin{itemize}
  \item When $k = 1$, we have $\mathcal{G}_{k} = \left\{ \mathcal{C}_{\left(1\right)}\right\}$. All possible $\mathcal{C}$'s are already in our generating set. 
  \item For $k = 2$, we have  $\mathcal{G}_{k} = \left\{\mathcal{C}_{\left(1\right)}, \mathcal{C}_{\left(2\right)}\right\}$. The partitions $\nu$ with $|\nu| = 2$ are $\left( 2 \right)$ and $\left( 1,1 \right)$. $\mcal{C}_{\left(2\right)}$ is already in our generating set. Let's check that $\mcal{C}_{\left(1,1\right)}$ is generated by $\mcal{G}_{2}$. 
  \bea
	\mcal{C}_{\left(1,1\right)} = \sum\limits_{\omega \in S_{10}\bigcap C_{\left(1,1\right)}}\omega  = \left( 1,2 \right) \left( 3,4 \right) + \left( 1,3 \right) \left( 2,4 \right) + \cdots + \left( 7,8 \right) \left( 9,10 \right).
  \eea
Using equation (\ref{eq:MacDMult}), we have 
  \bea
  	\mcal{C}_{\left(1\right)}\cdot\mcal{C}_{\left(1\right)} = \sum\limits_{\nu} a^{\nu}_{\left(1\right)\left(1\right)}\mcal{C}_{\nu}, \hspace{20pt} \left| \nu \right| \leq 2.
  \eea
  The terms for which $\left|\nu\right| < 2$ are already generated by $\mcal{G}_{1}$ which is contained in $\mcal{G}_{2}$. When $\left|\nu\right| = 2$, we need to sum over the $\nu$ for which $\nu \geq \lambda \bigcup \left(r\right)$. Here $\lambda = \left(1\right)$ and $\left(r\right) = \left(1\right)$. So $ \lambda \bigcup \left(r\right) = \left(1,1\right)$. We have:
	\bea
		\left(2\right) &\geq& \left(1,1\right)\\
		\left(1,1\right) &=& \left(1\right) \bigcup \left(1\right)
	\eea
	where the possible $\nu$'s are on the left hand side. Thus
	\bea
		\mcal{C}_{\left(1\right)}\cdot\mcal{C}_{\left(1\right)} = a\, \mcal{C}_{(2)} + b \,\mcal{C}_{\left(1,1\right)} + \sum\limits_{\nu\;\left(|\nu|\leq 1\right)} x_{\nu}\, \mcal{C}_{\nu}.\\
	\eea
	Checking this explicitly for $n=10$
	\bea
	\mcal{C}_{\left(1\right)}\cdot\mcal{C}_{\left(1\right)} = 3 \mcal{C}_{(2)} + 2 \,\mcal{C}_{\left(1,1\right)} + 45\mcal{C}_{\left(0\right)}.
	\eea
	The coefficient of $ \mcal{C}_{\left(1,1\right)}$ is always non-zero from equation (\ref{eq:coeffofC2}). This can also be seen from the diagrammatic algorithm described earlier. The algorithm for converting $ C_k$ into $T_{ k_1, k_2 \cdots }$ can also be used to multiply the $T$'s. The term $ \mcal{C}_{\left(1,1\right)}  $ results, in the diagrammatic algorithm, from the diagram with zero lines joining crosses from one  $ \mcal{C}_{\left(1,1\right)}  $ to the other. 
	\item $k = 3$, $\mathcal{G}_{k} = \left\{\mathcal{C}_{\left(1\right)}, \mathcal{C}_{\left(2\right)}, \mcal{C}_{\left(3\right)}\right\}$. The partitions with $\left| \nu \right| = 3$ are $\{ (3), (2,1) , (1,1,1) \}$. We need to check that $\cC$ labeled by each of these partitions is generated by $\mcal{G}_{3}$. All terms for which $|\nu|\leq 2$ are generated by $\mcal{G}_{2}$ which is contained in $\mcal{G}_{3}$. Now, $\mcal{C}_{(3)} \in \mcal{G}_{3}$ already. According to natural ordering
  \bea
  	(3) \geq (2,1) \geq (1,1,1).
	\label{eq:natorder3}
  \eea 
Thus, the next largest from $(3)$ is $(2,1)$. From equations (\ref{eq:MacDMult}) - (\ref{eq:coeffofC2}), we consider $\mcal{C}_{(2)} \cdot \mcal{C}_{(1)}$. Both $\cC_{(1)}$ and $\cC_{(2)}$ are contained in $\mcal{G}_{2}$. We see that $\mcal{C}_{(2)} \cdot \mcal{C}_{(1)}$ will contain $\mcal{C}_{(3)}$ and $\mcal{C}_{(2,1)}$. The only other terms will be $\mcal{C}_{\nu}$ such that $|\nu| \leq 2$, which are generated by $\mcal{G}_{2}$. The next largest partition is $(1,1,1)$. Thus, we consider $\mcal{C}_{(1,1)}\cdot \mcal{C}_{(1)}$. Again, both $\cC_{(1)}$ and $\cC_{(1,1)}$ are generated by $\mcal{G}_{2}$. From the ordering in (\ref{eq:natorder3}) and from equations (\ref{eq:MacDMult}) - (\ref{eq:coeffofC2}), $\mcal{C}_{(1,1)}\cdot \mcal{C}_{(1)}$ will contain $\mcal{C}_{(1,1,1)}$ and may contain $\mcal{C}_{(3)}$, $\mcal{C}_{(2,1)}$ along with $\mcal{C}_{\nu}$ for which $|\nu| \leq 2$. For $n = 10$,
  \bea
	\mcal{C}_{\left(1,1\right)}\cdot\mcal{C}_{\left(1\right)} = 2\, \mcal{C}_{(3)} + 3 \,\mcal{C}_{\left(2,1\right)} + 3\,\mcal{C}_{\left(1,1,1\right)} + 28\,\mcal{C}_{(1)}.
	\eea
	
	\item For $k = 4$, we have $\mathcal{G}_{k} = \left\{ \mathcal{C}_{\left(1\right)}, \mathcal{C}_{\left(2\right)}, \mcal{C}_{\left(3\right)}, \mcal{C}_{\left(4\right)}\right\}$. The natural ordering for partitions of 4 is a total ordering:
	\bea
		(4) \geq (3,1) \geq (2,2) \geq (2,1,1) \geq (1,1,1,1).
		\label{eq:natordn4}
	\eea
	Again, we need to check that $\cC$ labeled by each of these partitions is generated by $\mcal{G}_{4}$. All terms for which $|\nu|\leq 3$ are generated by $\mcal{G}_{3}$ which is contained in $\mcal{G}_{4}$. Now, $\mcal{C}_{(4)}$ is already contained in our generating set. We proceed to the next largest partition $(3,1)$. We get $\mcal{C}_{(3,1)}$ from $\mcal{C}_{(3)}\cdot \mcal{C}_{(1)}$. Both $\mcal{C}_{(3)}$ and $\mcal{C}_{(1)}$ are contained in $\mcal{G}_{3}$. $\mcal{C}_{(3)}\cdot \mcal{C}_{(1)}$ will contain $\mcal{C}_{(3,1)}$ and may contain $\mcal{C}_{(4)}$ as well as $\mcal{C}_{\nu}$ for which $|\nu| \leq 3$. For $n = 10$,
  \bea
	\mcal{C}_{\left(3\right)}\cdot\mcal{C}_{\left(1\right)} =  \mcal{C}_{(3,1)} + 5 \,\mcal{C}_{\left(4\right)} + 4\,\mcal{C}_{\left(1,1\right)} + 21\,\mcal{C}_{(2)}.
	\eea
	We continue in this way down the order in (\ref{eq:natordn4}) generating the $\cC$ operators labeled by each these partitions using $\cC$ operators generated by $\mcal{G}_{3}$. For $n = 10$,
	\bea
		\mcal{C}_{(2)} \cdot \mcal{C}_{(2)} &=& 240\,\cC_{(0)} + 22\,\cC_{(2)} + 8\,\cC_{(1,1)} + 5\,\cC_{(4)} + 2\,\cC_{(2,2)}.\\
		\mcal{C}_{(2,1)} \cdot \mcal{C}_{(1)} &=& 21\,\cC_{(2)} + 24\,\cC_{(1,1)} + 5\,\cC_{(4)} + 4\,\cC_{(3,1)} + 6\,\cC_{(2,2)} + 2\,\cC_{(2,1,1)}.\\
		\mcal{C}_{(1,1,1)} \cdot \mcal{C}_{(1)} &=& 15\,\cC_{(1,1)} + 2\,\cC_{(3,1)} + 3\,\cC_{(2,1,1)} + 4\,\cC_{(1,1,1,1)}.
 	\eea
	\item For $k = 5$, natural ordering is still a total ordering,
	\bea
		(5) \geq (4,1) \geq (3,2) \geq (3,1,1) \geq (2,2,1) \geq (2,1,1,1) \geq (1,1,1,1,1).
	\eea
	For $n=10$, we have
	\bea
		\cC_{(4)}\cdot \cC_{(1)} &=& 6\,\cC_{(5)} + \cC_{(4,1)} + 24\,\cC_{(3)} + 6\,\cC_{(2,1)}  \\
		\cC_{(3)}\cdot \cC_{(2)} &=& 6\,\cC_{(5)} + \cC_{(3,2)} + 28\,\cC_{(3)} + 12\,\cC_{(2,1)} + 112\,\cC_{(1)} \\
		\cC_{(3,1)}\cdot \cC_{(1)} &=& 6\,\cC_{(5)} + 5\,\cC_{(4,1)} + 3\,\cC_{(3,2)} + 2\,\cC_{(3,1,1)} + 15\,\cC_{(3)} + 15\,\cC_{(2,1)} \\&&+\, 12\, \cC_{(1,1,1)}\\
		\cC_{(2,2)}\cdot \cC_{(1)} &=& 3\,\cC_{(5)}+4\,\cC_{(3,2)} + \cC_{(2,2,1)} + 10\,\cC_{(2,1)}\\
		\cC_{(2,1,1)}\cdot \cC_{(1)} &=& 5\,\cC_{(4,1)} + 2\,\cC_{(3,2)} + 4\,\cC_{(3,1,1)} + 6\,\cC_{(2,2,1)} + 4\,\cC_{(2,1,1,1)}\\
		&& + 10\,\cC_{(2,1)} +\, 24\,\cC_{(1,1,1)} \\
		\cC_{(1,1,1,1)}\cdot \cC_{(1)} &=& 2\,\cC_{(3,1,1)} + 3\,\cC_{(2,1,1,1)} + 5\,\cC_{(1,1,1,1,1)} + 6\,\cC_{(1,1,1)}
	\eea
	\item For $k = 6$, natural ordering is no longer a total ordering. We have
	\bea
		(6) \geq (5,1) \geq (4,2) \geq \left\{ (4,1,1), (3,3) \right\} \geq (3,2,1) \geq \left\{ (3,1^3),(2^3) \right\} \geq (2,2,1,1) \geq (2,1^4) \geq (1^6)\nonumber\\
	\eea
The partitions $(4,1,1)$ and $(3,3)$ are incomparable according to natural ordering. Thus, $(3,3)$ will not appear in the product $\cC_{(4,1)}\cdot \cC_{(1)}$, and $(4,1,1)$ will not appear in the product  $\cC_{(3)}\cdot \cC_{(3)}$. Similarly for the incomparable $(3,1^3)$ and $(2^3)$, which are conjugates of $(4,1,1)$ and $(3,3)$. Calculating these products for $n=10$, we find
\bea
		\cC_{(4,1)}\cdot \cC_{(1)} &=& 7\,\cC_{(6)} + 6\,\cC_{(5,1)} + 3\,\cC_{(4,2)} + 2\,\cC_{(4,1,1)} + 10\,\cC_{(4)} + 15\,\cC_{(3,1)} + 12\,\cC_{(2,1,1)}.\\
		\cC_{(3)}\cdot \cC_{(3)} &=& 7\,\cC_{(6)} + 2\,\cC_{(3,3)} + 40\,\cC_{(4)} + 16\,\cC_{(3,1)} + 27\,\cC_{(2,2)} + 147\,\cC_{(2)} + 98\,\cC_{(1,1)}\nonumber\\ &&+ 1260\,\cC_{(0)}\\
		\cC_{(3,1,1)}\cdot \cC_{(1)} &=& 6\,\cC_{(5,1)} + 5\,\cC_{(4,1,1)} + 4\,\cC_{(3,3)} + 3\,\cC_{(3,2,1)} + 3\,\cC_{(3,1,1,1)} + 6\,\cC_{(3,1)} + 9\,\cC_{(2,1,1)}\nonumber\\ &&+ 24\,\cC_{(1,1,1,1)}\\
		\cC_{(2,2)}\cdot \cC_{(2)} &=& 7\,\cC_{(6)} + 5\,\cC_{(4,2)} + 3\,\cC_{(2,2,2)} + 25\,\cC_{(4)} + 8\,\cC_{(3,1)} + 26\,\cC_{(2,2)} + 8\,\cC_{(2,1,1)} \nonumber \\
		&&+ 70\,\cC_{(2)}.
\eea
	\item Assume that all $\cC$ labeled by partitions $\lambda$ for which $|\lambda | \leq k-1$ can be generated by $\mcal{G}_{k-1} = \left\{\mathcal{C}_{\left(1\right)}, \mathcal{C}_{\left(2\right)}, \cdots , \mcal{C}_{\left(k-1\right)}\right\}$
	\item Now consider the generating set $\mcal{G}_{k}$ and all partitions $\lambda$ such that $|\lambda| = k$. We show that all $\cC$ labeled by partitions of $k$ can be generated by $\mcal{G}_{k}$. According to natural ordering
	\bea
		(k)\geq (k-1,1) \geq (k-2,2) \geq (k-2,1,1), \cdots , (2,1^{k-2}) \geq (1^k).
	\label{eq:chainnatorderk}
	\eea
	We consider $\cC_{\lambda}\cdot \cC_{(r)}$ where both $\cC_{\lambda}$ and $\cC_{(r)}$ have $|\lambda| \leq k-1$ and $r \leq k-1$ and are thus generated by $\mcal{G}_{k-1}$. According to equations (\ref{eq:MacDMult}) - (\ref{eq:coeffofC2}), we sum over partitions $\nu$ such that $\nu \geq \lambda \bigcup \left(r\right)$ and $a^{ \lambda \bigcup \left(r\right)}_{\lambda \left(r\right)} > 0$.
	Now, $\cC_{(k)}$ is already contained in $\mcal{G}_{k}$. To generate $\cC_{(k-1,1)}$, we multiply $\cC_{(k-1)}\cdot \cC_{(1)}$. The term $\cC_{(k-1,1)}$ will appear with non-zero coefficient  and all partitions larger than $(k-1,1)$ may also appear. This only includes $\cC_{(k)}$, which is already contained in $\mcal{G}_{k}$. Next, to generate $\cC_{(k-2,2)}$, we multiply $\cC_{(k-2)}\cdot \cC_{(2)}$. The term $\cC_{(k-2,2)}$ will appear and partitions larger than $(k-2,2)$, which are $(k)$ and $(k-1,1)$, may also appear. However, $\cC_{(k)}$ is already contained in $\mcal{G}_{k}$, and $\cC_{(k-1,1)}$ is generated by $\mcal{G}_{k}$ as we have seen above. To generate $\cC_{(k-2,1,1)}$, we multiply $\cC_{(k-2,1)}\cdot \cC_{(1)}$. The term $\cC_{(k-2,1,1)}$ is sure to appear, while the partitions larger than $(k-2,1,1)$, i.e., $(k), (k-1,1)$ and $(k-2,2)$ may also appear. However, each one of these have already been shown to be generated by $\mcal{G}_{k}$. We may continue in this way proceeding one by one down the chain of partitions in (\ref{eq:chainnatorderk}). When we arrive at the smallest partition $(1^{k})$, we compute $\cC_{(1^{k-1})}\cdot \cC_{1}$. $\cC_{(1^k)}$ will be generated as well as all partitions larger may also appear. But each of these have, in turn, been shown to be generated by $\mcal{G}_{k}$ in the same way as described above. 
\end{itemize}
A small comment is in order. In proceeding down the list of partitions, we will arrive at a set of partitions that are mutually, or pairwise, incomparable according to natural ordering. To generate any one of these partitions, we still form a product of the form (\ref{eq:MacDMult}). Since we need to sum strictly over partitions that are larger than the one in question all incomparable partitions will not appear in the result. As an example, see $k = 6$.

We have successfully shown that $\mathcal{G}_{k}$ can generate any $\mathcal{C}$ labeled by partition $\lambda$ whose $|\lambda| \leq k$. In terms of the $T$ operators, this result means that $ \{T_{2}, \cdots , T_{k+1}\}$ is capable of generating any $T$ labeled by a partition $\lambda$ with branching number $k$ or less. This means that $\{ T_{2}, \cdots , T_{n} \}$ generates any central element $T_{\lambda}$ with branching number $n-1$ or less. However, the $T_{\lambda} \in \cZ(\mC ( S_n ) )$ with the largest branching number is $T_{n}$ with $B = n-1$, which is already contained in our generating set. Thus, $\{ T_{2}, \cdots , T_{n} \}$ can generate the centre of $\mC (S_{n})$.

{\bf Remark 3.2.2} Lastly, it is worth noting that $T_{n}$ can be expressed in terms of $\{ T_{2}, \cdots , T_{n-1} \}$. Using (\ref{eq:MacDMult}), the calculation of $\mathcal{C}_{n-2} \cdot \mathcal{C}_{1}$ yields $\mathcal{C}_{n-1}$ (which is $T_{n}$), $\mathcal{C}_{(n-2,1)}$, (which is $T_{(n-1,2)}$) and then $T_{\lambda}$ with a lower branching number than $n-1$, which can all be generated $\{ T_{2}, \cdots , T_{n-1} \}$. However, we exclude $T_{(n-1,2)}$ since it is labeled by a permutation of $n+1$. Thus, $T_{n}$ can be expressed in terms the set $ T_{2}, \cdots , T_{n-1} $.
\subsection{Generating sets for $ \cZ ( \mC ( S_n ) $ from irreducible representations.} 

A basis for the centre $ \cZ ( \mC ( S_n ) $  is given by projectors (orthogonal idempotents) 
\bea 
P_{R} ={  { d_R} \over n! } \sum_{ \sigma \in S_n } { \chi_R ( \sigma ) }  ~ \sigma 
\eea 
where $ \chi_R ( \sigma ) $ is the character in the  irreduciblel representation $ R$ of the permutation $ \sigma $. The $R$ correspond to Young diagrams. This is general fact about group algebras of finite groups, see e.g. 
\cite{GW}. 
These obey 
\bea 
&& P_R P_S  =\delta_{RS } P_R \cr 
&& \sum_{ R } P_R = 1 
\eea
where $1$ is the identity permutation and the identity in  the associative algebra $ \mC ( S_n )$. 
The number of these projectors is $p(n)$, the number of partitions of $n$. 
Taking the trace in an irrep $R$, and using orthogonality of characters, we have 
\bea 
\chi_S ( P_R )  = \delta_{ RS  } d_R
\eea
We make use of the following fact  (Lemma 2.1 in \cite{DLS1606}):

\noindent 
{\bf Lemma 3.3.1 } A linear combination $T = \sum_R a_R P_{R} $ with  all $a_R$ distinct generate the centre. To see this, we observe that 
\bea 
P_S   =  \prod_{S_1 \ne S } {  ( T - a_{S_1} )  \over  ( a_S - a_{ S_1} )   )  }  
\eea
which follows by expanding  the right hand side and using the projector equations. 
This is essentially a fact about the algebra of diagonal matrices: the algebra of diagonal matrices is generated by any diagonal matrix with distinct entries.

\subsection{Generating sets of cycle structures from lists of normalised characters } \label{sec:gensetsofcyclstrucandcharac}

We can expand any central element such as $T_2$ in terms of these projectors
\bea 
T_2 = \sum_{R } (a_2)^R P_R 
\eea
Take a trace in the representation $S$ on both sides 
\bea 
\chi_S ( T_2 ) = (a_2)^S d_S 
\eea
So we have 
\bea
T_2 = \sum_{ R } { \chi_R ( T_2 ) \over d_R } P_R 
\eea

From Lemma 3.3.1, this means that if the list of $ {\chi_R ( T_2) \over d_R } $ has no repetitions, 
then $T_2$ generates the centre. 

If we have a list of irreps $R$, where the normalized characters of $T_2$ are equal, then 
we can use another central element such as $T_3$. Within this block, we apply the Lemma 3.3.1 again. 
So to find out whether $T_2, T_3 $ generate the centre, we just need to look at the matrix
$ 2 \times p(n)$ matrix 
\bea 
( { \chi_R ( T_2 )\over d_R } ~, ~ { \chi_R ( T_3 ) \over d_R}  ) 
\eea
If no two rows are identical, then $T_2, T_3 $ generate the centre. 
We may apply Lemma 3.3.1 iteratively. More generally, we would like to consider the $ k \times p(n)$ matrix. The central elements $\{T_2 , T_3 , \cdots, T_k\}$ may each be expanded in terms of the projectors $P_{R}$, with their respective normalized characters being the expansion coefficients. If this list of normalized characters is distinct for each $R$ then no two rows in
\bea
( { \chi_R ( T_2 ) \over d_R } ~, ~ { \chi_R ( T_3 ) \over d_R} \cdots ~, ~ { \chi_R ( T_k ) \over d_R} ) 
\eea
are identical. According to Lemma 3.3.1, the $\{T_2 , T_3 , \cdots, T_k\}$ will generate the centre. For example if for $R_{1}$ and $R_{2}$, the list of normalized characters for $\{T_2 , T_3 , \cdots, T_{k-1}\}$ are identical then this list of $T_{i}$ and their respective powers are no longer linearly independent. They no longer generate the subspace of the centre spanned by $P_{R_{1}}$ and $P_{R_{2}}$. We now include one more central element $T_{k}$. If $ { \chi_R ( T_k ) \over d_R}$ is distinct for $R_{1}$ and $R_{2}$ then the set $\{T_2 , T_3 , \cdots, T_k\}$ are linearly independent and, according to Lemma 3.3.1, this list generates $\cZ ( \mC ( S_n ) )$.
It is interesting to study the sequence of the smallest $n$ values where the $\{T_2 , T_3 , \cdots, T_k\}$ fail to generate the centre of $ \mC ( S_n )$. Denote this sequence by $n_*(k)$. The problem of finding $ n_{*}(k) $ is a matter of understanding the degeneracies in the characters of $S_n$, using nice formulae for these characters available from the mathematics literature \cite{CGS2004,Lassalle2008}.
Using the discussion  in Section \ref{sec:CasimirsChargesMatInvts}  $n_*(k)$ is also the smallest value of $n$ where knowledge of the all the Casimir eigenvalues $ C_2 , C_3 , \cdots , C_k$ does not suffice to distinguish all the Young diagrams. 

\subsection{Generating the centre for $n=5$ with $T_{2}$}

The list of normalized characters for $ T_2$ is $\left\{ 10,5, 2, 0, -2, -5, -10 \right\} $. In this list, there are no degeneracies. Hence by the above argument, we expect $T_2$ to generate the centre. Below, we show this explicitly. 

We generate the following equations in Mathematica by taking successive powers of $T_{2}$. 

\bea
	\left( T_{2} \right)^{2} &=& 10 + 3 T_{3} + 2 T_{(2,2)}\cr
	\left( T_{2} \right)^{3} &=& 34 T_{2} + 16 T_{4} + 9 T_{(3,2)}\cr
	\left( T_{2} \right)^{4} &=& 340 + 207 T_{3} + 125T_{5} + 168 T_{(2,2)}\cr
	\left( T_{2} \right)^{5} &=& 2086 T_{2} + 1664 T_{4} + 1461 T_{(3,2)}\cr
	\left( T_{2} \right)^{6} &=& 20860 + 17703 T_{3} + 15625 T_{5} + 16672 T_{(2,2)}\cr
	\left( T_{2} \right)^{7} &=& 177094 T_{2} + 166656 T_{4} + 161469 T_{(3,2)}\cr 
	\left( T_{2} \right)^{8} &=& 1770940 + 1692687 T_{3} + 1640625T_{5} + 1666688T_{(2,2)}.
\eea
For example, after computing $\left(T_{2}\right)^{2}$, we count that the identity element $T_{1}$ appears ten times, $T_{3}$ appears 3 times and $T_{(2,2)}$, the formal sum of permutations having disjoint two-cycles, appearing twice. We may now invert this system of equations to solve for the other $T$ quantities. Solving for $T_{4}$ and $T_{(3,2)}$ in terms of $T_{2}$, we find
\bea
	T_{4} &=& -\frac{2603 }{2800}\left(T_{2}\right)^{3} + \frac{12987}{280000} \left(T_{2}\right)^{5} -\frac{103 }{280000}\left(T_{2}\right)^{7}\\
	T_{(3,2)} &=& \frac{331 }{525}\left(T_{2}\right)^{3}-\frac{3541 }{105000}\left(T_{2}\right)^{5} + \frac{29 }{105000}\left(T_{2}\right)^{7}
\eea
Next, we solve for $T_{1} =1 , T_{3}, T_{5}$ and $T_{(2,2)}$ in terms of $T_{2}$
\bea
	1 &=& \frac{62273}{208320}(T_{2})^{2} - \frac{45819 }{3472000}(T_{2})^{4} + \frac{6863
   }{27776000}(T_{2})^{6} - \frac{1}{666624}(T_{2})^{8} \\
	T_{3} &=& -\frac{541}{420} (T_{2})^{2} +\frac{25 }{336}(T_{2})^{4} - \frac{1}{1680} (T_{2})^{6} \\
	T_{5} &=&\frac{625}{10416}(T_{2})^{2} - \frac{125 }{6944}(T_{2})^{4} + \frac{43
  }{55552}(T_{2})^{6} - \frac{1}{166656}(T_{2})^{8},  \\
	T_{(2,2)} &=&\frac{195299}{208320}(T_{2})^{2} - \frac{31681}{694400}(T_{2})^{4} - \frac{1903
  }{5555200}(T_{2})^{6} + \frac{5}{666624}(T_{2})^{8}.
\eea
This shows that for $n=5$ each $T_{\mu}$ corresponding to a cycle-type $\mu$, may be written in terms of $T_{2}$. Thus, $T_{2}$ generates the centre $\mathcal{Z}\left(\mathbb{C}\left(S_{5}\right)\right)$.
\section{Dimensions and entropies associated with low order cycle structures }\label{sec:exptsdata}

From the previous section, we concluded that not all elements in $\mathcal{G}_{n}$ are needed to generate $\cZ ( \mC ( S_n ) )$. Indeed for some $n$, all we need is $T_{2}$. It is interesting to study this operator's ability to generate the centre as a function of $n$. In this section we present some data concerning $T_{2}$ as well as the combination of $T_{2}$ and $T_{3}$. We first discuss the codimension for the subspaces generated by these two central elements, and then we discuss the entropy related to the degeneracies of their normalized characters. 

\subsection{Co-dimensions as measures of uncertainty }\label{sec:codimdata}

We have proved  that $T_{2}$ generates the centre of $S_{n}$ if there are no repetitions or degeneracies in the list of its normalized character.  In this section, we refer to the normalized characters frequently and thus we define
\bea
	\frac{\chi_{R}\left( T_{k} \right)}{d_{R}} \equiv \widehat{\chi}^{R}_{k}.
\eea 
For a given $n$, if there are degeneracies in $\widehat{\chi}^{R}_{2}$ then we include the normalized character of $T_{3}$. If the list $\{\widehat{\chi}^{R}_{2},\widehat{\chi}^{R}_{3}\}$ has no repetitions then $T_{2}$ and $T_{3}$ generate the centre for that particular $n$. 

\begin{figure}[htb]
\centering
  \begin{tabular}{@{}cccc@{}}
    \includegraphics[width=.45\textwidth]{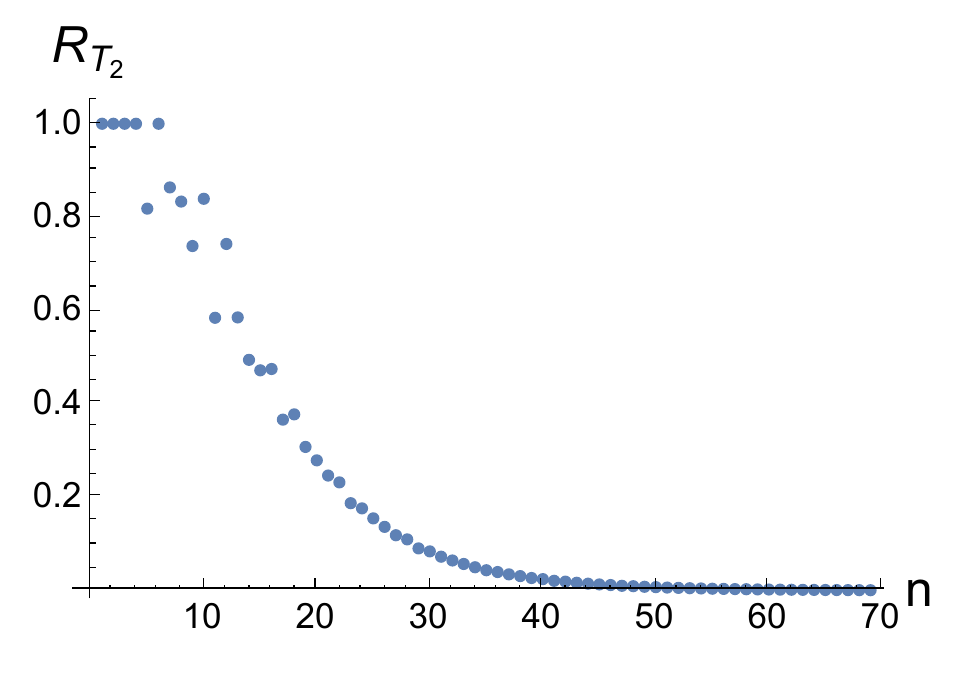} &
     \includegraphics[width=.45\textwidth]{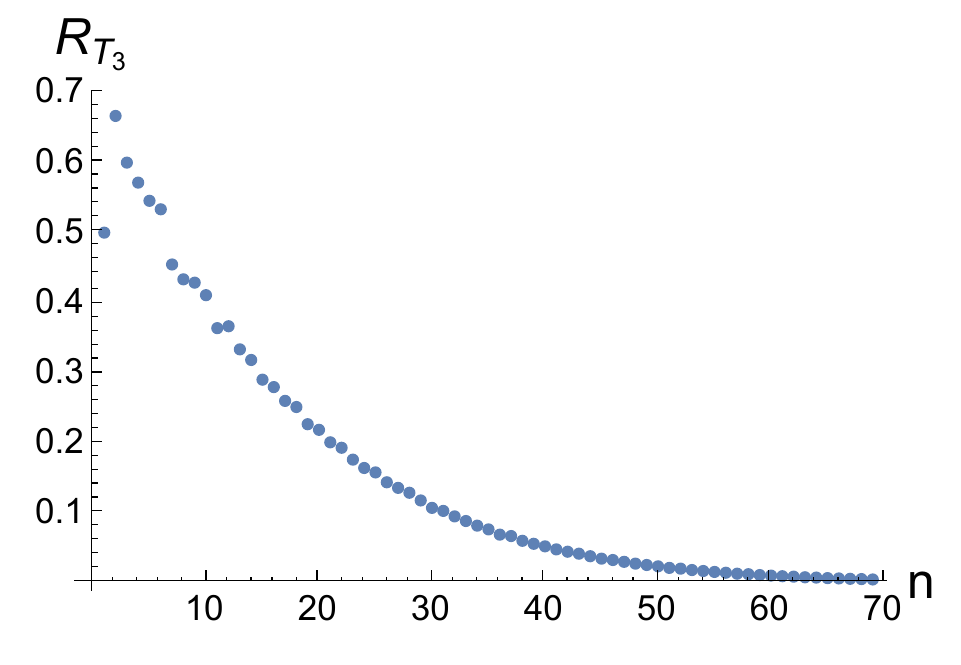}\\
    \includegraphics[width=.45\textwidth]{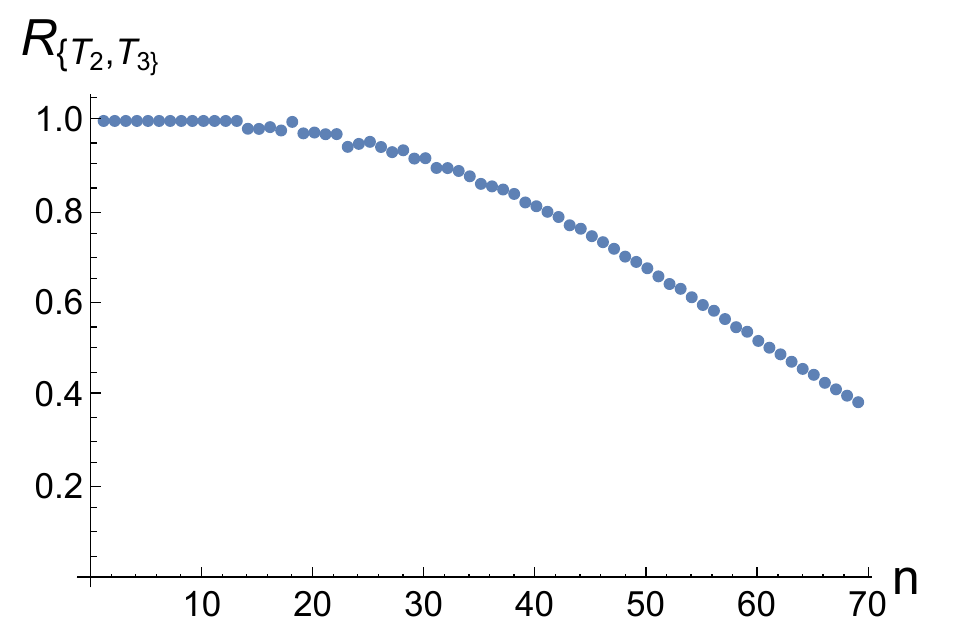} 
   \\
  \end{tabular}
  \caption{{\small{Figures showing relative dimension plots for $T_{2}, T_{3}$ and for $\{ T_{2},T_{3} \}$. }}}
  \label{RelativeDimplots}
\end{figure}

For $n = 1,2,3,4,5$, there are no repetitions in $\widehat{\chi}^{R}_{2}$, so $T_{2}$ does indeed generate $ \cZ ( \mC ( S_n ) ) $. For $n=6$, we encounter our first degeneracy. The list of normalized characters for $T_{2}$ is
\bea
	\{15, 9, 5, 3, 3, 0, -3, -3, -5, -9, -15\}
\eea
There are two repetitions in this list. Thus, $T_{2}$ no longer generates $ \cZ ( \mC ( S_n ) ) $. $T_{2}$ instead only generates a subspace of $ \cZ ( \mC ( S_n ) )$. The codimension  of a subspace is the difference between the dimension of the full space and the dimension of the subspace. We define $\mathrm{codim}_{T_{k}}\left(n\right)$ to be the difference of the dimension of $\cZ ( \mC ( S_n ) )$ and the dimension of the subspace generated by $T_{k}$. The codimension of a subspace generated by a 
 $T_{k}$ or a collection of $T_k$'s is a measure of how close the central element or collection of central elements is,  to generating the centre. For $ T_2$ at $n=6$, the number of distinct $ \widehat \chi^R_2$
is $9$, which is $2$ less  than $p(6)=11$, so  the codimension of the subspace generated  is equal to two. Interestingly, $T_{2}$ once again generates $ \cZ ( \mC ( S_n ) ) $ for $n=7$, producing a codimension of zero. From $n=8$ however, $T_{2}$ fails to generate $ \cZ ( \mC ( S_n ) ) $. The codimension data for $T_{2}$ for $n = 2$ to $n = 70$ is shown in table \ref{CoDimTable}.
These co-dimensions can be viewed as a measure of the uncertainty in the determination of Young diagrams with $n$ boxes when we only know the normalized character of $T_2$ : equivalently when we only know the second Casimir. 
\begin{table}[]
\begin{center}
\resizebox{18cm}{!} {
\begin{tabular}{|c|l*{17}{c}l|}
	\hline
	$n$ & 2 & 3 & 4&5&6&7&8&9&10&11&12&13&14&15 &16&17 \\
	$T_{2}$ codim  & 0&0& 0&0&2&0&3&5&11& 9& 32&26&56&89&122&156\\
	$T_{3}$ codim & 1 & 1 & 2 & 3 & 5 & 7 & 12 & 17 & 24 & 33 & 49 & 64 & 90 & 120 & 164 & 214\\
	$\{T_{2},T_{3}\}$ codim  & 0&0& 0&0&0&0&0&0&0& 0& 0&0&0&3&4&4\\
	\hline
	$n$ & 18 & 19 & 20&21&22&23&24&25&26&27&28&29&30&31 &32&33 \\
	$T_{2}$ codim  & 244& 305& 434& 571&755& 964 &1280&1613& 2059& 2599&3277& 4064& 5097& 6267& 7742& 9488\\
	$T_{3}$ codim & 285 & 367 & 485 & 619 & 801 & 1013 & 1298 & 1637 & 2052 & 2578 & 3214 & 3978 & 4945 & 6110 & 7492 & 9181\\
	$\{T_{2},T_{3}\}$ codim  & 8&1&17&20&29&36&89&98&111& 171& 253&292&460&556&856&1043\\
	\hline
	$n$ & 34 & 35 & 36&37&38&39&40&41&42&43&44&45&46&47 &48&49 \\
	$T_{2}$ codim  &11607& 14138& 17192&20776&25108&30214&36329&43486& 52053&62044&73908&87807& 104155&123263&145750&171880\\
	$T_{3}$ codim & 11224 & 13668 & 16606 & 20149 & 24278 & 29315 & 35258 & 42262 & 50636 & 60452 & 72049 & 85751 & 101917 & 120694 & 142862 & 168775 \\
	$\{T_{2},T_{3}\}$ codim  &1343& 1797& 2467& 3088& 3891& 4964& 6638& 8295& 10537& 13254& 17110&20958&26528&32990& 41070&51358\\
	\hline
	$n$ & 50 & 51 & 52&53&54&55&56&57&58&59&60&61&62&63 &64&65 \\
	$T_{2}$ codim  & 202537& 238158 & 279738 & 327980 & 384160 & 449171& 524666& 611871& 712853& 829335 & 963938& 1118836 & 1297411 & 1502656 & 1738693& 2009491 \\
	$T_{3}$ codim & 199031 & 234300 & 275590 & 323265 & 379108 & 443666 & 518646 & 605320 & 705914 & 821576 & 955785 & 1109954 & 1287924 & 1492280 & 1727867 & 1997539 \\
	$\{T_{2},T_{3}\}$ codim & 62802 & 77099 & 95467 & 117359 & 141406 & 173538 & 211477 & 254289 & \
309073 & 374251 & 444426 & 538166 & 643109 & 766460 & 914920 & 1088525 \\
	\hline
	$n$ & 66 & 67 & 68&69&70&&&&&&&&&&& \\
	$T_{2}$ codim &  2320397& 2676422 & 3084370 & 3550860 & 4084383&&&&&&&&&&&\\
	$T_{3}$ codim & 2307932 & 2662970 & 3070080 & 3535437 & 4068446&&&&&&&&&&&\\
	$\{T_{2},T_{3}\}$ codim  & 1286732 & 1530746 & 1807711 & 2130136 & 2507976&&&&&&&&&&&\\	
	\hline
\end{tabular}}
\end{center}
\label{CoDimTable}
\caption{{\small{Table showing the codimension data for $T_{2}$, for $T_{3}$, and then for $\{T_{2},T_{3}\}$ for $n = 2$ to $n=70$. 
}}}
\end{table}%
From this data, we can calculate the relative dimension for $T_{2}$
\bea
R_{T_{2}} = \frac{p\left(n\right) - \mathrm{codim}_{T_{2}}\left(n\right)}{p\left(n\right)},
\eea
where $p\left(n\right)$ is the dimension of $ \cZ ( \mC ( S_n ) ) $. The plot for this is shown in figure \ref{RelativeDimplots}.\\
We now discuss the codimension of the subspace generated by both $T_{2}$ and $T_{3}$. For $n=6$, the list of normalized characters of $T_{2}$ and $T_{3}$ is
{\small{
\bea
\Big\{\{15, 55\}, \{9, 31\}, \{5, 15\}, \{3, 19\}, \{3, 7\}, \{0, 10\}, \{-3,19\},
 \{-3, 7\}, \{-5, 15\}, \{-9, 31\}, \{-15, 55\}\Big\}.\nonumber\\
\eea}}
Since there are no repetitions in this list (i.e. each pair of numbers is unique in this list), these two central elements generate  $ \cZ ( \mC ( S_n ) ) $. Thus, the codimension for this subspace is zero. We find a zero codimension for all cases of $n$ up until $n = 15$. For $n=15$ the subspace generated by $\{T_{2},T_{3}\}$ has a codimension of 3. See table \ref{CoDimTable} for data for $n = 2$ to $n = 70$. The co-dimensions for $\{T_{2},T_{3}\}$ can again be viewed as a measure of the uncertainty in the determination of Young diagrams with $n$ boxes when we only know the normalized characters of $T_2$ and $T_3$ : equivalently when we only know the second and third Casimir. Figure \ref{RelativeDimplots} shows the relative dimension for $\{T_{2}, T_{3}\}$. We also plot the relative dimension for $\{T_{2},T_{3}\}$ in figure \ref{RelativeDimplots}. Note that the codimension of the space generated by $ \{T_{2}, \cdots, T_{n}\}$ is zero for all $n$. Furthermore, we note that the relative dimension plot for $\{T_{2}, \cdots, T_{n}\}$ would just be a straight line at $R_{\{T_{2}, \cdots, T_{n}\}} = 1$. 

It is natural to expect that the uncertainty of identifying the Young diagram is smaller when we know both $ \widehat{\chi}^{R}_{2}$ and $ \widehat{\chi}^{R}_{3}$ compared to when we only know one of these quantities. Furthermore, the higher the codimension of $T_{k}$ (or of some subset of $T$'s), the higher this uncertainty becomes. Thus, we expect
\bea\label{codimexpt} 
	\mathrm{codim}_{\{T_{2},T_{3}\}} &<& \mathrm{codim}_{T_{2}} \cr
	\mathrm{codim}_{\{T_{2},T_{3}\}} &<& \mathrm{codim}_{T_{3}}.
\eea
This is indeed reflected in the data of table \ref{CoDimTable} for all values of $n$ listed there. The relative sizes of $\mathrm{codim}_{T_{2}}$ and $\mathrm{codim}_{T_{3}}$ may also be studied. From the data, we see that 
\bea 
&&   \mathrm{codim}_{T_{2}} <  \mathrm{codim}_{T_{3}}  \hbox { for } n \le 25 \cr 
&& \mathrm{codim}_{T_{3}} <  \mathrm{codim}_{T_{2}}  \hbox { for } 26 \le n \le 70.
\eea
The behaviour of these codimensions for $n$ larger than 70 is an interesting problem. It is natural to conjecture that $\mathrm{codim}_{T_{3}} <  \mathrm{codim}_{T_{2}}$ persists for all $n$ higher than 26.

Note that AdS/CFT motivates the study of finite $N$ versions  of this codimension problem. 
There is a finite $N$ truncation of  $ \cZ ( \mC ( S_n ) )$, where we set to zero all the projectors 
$P_R$ with height $ l( R )$ constrained to satisfy $ l(R)  > N$. This subspace, which we denote $ \cZ_N ( \mC ( S_n ) ) $, is a proper subspace when $ N < n $ and forms a sub-algebra. Consider the set of generators $ \{ T_2 , \cdots , T_k \} $ 
for some $ k < n $. The particular case of $ k \sim N^{ 1/4} , n \sim N^2$ in the limit of large $N$ is of particular interest in connection with the information loss discussion of \cite{IntInfLoss}. $n \sim N^2  $ is the dimension of CFT operators which produce non-trivial deformations of the  AdS space-time. $k \sim N^{ 1/4}$ corresponds to the Planck scale cutoff. Calculating the codimensions in this regime of $ k , n $ is a very interesting problem for the future. 

\subsection{Average Entropies for uniform  probability distribution over values of charges }\label{sec:AvgEnt}

\begin{figure}[htb]
\centering
  \begin{tabular}{@{}cccc@{}}
    \includegraphics[width=.45\textwidth]{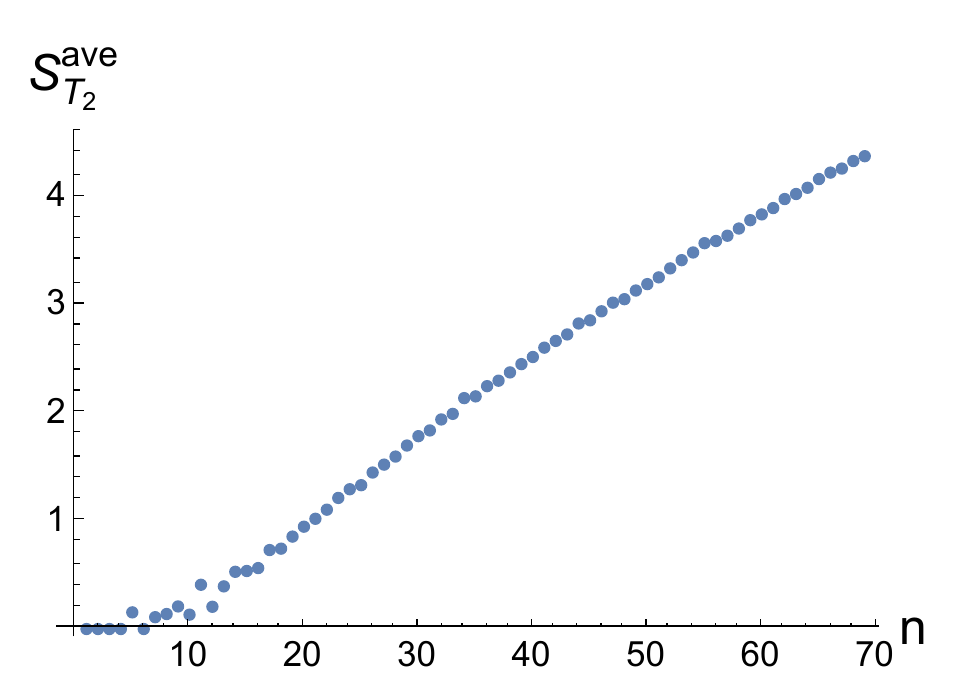} &
     \includegraphics[width=.45\textwidth]{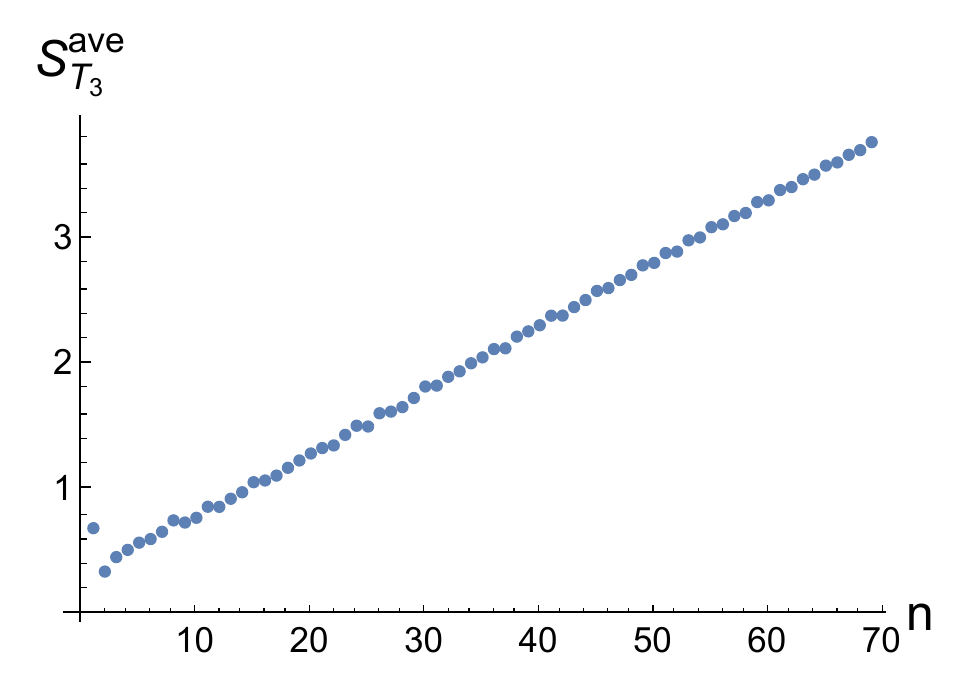}\\
    \includegraphics[width=.45\textwidth]{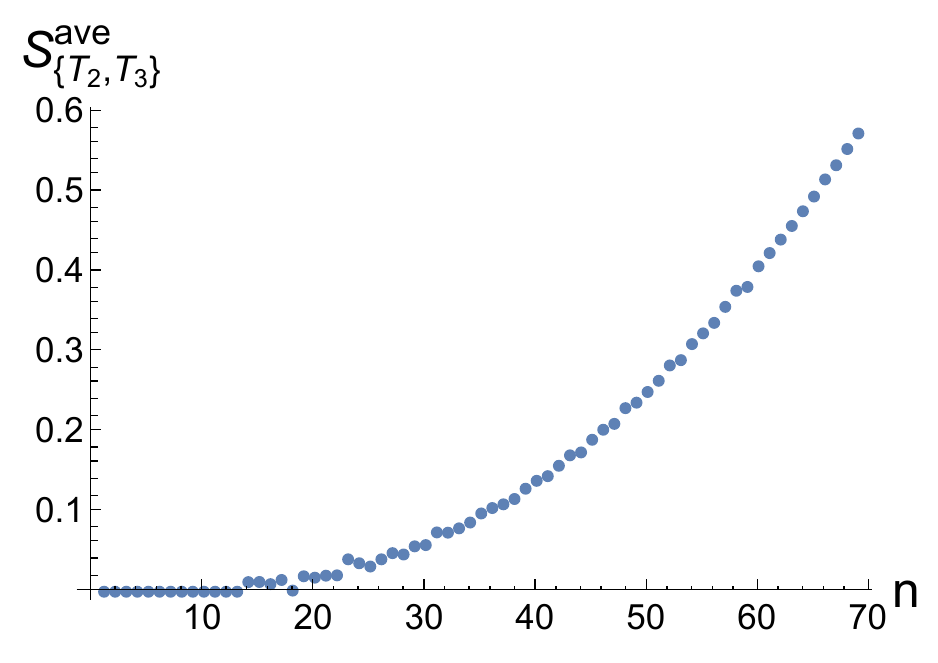} 
   \\
  \end{tabular}
  \caption{{\small{Figures showing the average entropy values $S^{ave}_{T_{2}}$, $S^{ave}_{T_{3}}$ and $S^{ave}_{\{T_{2},T_{3}\}}.$}}}
  \label{NewEntropyplots}
\end{figure}

Consider measuring the normalized character for the operator $T_{2}$. Given the discussion 
in Section \ref{sec:CasimirsChargesMatInvts}, in particular equation (\ref{CtoT}) this is equivalent to 
knowing the quadratic Casimir charge.   There will be a list of normalized characters generated over the Young diagrams. The value $v_{2}$ contained in this list occurs with multiplicity $M\left(v_{2}\right)$. Assuming that we have no knowledge about the half-BPS state beyond the dimension $n$ and the quadratic Casimir, this means that a total of $M\left(v_{2}\right)$ Young diagrams are equally likely. We thus have a uniform distribution over this subset of Young diagrams. We also have
\bea
	\sum\limits_{v_{2}}M\left(v_{2}\right) = p(n).
\eea
The Shannon entropy associated with this value of $v_{2}$, and the uniform distribution, is 
\bea
	S_{T_{2}}(v_{2}) = \log\left( M\left(v_{2}\right) \right).
\label{eq:ST2forv2}
\eea
This may be viewed as a measure of the uncertainty in our knowledge of the state $R$ when we only know  $\widehat{\chi}^{R}_{2} = v_{2}$. We may also take an average of these entropy values 
\bea\label{averageentropy} 
	S^{ave}_{T_{2}} = \frac{1}{N_{2}}\sum\limits_{v_{2}}\log\left(M\left(v_{2}\right)\right),
\eea
where the sum is over all distinct values of $\widehat{\chi}^{R}_{2} = v_{2}$ and $N_{2}$ is the total number of distinct $\widehat{\chi}^{R}_{2} $ values. This quantity may be viewed as a measure of the average uncertainty of the Young diagram when we only know the values of the $T_{2}$ normalized characters. We present data for $S^{ave}_{T_{2}} $ in table \ref{tab:T2T3NewEntropy} for $n = 2$ to $n = 70$. We also present a plot of this data in figure \ref{NewEntropyplots}. Similar data for $T_{3}$ is presented there as well. Note that the average entropy (\ref{averageentropy}) can also be viewed as an expectation value in a probability distribution over the values $v_2$, where all these values are equally probable, in other words, the uniform distribution over $v_2$. 

\begin{table}[]
\begin{center}
\resizebox{18cm}{!} {
\begin{tabular}{|c|l*{17}{c}l|}
	\hline
	$n$ & 2 & 3 & 4&5&6&7&8&9&10&11&12&13&14&15 &16&\\
	$S^{ave}_{T_{2}}$ &0.& 0.& 0.& 0.& 0.154033& 0.& 0.109444& 0.138629& 0.208835& 0.13273& \
0.409244& 0.205769& 0.394343& 0.529039& 0.536582 \\
	$S^{ave}_{T_{3}}$ & 0.693147& 0.346574& 0.462098& 0.51986& 0.577623& 0.606504& 0.664379& \
0.755526& 0.738199& 0.776171& 0.864573& 0.863525& 0.928273& 0.980178& \
1.06048\\
	$S^{ave}_{\{T_{2},T_{3}\}}$ &0.& 0.& 0.& 0.& 0.& 0.& 0.& 0.& 0.& 0.& 0.& 0.& 0.& 0.0120199& \
0.012214 \\
	\hline
$n$ & 17 & 18 & 19 &20 & 21 & 22& 23 & 24 & 25 & 26 & 27 & 28 & 29 & 30 & 31 \\
$S^{ave}_{T_{2}}$ & 0.563526& 0.730224& 0.743094& 0.854613& 0.946289& 1.01928& 1.10351& \
1.21204& 1.29252& 1.32994& 1.44694& 1.52004& 1.59453& 1.69658& 1.78313 \\	
$S^{ave}_{T_{3}}$ & 1.07416& 1.11325& 1.17569& 1.23406& 1.28993& 1.33327& 1.35432& \
1.43858& 1.51161& 1.50538& 1.61116& 1.62438& 1.66037& 1.73318& 1.82496\\
$S^{ave}_{\{T_{2},T_{3}\}}$ & 0.00946276& 0.0147087& 0.00141748& 0.0193172& 0.0175845& 0.0200677& \
0.0204703& 0.0405462& 0.0355927& 0.0315061& 0.040534& 0.0482671& \
0.046559& 0.0567626& 0.058324 \\ 
\hline
$n$ & 32 & 33 & 34 &35 & 36 & 37& 38 & 39 & 40 & 41 & 42 & 43 & 44 & 45 & 46 \\
$S^{ave}_{T_{2}}$ & 1.83615& 1.93773& 1.98932& 2.13505& 2.15106& 2.24552& 2.29611& \
2.37268& 2.44947& 2.51518& 2.60154& 2.66398& 2.72379& 2.82513& 2.85408  \\	
$S^{ave}_{T_{3}}$ & 1.83197& 1.90237& 1.94616& 2.01054& 2.05769& 2.12388& 2.12992& \
2.22336& 2.26533& 2.31421& 2.39071& 2.39148& 2.45929& 2.51554& 2.58844\\
$S^{ave}_{\{T_{2},T_{3}\}}$ & 0.0741798& 0.0738499& 0.0792513& 0.0865217& 0.0978702& 0.10465& \
0.109309& 0.115949& 0.128827& 0.138693& 0.14455& 0.1575& 0.17054& \
0.174216& 0.190026 \\
\hline
$n$ & 47 & 48 & 49 &50 & 51 & 52& 53 & 54 & 55 & 56 & 57 & 58 & 59 & 60 & 61 \\
$S^{ave}_{T_{2}}$ & 2.93773& 3.01755& 3.04966& 3.12983& 3.18908& 3.25141& 3.33456& \
3.41051& 3.48155& 3.56756& 3.58782& 3.63649& 3.70292& 3.78007& 3.83382  \\	
$S^{ave}_{T_{3}}$ & 2.61117& 2.67504& 2.71647& 2.79268& 2.81239& 2.89125& 2.90253& \
2.9925& 3.01555& 3.09754& 3.12063& 3.18674& 3.21088& 3.29807& 3.31192\\
$S^{ave}_{\{T_{2},T_{3}\}}$ & 0.202603& 0.209931& 0.229586& 0.236442& 0.249809& 0.263787& 0.282991& \
0.289623& 0.309702& 0.323097& 0.336236& 0.356257& 0.376416& 0.381171& \
0.406974 \\
\hline
$n$ & 62 & 63 & 64 & 65 & 66 & 67 & 68 & 69 & 70 &  &  &  &  &  &  \\
$S^{ave}_{T_{2}}$ & 3.89228& 3.97566& 4.02229& 4.07997& 4.16048& 4.22011& 4.25737& \
4.32714& 4.37174&  &  &  &  &  &\\ 
$S^{ave}_{T_{3}}$ & 3.39403& 3.41817& 3.48096& 3.51733& 3.58923& 3.61489& 3.67565& \
3.71282& 3.77678 &  &  &  &  &  &\\
$S^{ave}_{\{T_{2},T_{3}\}}$ & 0.423566& 0.440401& 0.457433& 0.47581& 0.494219& 0.515644& 0.533373& \
0.553679& 0.573106 &  &  &  &  &  & 
\\
\hline
\end{tabular}}
\end{center}
\label{tab:T2T3NewEntropy}
\caption{{\small{Table showing the average entropy values $S^{ave}_{T_{2}}$, $S^{ave}_{T_{3}}$ and $S^{ave}_{\{T_{2},T_{3}\}}.$}}}
\end{table}%

We may also consider the list of values $\{ v_{2}, v_{3} \}$ for $\widehat{\chi}^{R}_{2} $ and $\widehat{\chi}^{R}_{3} $. Denote the multiplicity of the pair $(v_{2},v_{3})$ by $M\left(v_{2},v_{3}\right)$. We have 
\bea
	\sum\limits_{v_{3}}M\left(v_{2},v_{3}\right) = M\left(v_{2}\right), \hspace{20pt} \sum\limits_{v_{2}}M\left(v_{2},v_{3}\right) = M\left(v_{3}\right),
\eea
where we sum over distinct values of $\widehat{\chi}^{R}_{3} = v_{3}$ in the first equation and over distinct values of $\widehat{\chi}^{R}_{2} = v_{2}$ in the second. $\log\left( M\left(v_{2},v_{3}\right) \right)$ is the entropy associated with the values $ ( v_{2} , v_{3} ) $ and the uniform distribution over the subset of $R$ corresponding to these values for $( \widehat{\chi}^{R}_{2}, \widehat{\chi}^{R}_{3}) $. Again we may compute the average of this entropy
\bea
	S^{ave}_{\{T_{2},T_{3}\}} = \frac{1}{N_{(2,3)}}\sum\limits_{v_{2},v_{3}}\log\left( M\left(v_{2},v_{3}\right) \right)
\eea
where $N_{(2,3)}$ is the total number of distinct values for $\{\widehat{\chi}^{R}_{2},\widehat{\chi}^{R}_{3}\}$. This entropy may be viewed as a measure of the uncertainty in identifying the exact Young diagram when we only know the values of the $T_{2}$ and $T_{3}$ normalized characters.

We expect that there should be a smaller uncertainty when we know both $\{\widehat{\chi}^{R}_{2},\widehat{\chi}^{R}_{3}\}$ compared to knowing just one of the normalized characters. In other words, we expect these average entropies to obey 
\bea 
&& S^{ave}_{\{T_{2},T_{3}\}}  < S^{ave}_{\{T_{2}\}}   \cr 
&& S^{ave}_{\{T_{2},T_{3}\}}  < S^{ave}_{\{T_{3}\}}
\eea
This is indeed compatible with the results in the table \ref{NewEntropyplots} and is also the trend we see in the comparison of the codimensions in (\ref{codimexpt}).  It is not a priori clear which   $S^{ave}_{\{T_{2}\}} $ or $ S^{ave}_{\{T_{3}\}} $ should be larger. The data shows that 
\bea 
&& S^{ave}_{\{T_{2}\}}  < S^{ave}_{\{T_{3}\}}  \hbox { for } n \le 31 \cr 
&& S^{ave}_{\{T_{3}\}}  < S^{ave}_{\{T_{2}\}}  \hbox { for } 32 \le n \le 70 
\eea
It is natural to conjecture, and would be interesting to prove that the trend visible for $ 32 \le n \le 70$ extends for all $ n \ge 32 $. 

We can define finite $N$ versions of these entropies by considering only Young diagrams with height no larger than $N$. In these finite $N$ ensembles of Young diagrams, we can define multiplicities of Young diagrams and derive entropies for specified values of $n$. It will be interesting to obtain estimates of 
the finite $N$ entropies for $ n \sim N^2 $, since this corresponds to classical solutions of supergravity. Also of particular interest, given the discussion \cite{IntInfLoss},  are the average entropies for  the sets $\{ \hat \chi^R_{ 2} , \hat \chi^R_3 , \cdots , \hat \chi^{ R}_{ k } \} $ where $ k \sim N^{ 1/4}$.   

\subsection{Average Entropies for multiplicity-weighted probability distributions over values of charges  }\label{sec:AvgEntUnifYoung}

In the discussion of the average entropy over a set of known charges, we used a uniform distribution over the spectrum  of charges. We found that the average entropies thus calculated satisfied inequalities which were similar to those satisfied by the co-dimensions, lending support to the idea that both the co-dimensions and the average entropies are sensible measures of the information available from knowing a set of charges - in particular, the inequalities reflect the fact that knowing more charges reduces the uncertainty.  The comparisons of the information available by knowing just one charge. e.g. $ T_2$ versus $T_3$ depends on how one measures the information, whether it is through codimensions or average entropies. The data is compatible with the conjecture that at large $n$, a definite pattern emerges: there is more information in knowing $T_3$ rather than $T_2$, whether this is measured by looking at the codimensions or average entropies. 

Given a set of charges, and an associated multiplicity of Young diagrams states, 
there is yet another interesting way to measure the information or conversely the uncertainty 
associated with that probability distribution. 
Suppose we take the set of values of $ \wchi^R_2$ at a given $n$. Let these values form the set 
 $ \cV_2$. 
 The multiplicity of a given value $v_{2} \in \cV_2$ is 
\bea 
M ( v_{2} ; n ) =  \sum_{ R \vdash n  } \delta ( v_{2} , \wchi^R_2 ).  
\eea
Let $N_{2}$ be the size of the set $\cV_2$. The probability of having value $v_{2}$ is 
\bea 
P ( v_{2}   ;n ) = { M ( v_{2} ; n ) \over p ( n )  } 
\eea
since the total number of Young diagrams is $ p(n)$. 
\bea 
\sum_{ v_{2} \in \cV_2 } P ( v_{2}  ; n )  = 1 
\eea
Now consider the Shannon entropy for this probability distribution 
\bea\label{shannonsep}  
&& S ( T_2 ; n ) = -  \sum_{ v_{2} }  P ( v_{2} ; n ) \log P ( v_{2} ; n )  \cr 
 && = - { 1 \over p(n ) } \sum_{ v_{2} } M ( v_{2} ; n ) \log M ( v_{2} ; n ) + { 1 \over p(n ) } \sum_{ v_{2} }  M ( v_{2} ; n ) \log p(n) \cr 
 && = \log ( p(n) ) -    \sum_{ v_{2} } { M ( v_{2} ; n )\over p(n ) }   \log M ( v_{2} ; n )  
\eea
This entropy has an interesting interpretation in a quantum information setting involving quantum measurement and classical communication. Suppose we have a density matrix for the Hilbert space of Young diagram states, where the diagonal part of the density matrix is 
\bea\label{rhodiag}  
\rho_{\rm{diag} } = \sum_{  n = n_0  }^{ n_1 } { 1 \over ( n_1 - n_0 + 1 )}   \sum_{ R  \vdash n } \frac{1}{p(n)}  | R \rangle \langle R |
\eea
for a uniform probability distribution over Young diagram states with energies between $ n_0 $ and $n_1$. The states $|R \rangle  $ have unit norm.  Suppose observer $A$ measures the energy $n$ and the set of charges $\{ T_2, \cdots , T_n \}$ to determine the exact $R$.  Given the form of  (\ref{rhodiag})
we have a uniform probability distribution over $R$. The Observer $A$ communicates the information to Observer $B$ of the energy $n$, but to observer $C$ the more detailed information of $n $, along with the eigenvalues of $C_2$, or equivalently the normalized character of $T_2$. The first term  in (\ref{shannonsep}) is a measure of 
the uncertainty open to $B$, who knows that the Young diagram is one of $p(n)$ 
but has no further information \footnote{This  informal discussion of uncertainty can be made more precise by using the Shannon Noiseless Coding Theorem \cite{NaC,grunwald2004shannon}}. The second term 
\bea\label{secterm} 
\sum_{ v_{2} } { M ( v_{2} ; n )\over p(n ) }   \log M ( v_{2} ; n ) 
\eea
is a measure of the uncertainty $ \log M ( v_{2} ; n )$ for each $ v_{2} $ value averaged over the different $v_{2}$ values according to a probability with which $v_{2}$ occurs in the measurements of $A$. The difference is a measure of the reduction in uncertainty, due to the additional information available to $C$ compared to $B$.

 Equation (\ref{secterm}) provides the relation between the entropy $S ( T_2 ; n )$ above and the entropy $S_{T_{2}}(v_{2})$ defined in equation (\ref{eq:ST2forv2}) in section \ref{sec:AvgEnt}. We can view (\ref{secterm}) as an expectation value of $S_{T_{2}}(v_{2})$ taken over the probabilities $M ( v_{2} ; n )\over p(n )$. Thus,
\bea
	S ( T_2 ; n ) = \log ( p(n) ) - \mathbb{E}\left( S_{T_{2}}(v_{2}) \right).
\eea

\begin{figure}[htb]
\centering
  \begin{tabular}{@{}cccc@{}}
    \includegraphics[width=.45\textwidth]{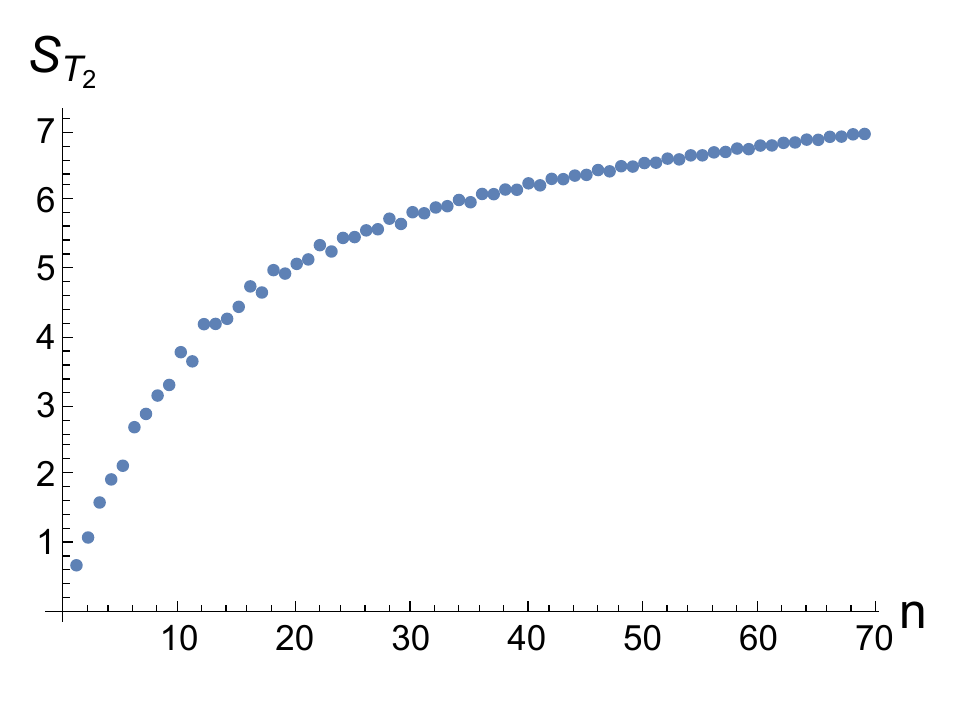} &
     \includegraphics[width=.45\textwidth]{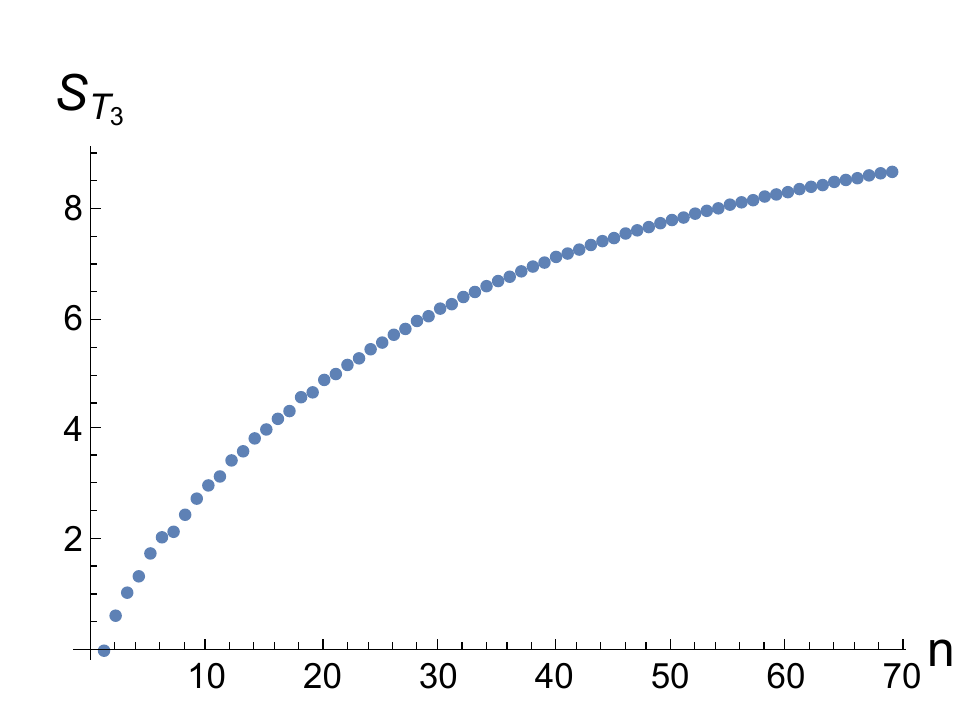}\\
    \includegraphics[width=.48\textwidth]{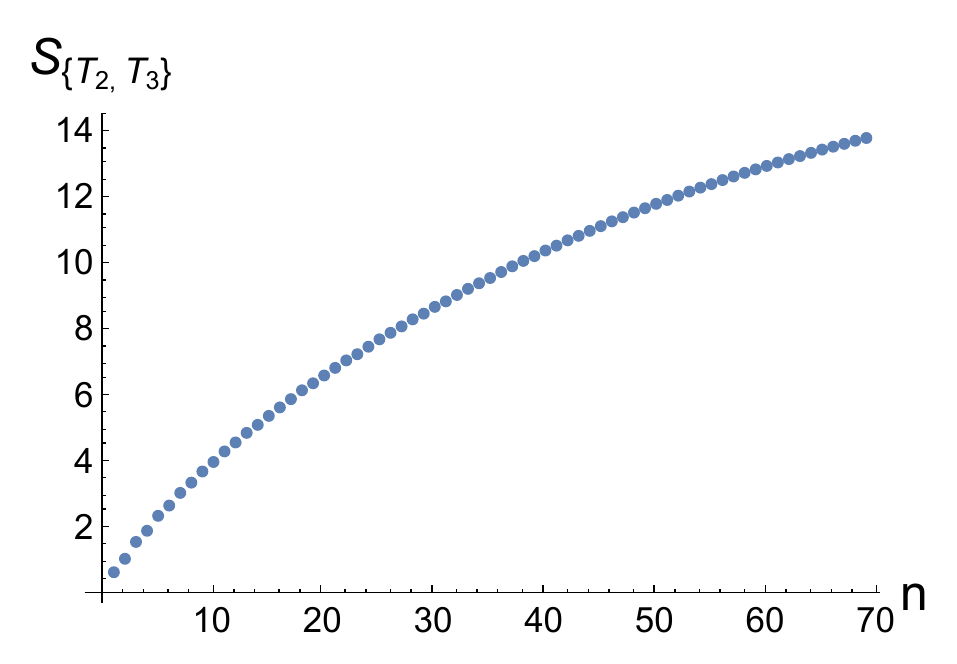} 
   \\
  \end{tabular}
  \caption{\small{Figures showing entropy $S_{T_{2}}, S_{T_{3}}$ and for $S_{\{T_{2},T_{3}\}}$. 
  }}
  \label{EntropyT2T3}
\end{figure}

It is interesting to plot the entropy (\ref{shannonsep}) as a function of $n$ using data we already have, which is motivated by questions such as:  What is this entropy as a function of $n$? How does it behave at large $n$? This data is presented in table \ref{tab:T2T3Entropy}.
\begin{table}[]
\begin{center}
\resizebox{18cm}{!} {
\begin{tabular}{|c|l*{17}{c}l|}
	\hline
	$n$ & 2 & 3 & 4&5&6&7&8&9&10&11&12&13&14&15 &16&\\
	$S_{T_{2}}$ & 0.693147& 1.09861 & 1.60944 & 1.94591 & 2.14584 & 2.70805 & 2.902 & 3.17015 & 3.32476 & 3.80255 & 3.66872 & 4.21163 & 4.2155 & 4.29012 & 4.46491 \\
	$S_{T_{3}}$ & 0 & 0.636514& 1.05492& 1.35178& 1.76776& 2.06111& 2.16129& 2.47089& 2.76364& 3.00383& 3.16641& 3.45826& 3.62378& 3.85848& 4.02057\\
	$S_{\{T_{2},T_{3}\}}$ & 0.693147& 1.09861& 1.60944& 1.94591& 2.3979& 2.70805& 3.09104& 3.4012& 3.73767& 4.02535& 4.34381& 4.61512& 4.90527& 5.14685& 5.41841 \\
	\hline
$n$ & 17 & 18 & 19 &20 & 21 & 22& 23 & 24 & 25 & 26 & 27 & 28 & 29 & 30 & 31 \\
$S_{T_{2}}$ & 4.76269& 4.6738& 4.99981& 4.94975& 5.09142& 5.15615& 5.36429& 5.27195& 5.47018& 5.482& 5.58016& 5.59657& 5.75089& 5.6739& 5.84508 \\	
$S_{T_{3}}$ & 4.21406& 4.35544& 4.6068& 4.69534& 4.92155& 5.02822& 5.19344& 5.31327& 5.47841& 5.60027& 5.74219& 5.84809& 5.99283& 6.07895& 6.21713\\
$S_{\{T_{2},T_{3}\}}$ & 5.67506& 5.92444& 6.19158& 6.40336& 6.63889& 6.86859& 7.09512& 7.28201& 7.50869& 7.73209& 7.92885& 8.12253& 8.33612& 8.50813& 8.71314 \\ 
\hline
$n$ & 32 & 33 & 34 &35 & 36 & 37& 38 & 39 & 40 & 41 & 42 & 43 & 44 & 45 & 46 \\
$S_{T_{2}}$ & 5.82958& 5.91412& 5.93302& 6.02386& 5.99058& 6.11159& 6.10852& 6.17636& 6.17289& 6.26647& 6.23725& 6.33206& 6.32741& 6.37935& 6.39015  \\	
$S_{T_{3}}$ & 6.2982& 6.42812& 6.51941& 6.62493& 6.71789& 6.79649& 6.89425& 6.98144& 7.05408& 7.15678& 7.2181& 7.28986& 7.37349& 7.4435& 7.49904\\
$S_{\{T_{2},T_{3}\}}$ & 8.87932& 9.07243& 9.25758& 9.42605& 9.58561& 9.76661& 9.93823& 10.1018& 10.2462& 10.4154& 10.5627& 10.7234& 10.859& 11.0096& 11.152 \\
\hline
$n$ & 47 & 48 & 49 &50 & 51 & 52& 53 & 54 & 55 & 56 & 57 & 58 & 59 & 60 & 61 \\
$S_{T_{2}}$ & 6.46051& 6.44134& 6.5165& 6.51129& 6.56246& 6.56774& 6.62732& 6.61593& 6.67486& 6.67503& 6.71835& 6.72465& 6.77441& 6.76538& 6.81789  \\	
$S_{T_{3}}$ & 7.58222& 7.64015& 7.69895& 7.76954& 7.82713& 7.87246& 7.94326& 7.99421& 8.03952& 8.10671& 8.14905& 8.18788& 8.25452& 8.29205& 8.33362\\
$S_{\{T_{2},T_{3}\}}$ & 11.2967& 11.4233& 11.5622& 11.6938& 11.8266& 11.9438& 12.0728& 12.1975& 12.3155& 12.4226& 12.5447& 12.6535& 12.7626& 12.8684& 12.973 \\
\hline
$n$ & 62 & 63 & 64 & 65 & 66 & 67 & 68 & 69 & 70 &  &  &  &  &  &  \\
$S_{T_{2}}$ & 6.81972& 6.85732& 6.86305& 6.90441& 6.90174& 6.94541& 6.94837& 6.98194& 6.98709&  &  &  &  &  &\\ 
$S_{T_{3}}$ & 8.3895& 8.43081& 8.46249& 8.52008& 8.55551& 8.58719& 8.63838& 8.67634& 8.70224 &  &  &  &  &  &\\
$S_{\{T_{2},T_{3}\}}$ & 13.076& 13.1789& 13.2702& 13.3668& 13.4665& 13.5534& 13.6395& 13.7326& 13.816 &  &  &  &  &  & 
\\
\hline
\end{tabular}}
\end{center}
\label{tab:T2T3Entropy}
\caption{{\small{Table showing the entropy data for $T_{2}$, $T_{3}$ and $\{ T_{2}, T_{3} \}$.
}}}
\end{table}%
Next consider $ T_2 , T_3 , \cdots , T_k $. We want to think about the values 
$ \{  v_{2} , v_{3} , \cdots v_{k} \} $ of the normalized characters 
$  \wchi^R_2 ,   \wchi^R_3 , \cdots ,  \wchi^R_k $. This vector of normalized characters 
lives in the space $ \cV_{\{ 2,3, \cdots,k \}}$. 
The size of this set is 
$ N_{(2,3,\cdots, k)}$. The multiplicity of a given value-set is 
\bea 
M (  v_{2}, \cdots , v_{k}  ; n ) 
=  \sum_{ R \vdash n } 
\delta ( v_{2} , \wchi^R_2 )  \delta ( v_{3} , \wchi^R_3 ) 
\cdots  \delta ( v_{k} , \wchi^R_k ).
\eea  
Now we can define an entropy for this set of generators 
\bea 
&& \hspace{-26pt} S ( k ; n ) = S ( T_2 , \cdots , T_k ; n ) = 
- { 1 \over p(n) }  \sum_{ v_{2}} \sum_{ v_{3}  } \cdots \sum_{ v_{k}  } \Big[
M (  v_{2} , \cdots , v_{k}  ; n ) 
\log M ( v_{2} ,v_{3} , \cdots , v_{k}  ; n ) \cr 
&& \qquad \qquad \qquad \qquad \qquad  - M ( v_{2}, v_{3} , \cdots , v_{k}  ; n ) 
\log p(n) \Big]
\eea
Entropy data for $\{ T_{2}, T_{3}\}$ is also found in table \ref{tab:T2T3Entropy}. In addition, the entropy data in this table is plotted in figure \ref{EntropyT2T3}.

These entropies demonstrate the following behaviours. For all values of $n$ listed in the table \ref{tab:T2T3Entropy}
\bea
	S_{T_{2}} \leq S_{\{T_{2},T_{3}\}}\cr
	S_{T_{3}} < S_{\{T_{2},T_{3}\}}.
\eea
The only $n$ values for which $S_{T_{2}} = S_{\{T_{2},T_{3}\}}$ are the ones where $T_{2}$ generate the centre: $n = 2,3,4,5$ and $7$. This is exactly as expected from our interpretation of these entropies in terms of information gained by knowing in addition to the energy $n$, the specified charges. We generically gain more information from knowing more charges,  unless the more limited set of charges
is sufficient to determine the Young diagram entirely.

When analyzing the relative sizes between $S_{T_{2}}$ and $S_{T_{3}}$ the data indicates
\bea 
&& S_{T_{3}}  < S_{T_{2}}  \hbox { for } n \le 23 \cr 
&& S_{T_{2}}  < S_{T_{3}}  \hbox { for } 24 \le n \le 70.
\eea
Again, it is natural to again conjecture that $S_{T_{2}}  < S_{T_{3}}$ for all $n$ larger than 24. 
If this conjecture, along with the corresponding conjectures in sections \ref{sec:codimdata} and
 \ref{sec:AvgEnt}, are true, they would support the plausible conclusion that different measures of uncertainty - co-dimensions and variations in the choice of entropy function - give the same ranking of information provided by different conserved charges in the limit of large $n$.

The above entropies are relevant to AdS/CFT questions when $ N >  n$. 
We can also define finite $N$ entropies, where $ n > N $, 
motivated by the discussion of \cite{IntInfLoss}. 
In this case, we are interested in Young diagrams with no more than $N$ rows, which we express as
$ l(R ) \le N$, using $l(R)$ to refer to the vertical length of the Young diagram. 

\bea 
M (  v_{2} , \cdots , v_{k}  ; n , N  ) 
= \sum_{ \substack  { R \vdash n \\ l(R ) \le N }  } 
\delta ( v_{2} , \wchi^R_2 )  \delta ( v_{3} , \wchi^R_3 ) 
\cdots  \delta ( v_{k} , \wchi^R_k )
\eea

The associated entropies are 
\bea 
&& S ( k ; n , N ) = - { 1 \over p(n , N ) }  \sum_{ v_{2}} \sum_{ v_{3}  } \cdots \sum_{ v_{k}  } \Big[
M (  v_{2} , \cdots , v_{k} ; n , N  ) 
\log M ( v_{2} , v_{3} , \cdots , v_{k} ; n , N  ) \cr 
&& \hspace{60pt}  - M ( v_{2} , v_{3} , \cdots , v_{k}  ; n , N  ) 
\log p(n , N ) \Big].
\eea
Here $ p ( n , N ) $ is the number of Young diagrams with no more than $N$ rows. 
Of particular interest, from the discussion in \cite{IntInfLoss}, is the large  $N$ behaviour of 
\bea 
S ( k = N^{ 1/4} , n = N^2 , N ) 
\eea
$k = N^{ 1/4} $ corresponds to the Planck scale, while $ N^2$ is the dimension of CFT operators which cause a significant backreaction in the geometry. 
We leave a systematic computation and discussion of these finite $N$ version of the co-dimensions and entropies we have discussed for the future. 


\section{ Content  distribution functions  and diophantine equations } \label{sec:CDFs} 

In section 3 we considered the problem of generating $\cZ(\mC(S_n))$ using central elements $\{ T_2, T_3 , \cdots T_k\} $.  We used results in the representation theory of  $\mC ( S_{n}) $ to show that this question can be answered by inspection of  the normalized characters of the $T_{i}$. This allowed us to generate interesting data on the subspaces generated by $T_2$, by $T_3$ and the pair $ \{ T_2 , T_3 \}$. In this section, we once again rephrase the problem in terms of the so-called content polynomials, which 
have been used to produce elegant expressions for normalized characters in \cite{CGS2004,Lassalle2008}. 
We show that the normalized characters of  $\{ T_2  , T_3 , \cdots , T_k \} $ can be expressed in terms of the first $k$ content polynomials. One can then reformulate the problem of generating the centre, and distinguishing all Young diagrams, in terms of these polynomials. This new formulation is used to write simple code in  Mathematica to determine the values of $n$ for which the first $k$ polynomials distinguish all Young diagrams (see Appendix \ref{ref:MTCA}). In section \ref{sec:CDFs} we define the content distribution function (CDF) for a Young diagram. We prove that each Young diagram can be uniquely specified by its corresponding CDF. The content polynomials can be expressed as moments of the distribution functions, analogous to moments of a probability distribution. We show that knowledge of all $n$ moments for a Young diagram uniquely determines its CDF. We show that the event of two diagrams having $k$ degenerate moments ($k<n$) translates to a set of $k$ vanishing moment equations in the difference of the two respective CDFs. Lastly we provide some examples of CDFs and CDF plots for degenerate diagrams at $n=6, 15, 24$. We explain a visualization of the CDF plots in terms of segmented open strings with Dirichlet boundary conditions.

\subsection{ Content polynomials and normalized characters }\label{sec:contpolynormcharacters}

An important result in \cite{CGS2004} (Lemma 3.1) is 
\bea
{ \chi_R ( T_{ k } )  \over d_R } =  \la c^{k-1} ( R ) \ra +
  \sum_{ l = 0 }^{ k-2} \sum_{ \lambda \vdash l     } \Omega_{ k -1 }^{ \lambda }  ( n ) 
  \la c^{ \lambda_1} ( R ) \ra \la c^{ \lambda_2} ( R )  \ra \cdots   \la c^{ \lambda_r} ( R )  \ra
  \label{eq:tricontpolynormcharc}
\eea
We are summing over partitions  $ \lambda = ( \lambda_1 , \lambda_2 , \cdots , \lambda_r ) $ where 
$ \lambda_1 + \lambda_2 + \cdots + \lambda_r = l \le ( k-2) $. The $\la c^{ k  } ( R ) \ra $ are defined as
\bea 
\la c^{ k } ( R ) \ra  = \sum_{ ( i , j ) \in R } ( j - i )^k 
\eea
where we are summing over the coordinates of the boxes of a Young diagram: $i$ is the row number and $j$ is the column number. For example the Young diagram with row lengths $ [ 4,3,1] $ has boxes with 
$ ( i , j ) \in \{ ( 1,1)  , ( 1, 2 ) , ( 1, 3 ) , ( 1, 4 ) , ( 2, 1) , ( 2,2 ) , (2,3) , (3,1) \} $. The number $j-i$ is called the content of the box labeled by $(i,j)$. $ \Omega_{ k-1 }^{ \nu } (n)$ is a polynomial of degree at most $ ( k-1 - l ) /2 +1 - r $. Equation (\ref{eq:tricontpolynormcharc}) implies a  triangular structure in the linear transformation  between the normalized characters and the content polynomials. Knowing all the content polynomials up to $k-1$ means that we know the normalized character for $T_{k}$. For $k=2$ we have $\widehat{\chi}^{R}_{2} = \la c(R)  \ra$. For $k=3$, we have $\widehat{\chi}^{R}_{3} = \la c^{2}(R)\ra + \Omega^{1}_{2}(n)\la c(R)\ra$. Lastly, for $k=4$, 
\bea
	\widehat{\chi}^{R}_{4} = \la c^{3}(R)\ra + \Omega^{1}_{3}(n)\la c(R) \ra + \Omega^{(2,0)}_{3}(n)\la c^{2}(R) \ra +  \Omega^{(1,1)}_{3}(n)\la c(R) \ra^{2}.
\eea
It is immediately apparent that degeneracies in the normalized characters $\widehat{\chi}^{R}_{2} $, $\widehat{\chi}^{R}_{3}$ and $\widehat{\chi}^{R}_{4}$ translate to degeneracies in the content polynomials $\la c(R)\ra, \la c^{2}(R) \ra$ and $\la c^{3}(R) \ra$. 


\subsection{Computing with content polynomials } \label{sec:compwcontpoly}

If we are given the $T_2 , \cdots , T_{ k } $ what is the smallest $n$, where these fail to generate the centre of $ \mC ( S_n )$? This question can be reformulated in terms of the content polynomials. Given $\{ \la c\left( R \right) \ra, \la c^{2}\left( R \right) \ra, \cdots , \la c^{k}\left( R \right) \ra \}$, what is the smallest $n$ such that the sequence of  these lists as $R$ runs over  Young diagrams with $n$ boxes 
has degeneracies, i.e. multiple $R$ have the same list.  
The experimental answer is for $ k $ starting from $1$, is displayed in table \ref{firstnwithk}. 
\begin{table}[h!]
\begin{center}
\begin{tabular}{|c|c|}
	\hline
	$k$ & first $n$\\
	\hline
	1 & 6\\
	\hline
	2 & 15\\
	\hline
	3 & 24\\
	\hline
	4 & 42\\
	\hline
	5 & 80\\
	\hline
\end{tabular}
\end{center}
\caption{\small{Table showing the smallest values of $n$ for which $ T_{2}, \cdots, T_{k}$ fail to generate the centre $ \mC ( S_n )$. We give the $n$ values for $k$ from 1 to 5. This table was generated by computing the content polynomials for each irrep $R$ at fixed $n$. Degeneracies in the content polynomials translate to degeneracies in the normalized characters.}}
\label{firstnwithk}
\end{table}%

For up to $n=5$, the first content polynomial $\langle c\left( R\right) \rangle$, where $R$ labels the Young diagram, is able to distinguish all $R \vdash 5 $. The first time $\langle c\left( R\right) \rangle$ fails to distinguish all Young diagrams is at $n=6$. However, the set $\{ \langle c\left( R\right) \rangle, \langle c^{2}\left( R\right) \rangle \}$ is unique to each $R\vdash 6$. These two polynomials together are then able to distinguish all Young diagrams for $n=6$ up to $n=14$. The first $n$ for which this set has a multiplicity, and thus fails to distinguish the Young diagrams is at $n=15$. This set is no longer unique to each $R$. We then extend the set to $\{ \langle c\left( R\right) \rangle, \langle c^{2}\left( R\right) \rangle ,  \langle c^{3} \left( R\right) \rangle \}$, which then works for up to $n = 24$ where it fails for the first time. It is remarkable that just five numbers, $\{ \langle c ( R ) \rangle ,\langle c^{2} ( R ) \rangle,\langle c^{3}( R )\rangle  ,\langle c^{4}(R )  \rangle,\langle c^{5} (R)  \rangle\}$, are able to distinguish all Young diagrams corresponding to partitions of $n$ up to $n = 79$. The first time these five polynomial values fail is for $n=80$. 
It is instructive to look at these multiplicities and at the Young diagrams which have these degenerate content polynomials. The degeneracies (defined as multiplicities minus one, i.e. difference in length of the list of content vectors and the length of the list with repetitions removed) for $ k =1$ at $ n = 6 , 7, 8  , 9 $ are 
 \bea 
 2, 0 , 3 , 5 
 \eea
 The degeneracies for $ k  =2 $ at $ n = 15 , 16 , 17 ,18   $ are 
 \bea 
 3,4,4,8 
 \label{eq:keq2deg}
 \eea
 The degeneracies at $ k =3$ for $ n = 24 , 25 , 26 , 27, 28 $ are 
 \bea 
 1,0,0, 2, 2 
  \label{eq:keq3deg}
 \eea
 The degeneracies for $ k =4 $ at $ n = 42 , 43 , 44 , 45 , 46  $ are 
 \bea 
 2,2,6, 8 , 8 
 \eea
 The degeneracies for $ k =5$ at $ n = 80 , 81 , 82, 83 , 84 , 85 $ are 
 \bea 
 3, 0 , 2, 2, 11 , 12 
 \eea
It is interesting that the degeneracies start off very low. It is useful to look at the 
degenerate Young diagrams which share the same set of Casimirs, when we are at these thresholds of 
distinguishability.

\subsection{ Content distribution functions  }  \label{subsec:CDFs}

Every box in a Young diagram has a content $c$ given by $j-i$.
For example at $ n =3$, we have 3 Young diagrams which can be described 
in terms of their row lengths $ [ 3 ] , [2,1] , [1^3 ] $. 
The contents are $ \{ 0 , 1, 2 \} , \{ 0 , 1, -1\} , \{ 0 , -1 , -2 \} $. 
These can be plotted on a content axis labelled $c$, and the content multiplicities 
$ f(c)$. The respective content multiplicities are given by $ ( c , f(c) ) $
\bea 
&& [ 3 ] : ( 0 , 1 ) , ( 1,1 ) , ( 2 , 1 ) \cr 
&& [ 2 ,1 ] : ( 0 , 1 ) , ( 1 , 1 ) , ( -1 , 1 )  \cr 
&& [1^3 ] : ( 0,1 ) ,  (-1 , 1 ) , ( -2 , 1 ) 
\eea
The content distribution functions can be visualized as piecewise linear or segmented open strings defined in the content space from $-n$ to $n$. Each segment has slope -1, 0 or 1. The open string has Dirichlet boundary conditions of zero at its ends. This means that $f(\pm n) = 0$. 

{ \bf Proposition 5.3.1 } A Young diagram is uniquely specified by its content distribution function $f\left(c\right)$.\\

We begin this proof by noting that a Young diagram has a depth $d$. This is the number of boxes along the diagonal at $45$ degrees when to the horizontal. All the boxes along the diagonal have content $0$. Alongside 
$d$, 
there is a set of parameters 
\bea 
 ( d;  k_1^+ , l_1^+ , \cdots , k_{ p-1}^+ , l_{ p-1}^+ , k_p^+ ; k_1^- , l_1^{-} , \cdots , k_{ q-1}^{ -} , l_{ q-1}^- , k_q^- )
 \eea
These parameters are illustrated in Figure \ref{fig:Youngdiaparameters}. Going up from corner of the deepest box with content $0$, we have $k_1^+$ boxes before we reach a corner, then we go $ l_1^+$ steps horizontally to get to the next corner, then $k_2^+$ steps vertically to the next corner, then $ l_2^+$ steps horizontally. This continues until we have $k_p^+$ steps up for some positive $p$. Similarly going to the left from corner of the deepest box of content zero, we have $l_1^-$ steps to the next corner, going down from there we have $l_1^-$ steps to the next corner and so forth, until we get to $l_q^-$ steps left for some positive $q$. 
\begin{center}
\begin{figure}[h!]
\begin{tikzpicture}[semithick, >=Stealth]
\coordinate (A0) at (-3.5,0);
\filldraw[black] (A0) circle (0pt) ;
\draw (0,0) -- (5,0) -- (5,5) -- (0,5) -- cycle ;
\coordinate (A1) at (5,1);
\coordinate (A2) at (6,1);
\coordinate (A3) at (6,2);
\coordinate (A4) at (7,2);
\coordinate (A5) at (7,3);
\draw (A1) -- (A2) -- (A3) -- (A4) -- (A5);
\path (A5) + (45:0.25) coordinate (A6); 
\path (A5) + (45:0.5) coordinate (A7); 
\path (A5) + (45:0.75) coordinate (A8); 
\filldraw[black] (A6) circle (0.65pt) ;
\filldraw[black] (A7) circle (0.65pt) ;
\filldraw[black] (A8) circle (0.65pt) ;
\coordinate (A9) at (7.75,3.75);
\coordinate (A10) at (8.75,3.75);
\coordinate (A11) at (8.75,4.5);
\coordinate (A12) at (9.75,4.5);
\coordinate (A13) at (9.75,5);
\draw (A9) -- (A10) -- (A11) -- (A12) -- (A13);
\draw (5,5) -- (A13);

\coordinate (B1) at (4,0);
\coordinate (B2) at (4,-1);
\coordinate (B3) at (3,-1);
\coordinate (B4) at (3,-2);
\coordinate (B5) at (2,-2);
\draw (B1) -- (B2) -- (B3) -- (B4) -- (B5);
\path (B5) + (225:0.25) coordinate (B6); 
\path (B5) + (225:0.5) coordinate (B7); 
\path (B5) + (225:0.75) coordinate (B8); 
\filldraw[black] (B6) circle (0.65pt) ;
\filldraw[black] (B7) circle (0.65pt) ;
\filldraw[black] (B8) circle (0.65pt) ;
\coordinate (B9) at (1.35,-2.65);
\coordinate (B10) at (0.5,-2.65);
\coordinate (B11) at (0.5,-3.5);
\coordinate (B12) at (0,-3.5);
\draw (B9) -- (B10) -- (B11) -- (B12);
\draw (0,0) -- (B12);

\draw[<->] (0,5.25) -- (5,5.25) node [midway, above] {$d$}; 
\draw[<->] (-0.25,5) -- (-0.25,0) node [midway, left] {$d$}; 
\draw[->] (5.1,0) -- (5.1,0.9) node [near start, right] {\small{$k^{+}_{1}$}}; 
\draw[->] (5.15,0.9) -- (5.9,0.9) node [near end, below right] {\small{$l^{+}_{1}$}}; 
\draw[->] (6.1,1) -- (6.1,1.9) node [near start, right] {\small{$k^{+}_{2}$}}; 
\draw[->] (6.16,1.9) -- (6.9,1.9) node [near end, below right] {\small{$l^{+}_{2}$}}; 
\draw[->] (8.85,3.75) -- (8.85,4.4) node [at start, right] {\small{$k^{+}_{p-1}$}}; 
\draw[->] (8.85,4.4) -- (9.75,4.4) node [at end, below right] {\small{$\!\!l^{+}_{p-1}$}}; 
\draw[->] (9.9,4.5) -- (9.9,5) node [midway, right] {\small{$k^{+}_{p}$}}; 

\draw[->] (5,-0.1) -- (4.1,-0.1) node [at start, below ] {\small{$l^{-}_{1}$}};
\draw[->] (4.1,0) -- (4.1,-0.9) node [near end, right] {\small{$k^{-}_{1}$}}; 
\draw[->] (4,-1.1) -- (3.1,-1.1) node [near start, below right] {\small{$l^{-}_{2}$}}; 
\draw[->] (3.1,-1.1) -- (3.1,-1.9) node [near end, right] {\small{$k^{-}_{2}$}}; 
\draw[->] (1.35,-2.75) -- (0.5,-2.75) node [at start, below] {\small{$\;l^{-}_{q-1}$}}; 
\draw[->] (0.6,-2.75) -- (0.6,-3.5) node [near end, below right] {\small{$k^{-}_{q-1}$}}; 
\draw[->] (0.5,-3.6) -- (0,-3.6) node [midway, below] {\small{$l^{-}_{q}$}}; 
\coordinate (FarLeft) at (-3,-10.5);
\coordinate (Origin) at (4.5,-10.5);
\coordinate (FarRight) at (13.25,-10.5);
\coordinate (YaxisTop) at (4.5,-4);
\draw[->] (FarLeft) -- (Origin) -- (FarRight);
\draw[->] (Origin) -- (YaxisTop);

\coordinate (C9) at (-0.5,-10.5);
\filldraw[draw=black] (-0.45,-10.62) rectangle ++(0.01,0.25);
\node at (C9) [below] {\small{$\big(\!-l^{-}_{1}\!\!-\!k^{-}_{1}\!\!-\!\cdots$}};
\path (C9) + (-90:0.65) coordinate (C10); 
\node at (C10) [below] {\small{$-k^{-}_{q-1}\!\!-\!l^{-}_{q}+1\big)$}};

\coordinate (Contd) at (4.5,-5);
\node at (Contd) [above right] {\small{$d$}};
\coordinate (Conteq1) at (4.5,-9.55);
\node at (Conteq1) [right] {\small{$1$}};

\draw (Contd) -- ++(-60:2) coordinate (k1plus);
\draw (k1plus) -- ++(0:1.45) coordinate (k1plusl1);
\draw (k1plusl1) -- ++(-60:1.5) coordinate (k1plusl1plusddd);

\filldraw[black] (Contd) circle (0.65pt) ;
\filldraw[black] (Conteq1) circle (0.65pt) ;

\path (k1plusl1plusddd) + (-60:0.2) coordinate (dot4); 
\path (k1plusl1plusddd) + (-60:0.4) coordinate (dot5); 
\path (k1plusl1plusddd) + (-60:0.6) coordinate (dot6); 
\path (k1plusl1plusddd) + (-60:0.8) coordinate (dot7); 
\filldraw[black] (dot4) circle (0.65pt) ;
\filldraw[black] (dot5) circle (0.65pt) ;
\filldraw[black] (dot6) circle (0.65pt) ;
\draw (dot7) -- ++(0:1.25) coordinate (k1pll1pldddlpm1);
\draw (k1pll1pldddlpm1) -- ++(-60:2.05) coordinate (sumofall);
\draw[ultra thick] (sumofall) -- ++(0:1.5) coordinate (posofn);
\path (k1pll1pldddlpm1) + (-60:1) coordinate (NewCoord);
\path (NewCoord) + (-90:1) coordinate (NewCoord2); 
\node at (NewCoord2) [below] {\small{$\big(k^{+}_{1}\!\!+\!l^{+}_{1}\!\!+\!\cdots$}};
\path (NewCoord) + (-90:1.05) coordinate (lastdash); 
\path (NewCoord) + (-90:1.5) coordinate (belowsumofall); 
\node at (belowsumofall) [below] {\small{$k^{+}_{p-1}\!\!+\!l^{+}_{p-1}\!\!+\!k^{+}_{p}-1\big)$}};
\filldraw[draw=black] (lastdash) rectangle ++(0.01,0.25);
\path (k1plus) + (-90:3.85) coordinate (k1pllabel); 
\path (k1plusl1) + (-90:3.85) coordinate (k1pll1label); 
\node at (k1pllabel) [below] {\small{$k^{+}_{1}$}};
\path (k1pllabel) + (-90:0.03) coordinate (dash1);
\path (k1pll1label) + (-90:0.03) coordinate (dash2);  
\filldraw[draw=black] (dash1) rectangle ++(0.01,0.25);
\filldraw[draw=black] (dash2) rectangle ++(0.01,0.25);
\node at (k1pll1label) [below] {\small{$k^{+}_{1}\!\!+\!l^{+}_{1}$}};
\path (posofn) + (-90:0.12) coordinate (dashofn);
\filldraw[draw=black] (dashofn) rectangle ++(0.01,0.25);
\node at (dashofn) [below] {\small{$n$}};

\draw (Contd) -- ++(240:2.25) coordinate (l1minus);
\draw (l1minus) -- ++(180:1.25) coordinate (l1mminusk1m);
\draw (l1mminusk1m) -- ++(240:1.45) coordinate (l1mmk1mddd);

\path (l1mmk1mddd) + (240:0.2) coordinate (dot7); 
\path (l1mmk1mddd) + (240:0.4) coordinate (dot8); 
\path (l1mmk1mddd) + (240:0.6) coordinate (dot9); 
\path (l1mmk1mddd) + (240:0.8) coordinate (dot10); 
\filldraw[black] (dot7) circle (0.65pt) ;
\filldraw[black] (dot8) circle (0.65pt) ;
\filldraw[black] (dot9) circle (0.65pt) ;
\draw (dot10) -- ++(180:1) coordinate (l1mmk1mmdddkqm1);
\draw (l1mmk1mmdddkqm1) -- ++(240:1.85) coordinate (sumofallminus);
\draw[ultra thick] (sumofallminus) -- ++(180:1.4) coordinate (posofmn);
\filldraw[draw=black] (-2.35,-10.62) rectangle ++(0.01,0.25);
\coordinate (C13) at (-2.35,-10.5);
\node at (C13) [below] {\small{$-n$}};

\end{tikzpicture}
\caption{\small{Above: Young diagram being specified by the parameters $ d , k_i^{ +} , l_i^+ , k_i^- , l_i^- $. Below: a typical content distribution plot. The content of the boxes along the main diagonal is $c = 0$ and its multiplicity is $d$. The content of the box at the top right most corner of the diagram is $k^{+}_{1}+l^{+}_{1} + \cdots k^{+}_{p}-1$ and its multiplicity is 1. Similarly, the content of the box at the bottom left most corner is $l^{-}_{1}-k^{-}_{1} - \cdots l^{-}_{q}+1$ also with multiplicity of 1. At the end points of $n$ and $-n$, the CDF open string is at zero for all Young diagrams.} }
\label{fig:Youngdiaparameters}
\end{figure}
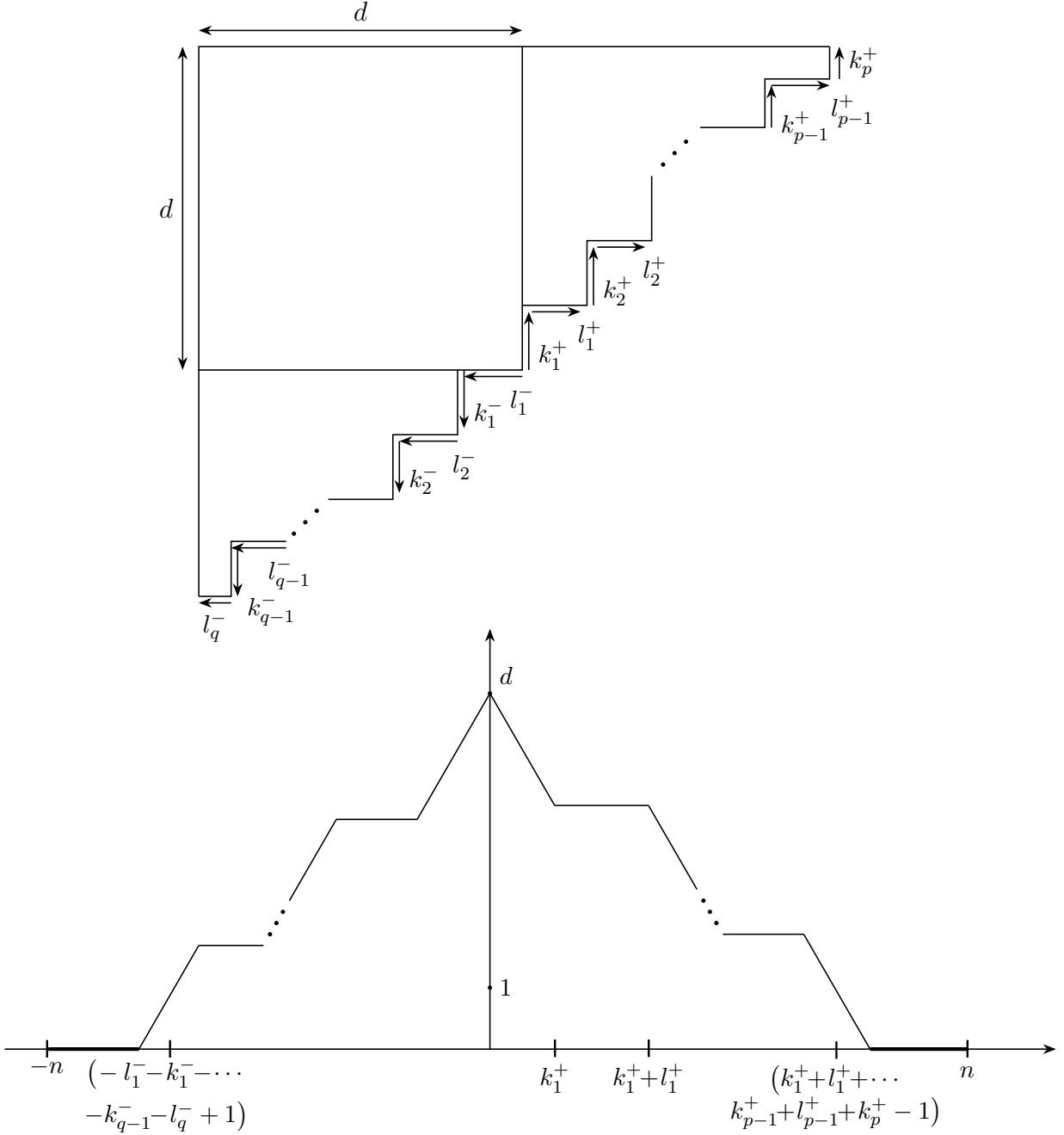
\end{center}
These parameters uniquely specify the Young diagram. Note that 
\bea 
&& d = k_1^+ + k_2^+ + \cdots + k_p^+ \cr 
&& d = l_1^- + l_2^- + \cdots + l_q^- 
\label{eq:deqsumofksandls}
\eea
The content distributions consist of segments of slope $0, -1 , 1$. Define three functions of the content $c$, with parameters $ k , a , b$.
$a , b $ are integers with $ b \ge a$. $k$ is a positive integer equal to the value of the function at $a$ so that  $ \Theta ( k , a , b ; c=a ) = k$. The three functions are
\bea 
&& \Theta_0  ( k , a , b ;  c ) = k \sum_{ i =a }^{  b }  \delta ( c , i  ) \cr 
&& \Theta_- ( k , a , b ; c ) =  \sum_{ i =a}^{ b }  ( k + a - i  ) \delta ( c , i  ) \cr 
&& \Theta_+ ( k , a, b ; c ) = \sum_{ i =a }^{ b } ( k - a  +  i ) \delta ( c , i ) 
\eea
The content distribution function is easily written in terms of these.
\bea 
f ( c )  = f_+ (c) + f_- ( c) 
\eea 
where $ f_+ , f_-$ are given below. 
\bea 
f_{+}(c) &=& \Theta_{-}\big( d,0,k^{+}_{1}-1;c \big) +\nonumber\\
&&  \Theta_{0}\big( d-k^{+}_{1},k^{+}_{1},k^{+}_{1}+l^{+}_{1}-1;c \big)+ \Theta_{-}\big( d-k^{+}_{1},k^{+}_{1}+l^{+}_{1},k^{+}_{1}+l^{+}_{1}+k^{+}_{2}-1;c \big)  \nonumber\\
&& + \cdots + \nonumber\\
&&\Theta_{0}\Big( d-\sum^{p-1}_{i=1}k^{+}_{i},\sum^{p-2}_{i=1}(k^{+}_{i}+l^{+}_{i})+k^{+}_{p-1}, \sum^{p-1}_{i=1}(k^{+}_{i}+l^{+}_{i})-1;c \Big) + \nonumber\\
&&\Theta_{-}\Big( d-\sum^{p-1}_{i=1}k^{+}_{i},\sum^{p-1}_{i=1}(k^{+}_{i}+l^{+}_{i}), \sum^{p-1}_{i=1}(k^{+}_{i}+l^{+}_{i})+k^{+}_{p}-1;c \Big) + \nonumber\\
&&\Theta_{0}\Big( d-\sum^{p}_{i=1}k^{+}_{i},\sum^{p-1}_{i=1}(k^{+}_{i}+l^{+}_{i})+k^{+}_{p}, n;c \Big),
\eea
and
\bea 
f_{-}(c) &=& \Theta_{+}\big(d-l^{-}_{1},-l^{-}_{1},-1;c \big) +  \Theta_{0}\big( d-l^{-}_{1},-l^{-}_{1}-k^{-}_{1},-l^{-}_{1}-1;c \big) \nonumber\\
&&  \Theta_{+}\big( d-l^{-}_{1}-l^{-}_{2},-l^{-}_{1}-k^{-}_{1}-l^{-}_{2},-l^{-}_{1}-k^{-}_{1}-1;c \big)\nonumber\\
&&+ \Theta_{0}\big( d-l^{-}_{1}-l^{-}_{2},-l^{-}_{1}-k^{-}_{1}-l^{-}_{2}-k^{-}_{2},-l^{-}_{1}-k^{-}_{1}-l^{-}_{2}-1;c \big)  \nonumber\\
&& + \cdots + \nonumber\\
&&\Theta_{0}\Big( d-\sum^{q-1}_{i=1}l^{-}_{i}, -\sum^{q-1}_{i=1}(l^{-}_{i}+k^{-}_{i}), -\sum^{q-2}_{i=1}(l^{-}_{i}+k^{-}_{i})-l^{-}_{q-1}-1;c \Big) + \nonumber\\
&&\Theta_{+}\Big( d-\sum^{q}_{i=1}l^{-}_{i},-\sum^{q-1}_{i=1}(l^{-}_{i}+k^{-}_{i})-l^{-}_{q}, -\sum^{q-1}_{i=1}(l^{-}_{i}+k^{-}_{i})-1;c \Big) +\nonumber\\
&& \Theta_{0}\Big( d-\sum^{q}_{i=1}l^{-}_{i}, -n, -\sum^{q-1}_{i=1}(l^{-}_{i}+k^{-}_{i})-l^{-}_{q};c \Big).
\eea
The Young diagram in equation (\ref{eq:YDexamplefplus}) illustrates an example of the counting in $f_{+}(c)$ and $f_{-}(c)$. Here, $k^{+}_{1} = 3, k^{+}_{2} = 2,k^{+}_{3} = 3, l^{+}_{1} = 5, l^{+}_{2} = 5$. Then $l^{-}_{1} = 2, l^{-}_{2} = 3, l^{-}_{3} = 2, l^{-}_{4} = 1, k^{-}_{1} = 2, k^{-}_{2} = 2, k^{-}_{3} = 3$. The function $\Theta_{-}$ will count the content multiplicity of the shaded regions to the right of, and including, the main diagonal (whose content is zero). It counts content multiplicity from $c=0$ to $k^{+}_{1}-1$, (i.e. from $c=0$ to $c=2$). Then it counts the multiplicity from $k^{+}_{1}+l^{+}_{1}$ to $k^{+}_{1}+l^{+}_{1}+k^{+}_{2}-1$, (i.e. from $c=8$ to $c=9$). Lastly, $\Theta_{-}$ evaluates the multiplicity for contents $k^{+}_{1}+l^{+}_{1}+k^{+}_{2}+l^{+}_{2}$ to $k^{+}_{1}+l^{+}_{1}+k^{+}_{2}+l^{+}_{2} + k^{+}_{3}-1$, (i.e. from $c=15$ to $c=17$). The shaded regions below the main diagonal are handled by the function $\Theta_{+}$. For this example, it will count the contents $-1$ to $-l^{-}_{1}$ (i.e. from $c=-1$ to $c=-2$), then $-l^{-}_{1}-k^{-}_{1}-1$ to $-l^{-}_{1}-k^{-}_{1} - l^{-}_{2}$ (i.e. from $c=-5$ to $c=-7$) and then $-l^{-}_{1}-k^{-}_{1}-l^{-}_{2}-k^{-}_{2}-1$ to $-l^{-}_{1}-k^{-}_{1}-l^{-}_{2}-k^{-}_{2}-l^{-}_{3}$ (i.e. from $c=-10$ to $c=-11$). Lastly, the content multiplicity of the unshaded regions are evaluated by the function $\Theta_{0}$.
\bea
	&&\ytableausetup{boxsize=1.2em}
	\nonumber\\
	&&\begin{ytableau}
          *(black!30) 0&*(black!30) & *(black!30)& &&&&&*(black!30)&*(black!30)&&&&&&*(black!30)&*(black!30)&*(black!30)\\
        *(black!30)& *(black!30)0& *(black!30)& *(black!30)& &&&&&*(black!30)&*(black!30)&&&&&&*(black!30)&*(black!30)\\
        *(black!30) &*(black!30)& *(black!30)0& *(black!30)&*(black!30) &&&&&&*(black!30)&*(black!30)&&&&&&*(black!30)\\
         &*(black!30)&*(black!30)& *(black!30)0&*(black!30)&*(black!30)&&&&&&*(black!30)&*(black!30)\\
         && *(black!30)&*(black!30) &*(black!30)0&*(black!30)&*(black!30)&&&&&&*(black!30)\\
        *(black!30)&& & *(black!30)&*(black!30)&*(black!30)0&*(black!30)&*(black!30)\\
         *(black!30)&*(black!30)& & &*(black!30)&*(black!30)&*(black!30)0&*(black!30)\\
          *(black!30)&*(black!30)& *(black!30)&&&*(black!30)&*(black!30)&*(black!30)0\\
          &*(black!30)&*(black!30)&*(black!30)&&\\
          &&*(black!30) &*(black!30)&*(black!30)&\\
          *(black!30)&&\\
          *(black!30)&*(black!30)&\\
          \\
          \\
          \\
\end{ytableau}
\label{eq:YDexamplefplus}
\eea
The CDF $f(c)$ is defined from $-n$ to $n$. For a generic Young diagram $f(c) = 0$ from 
$	\sum\limits^{p-1}_{i=1}(k^{+}_{i}+l^{+}_{i}) + k^{+}_{p} $ to $ n$.
Since $d= \sum_{p}^{i=1} k^{+}_{i}$ (as seen from equation (\ref{eq:deqsumofksandls})), the last $\Theta_{0}$ function in $f_{+}(c)$ is zero over this range. The slope of $f(c)$ over this range is also zero. For the special case of the Young diagram being a single row of $n$ boxes, this $\Theta_{0}$ is defined only at $c=n$, and is still equal to zero. However the slope of $f(c)$ from $c=n-1$ to $c=n$ is $-1$. This last $\Theta_{0}$ function, even though it is equal to zero, is included in our definition so that the CDF is indeed defined up to $n$. Similarly for the last $\Theta_{0}$ function in $f_{-}(c)$. Generically, 
 $f(c) = 0$ from 
$	-\sum\limits^{q-1}_{i=1}(l^{-}_{i}+k^{-}_{i}) - l^{-}_{q} $  to $ -n$.

Again from (\ref{eq:deqsumofksandls}), $d= \sum_{q}^{i=1} l^{-}_{i}$. Thus, the last $\Theta_{0}$ function in $f_{+}(c) = 0$ over this range. The CDF's slope is zero here in this region. In the special case of a single column of $n$ boxes the slope is $+1$ from $1-n$ to $-n$.

The Young diagram is uniquely determined by the parameters $ d , k_i^{ +} , l_i^+ , k_i^- , l_i^- $ which determine the steps from the box at the end of the leading diagonal to the corners. In turn the content distribution function is uniquely determined by the location of 
the kinks. The correspondence between the kinks in the CDF and the corners in the Young diagram are given by the equations above.

{ \bf Proposition  5.3.2} The content distribution functions for Young diagrams with $n$ boxes 
are uniquely determined by the $n$ moments 
\bea 
\sum_{ c } c^k f(c ) = \langle c^k \rangle  \hbox{ for } k \in \{ 0 , 1, \cdots , n-1 \}  
\eea
$ \langle c^0 \rangle = n $. 
This follows from the fact we have proved in section \ref{sec:T1toTngens} that $ T_2 , \cdots , T_n $ generate the centre of $ \mC ( S_n )$, 
and that these in turn are determined by the above moments.

Experimental data in fact shows that the number of moments needed is in fact lower than $n$. Indeed, we have already proved (Remark 3.2.2) 
that for any $n$ that $ \{ T_2 , \cdots , T_{ n-1} \} $ suffice to generate the centre, 
so the following stronger theorem is true.\\ 
{ \bf Proposition 5.3.3 } The content distribution functions for Young diagrams with $n$ boxes 
are uniquely determined by the $n-1$ moments 
\bea 
\sum_{ c } c^k f(c ) = \langle c^k \rangle  \hbox{ for } k \in \{ 0 , 1, \cdots , n-2 \}  
\eea

Some properties of content distribution functions. They are valued in the positive integers and have the following properties. 

\begin{figure}[htb]
\centering
  \begin{tabular}{@{}cccc@{}}
    \includegraphics[width=.45\textwidth]{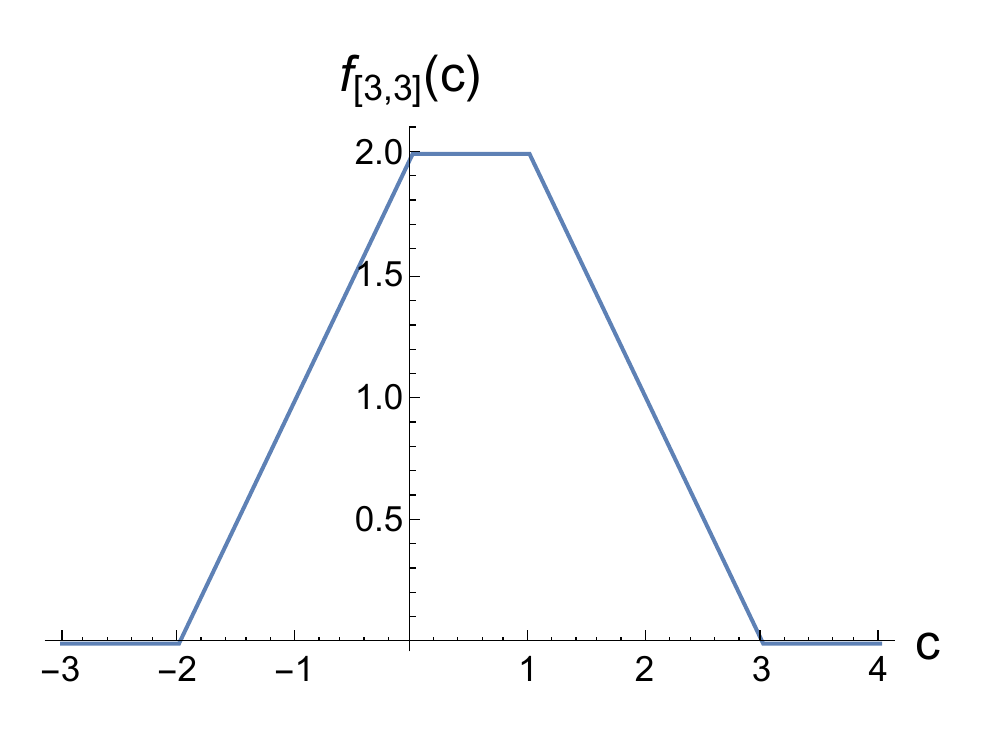} &
    \includegraphics[width=.48\textwidth]{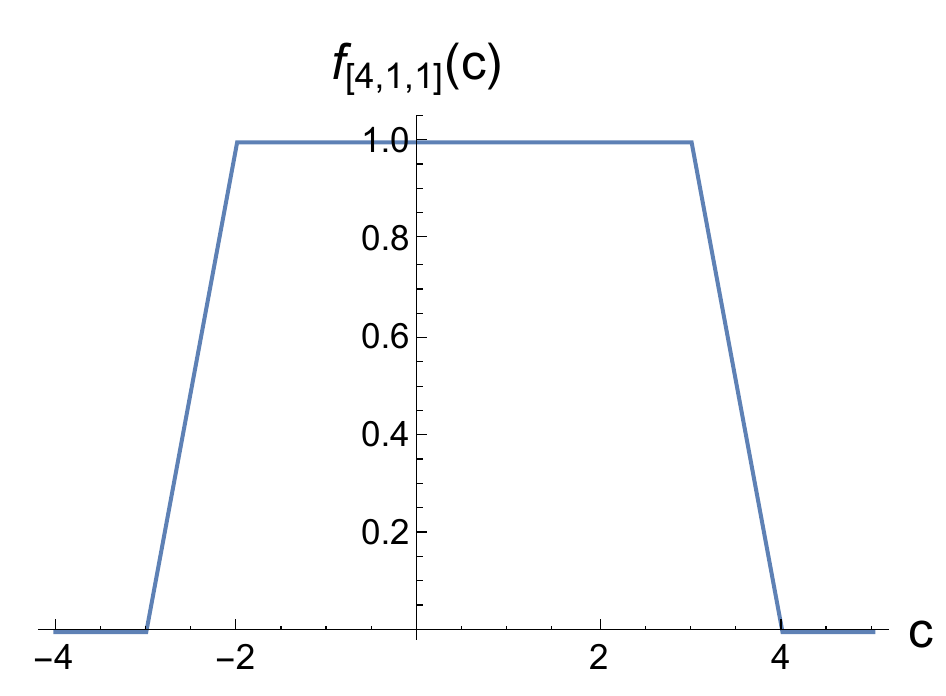} &
   \\
  \end{tabular}
  \caption{The CDF plots for the partitions degenerate in $\la c \ra$ at $n=6$. }
  \label{n6CDFs}
\end{figure}

\begin{enumerate} 

\item 

\bea\label{defCDF1}  
&& f ( c +1 ) - f ( c ) \in \{ 0 , 1 \}   \hbox { for } c \ge 0    \cr 
&& f ( c ) - f ( c -1  )  \in \{ 0 , 1 \} \hbox { for } c \le 0 
\eea

\item 

\bea\label{defCDF2}  
f ( c )  \hbox{ is a connected curve, extending from}   [ - c_1 , c_2 ] \hbox{, where }  c_1 , c_2 \ge 0 
\eea 

\item Boundary values 
\bea\label{defCDF3}  
&&  f ( -c_1 + 1  ) = 1 \cr  
&& f ( r_1 - 1  ) = 1 
\eea
Note that $ c_1 $ is  the number of non-zero row lengths, or  the length of the first column.
$r_1$ is the length of the first row. 

\item It follows from the above properties that the maximum value of $f$ occurs at $ 0$, and is equal to the depth of the Young diagram (i.e. the length of the diagonal ). 

\item It follows from the above that the maximum $ f_0 $ is $ \lfloor { \sqrt { n } }  \rfloor $. 
\end{enumerate} 
The problem of generating the centre and distinguishing all Young diagrams at given $n$ may now be recast in the language of the content distribution functions and moments. Given functions obeying the conditions in (\ref{defCDF1})  (\ref{defCDF2}) (\ref{defCDF3}), what is the smallest positive integer $n = \la c^0 \ra $, such that the equations 
\bea 
 \langle  c^0 \rangle_{ f } & =  & \langle c^0 \rangle_{ f'} \cr 
& \vdots &  \cr 
\langle c^k \rangle_f  & = &  \langle c^{ k } \rangle_{ f'}   
\eea
imply that $ f = f'$ for all $ m < n$. This is a reformulation of the results of section \ref{sec:compwcontpoly} now in terms of CDFs and moments. The data presented in table \ref{firstnwithk} are the smallest integers where the equality of moments up to $k$ allows a multiplicity of content distribution functions.
For $ n =6$, there are two degenerate Young diagram pairs. $ [ 4,1,1] $ and $ [ 3,3 ] $ both have $ < c > =3$. Here we are describing the Young diagrams by their row lengths. 
The content distributions are 
\bea 
[ 4,1,1 ] \rightarrow f ( c ) & = &  1 \hbox{ for  the range }   [ -2 , 3 ] \cr 
 [ 3,3 ] \rightarrow f ( c ) & = &  2 \hbox{ for }  \{ 0 , 1 \} \cr 
                               & = &  1 \hbox { for }  \{ -1 , 2 \} .
\eea
The CDF plots for these two partitions are displayed in figure \ref{n6CDFs}.
Define the difference between the two CDFs for $[4,1,1]$ and $[3,3]$: $ \Delta_{i} = f_i - f'_i $ defined over $ i \in [ -2 , 3 ] $. This six dimensional vector is 
\bea 
\{ -1 , 0  , 1, 1, 0 , -1 \}. 
\eea
Obviously it satisfies $ \sum_{ i } \Delta_i = 0 $ and $ \sum_i i \Delta_i = 0 $. Another degenerate pair of diagrams at $ n =6$ which satisfy the same equations is 
\bea 
 [ 3  , 1^3 ]  & \rightarrow & \langle  c \rangle  = -3 \cr  
 [ 2^3 ] &  \rightarrow  & \langle  c \rangle  = -3 
\eea
These are conjugates of the $ [ 4, 1,1 ] , [ 3,3 ] $ degenerate pair. From this simple example, it seems useful to study the data in terms of these content distribution functions, and their differences. 
Given any two content distribution functions $ f_i $ and $ f_i'$, define 
\bea 
\Delta_i  = f_i  - f_i'  
\eea
These are positive or negative integers. 
These differences have the properties that 
\bea 
\Delta_{ i +1 } - \Delta_i \in \{ 0 , -1 , 1 \} 
\eea
Existence of degenerate moments implies that the following equations have non-tivial solutions 
\bea 
\sum_{ i } \Delta_i &= &  0  \cr 
\sum_{ i } i \Delta_i &= &  0 \cr 
&\vdots  & \cr 
\sum_{ i } i^k \Delta_i & = & 0 
\eea
This looks like a function with a set of vanishing moments up to $k$. For low values of $n$, there are 
no solutions for  $ \Delta_i$ to these equations. Functions with vanishing moments up to a certain maximum are studied in the literature on wavelets (see for example \cite{Patilkulkarni2013, Stephane2009, Keinert2003}). Discrete wavelets are also an active area of research. So this could be a way to approach an answer to our question.

The problem of extending the sequence $n_{*}( k ) $ displayed in table \ref{firstnwithk} to larger values of $k$ is very interesting. It is indeed plausible that the CDFs could play a major role in this regard. The degeneracies that arise at $ n_*(k)$ capture the failure of the first $k$ to distinguish the Young diagrams. These degeneracies are related to the existence of vanishing moment equations for differences of CDFs. Some of these solutions are exhibited using the computer generated data in section \ref{sec:n1524examples}. The CDFs may lead to new approaches for determining $n_{*}( k)$ drawing on 
number theory (diophantine equations) and probability theory.

\subsection{Degenerate CDF examples at $n=15$ and $n=24$.}\label{sec:n1524examples}

At $n=15$, there are three degenerate pairs for $\{\la c \ra , \la c^2 \ra\}$. See table \ref{CoDimTable} and equation (\ref{eq:keq2deg}). Below, we list these partitions, give their values for $\{\la c \ra , \la c^2 \ra\}$, as well as their CDFs. Lastly, we construct $\Delta$ for each degenerate pair.
\bea 
&& \{ [ 6 , 3 ,  2^3 ] , [ 5^2 , 2 , 1^3 ] \}  \hbox { have } \{\la c \ra , \la c^2 \ra\} = \{ 0 , 100 \} \cr 
&& \{ [ 7 , 4 , 2^2 ] ,  [ 6^2 , 1^3 ] \} \hbox { have } \{\la c \ra , \la c^2 \ra\} = \{ 15 , 115 \} \cr 
&& \{ [ 5 , 2^5 ] , [ 4^2 , 2^2 , 1^3 ] \} \hbox{have} \{\la c \ra , \la c^2 \ra\} = \{ -15 , 115 \} 
\eea

\begin{figure}[h!]
\centering
  \begin{tabular}{@{}cccc@{}}
    \includegraphics[width=.45\textwidth]{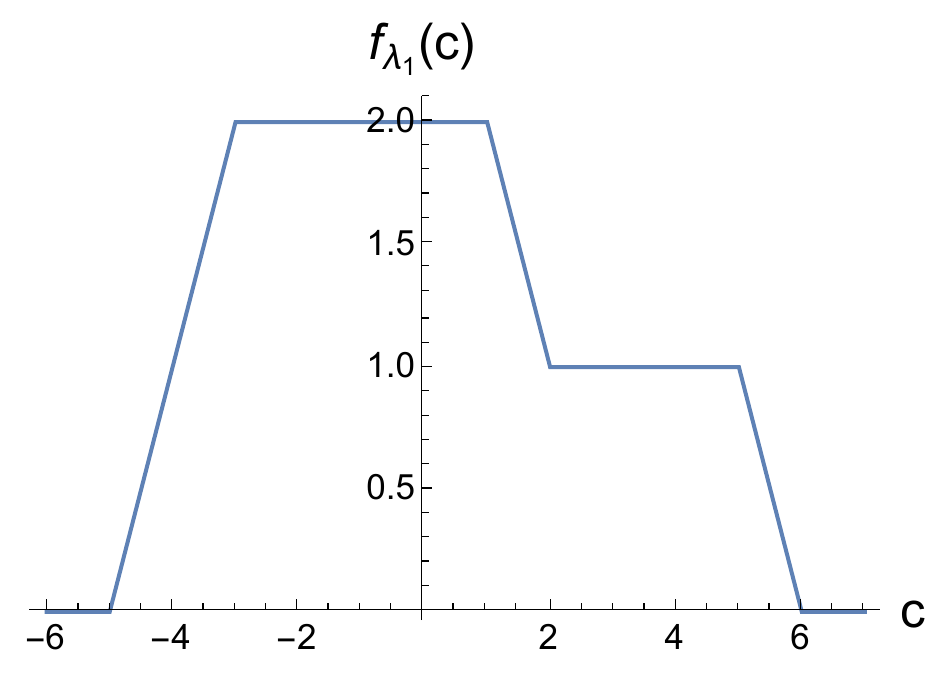} &
    \includegraphics[width=.48\textwidth]{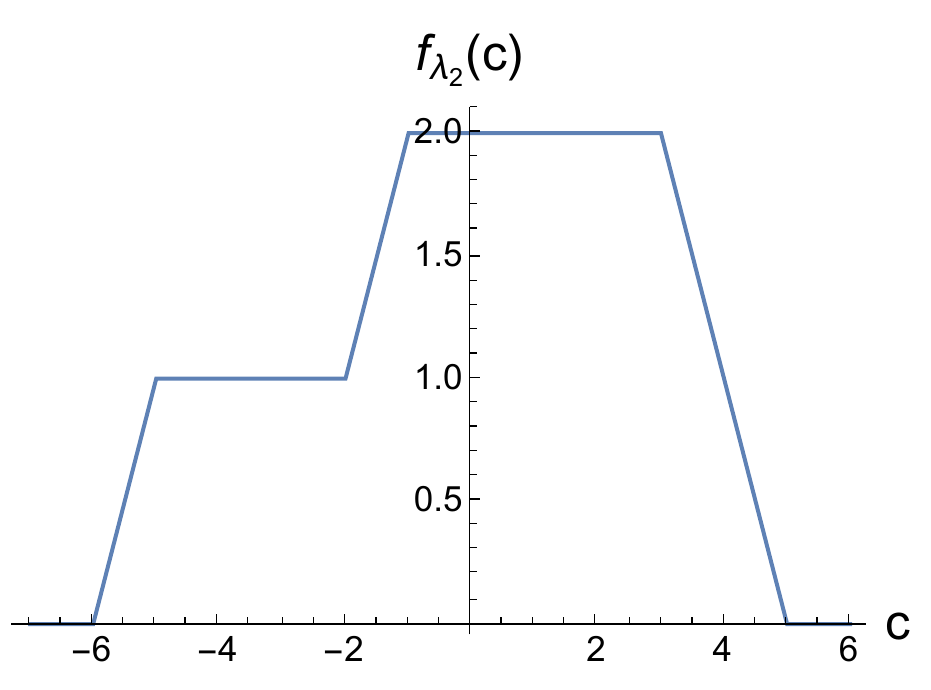} &\\
      \includegraphics[width=.45\textwidth]{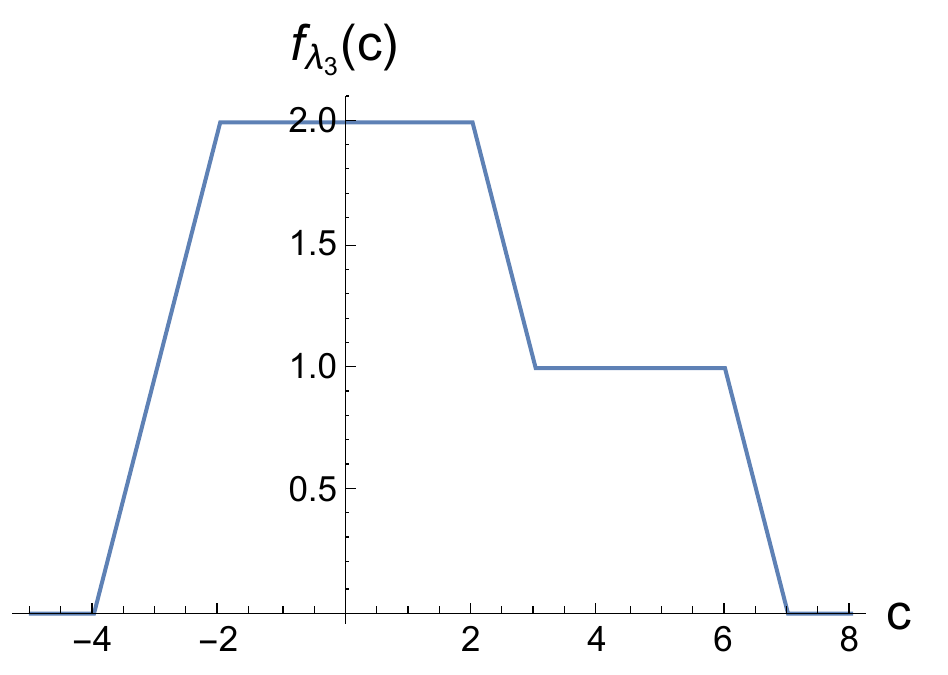} &
    \includegraphics[width=.48\textwidth]{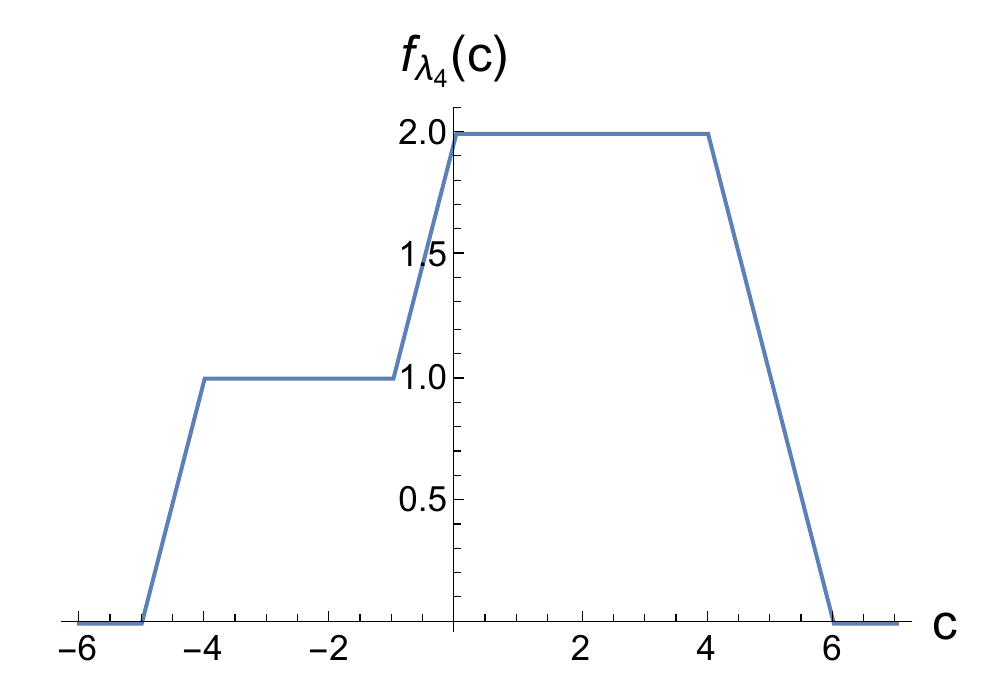} &\\
      \includegraphics[width=.45\textwidth]{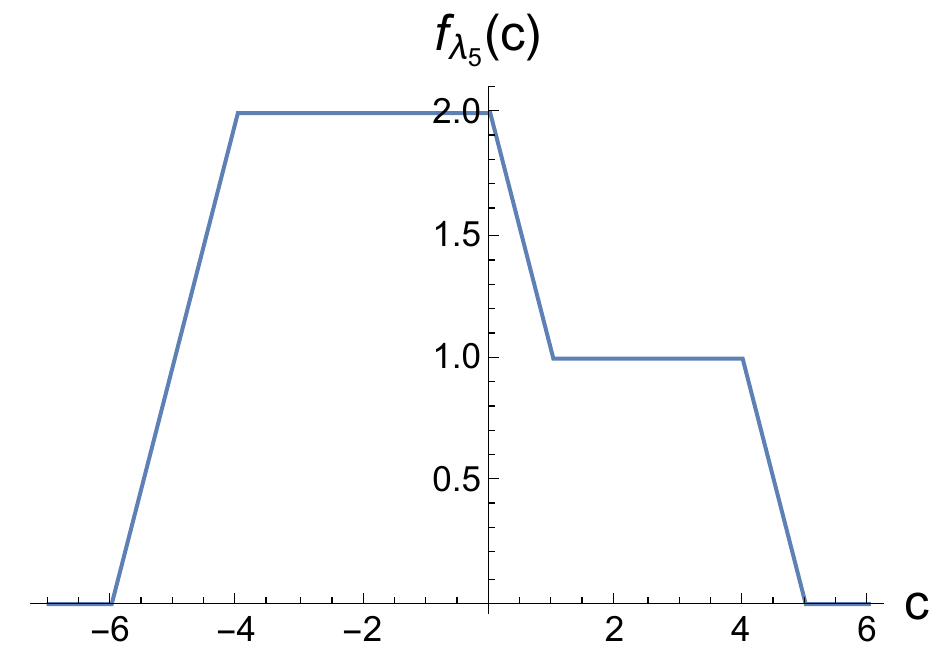} &
    \includegraphics[width=.48\textwidth]{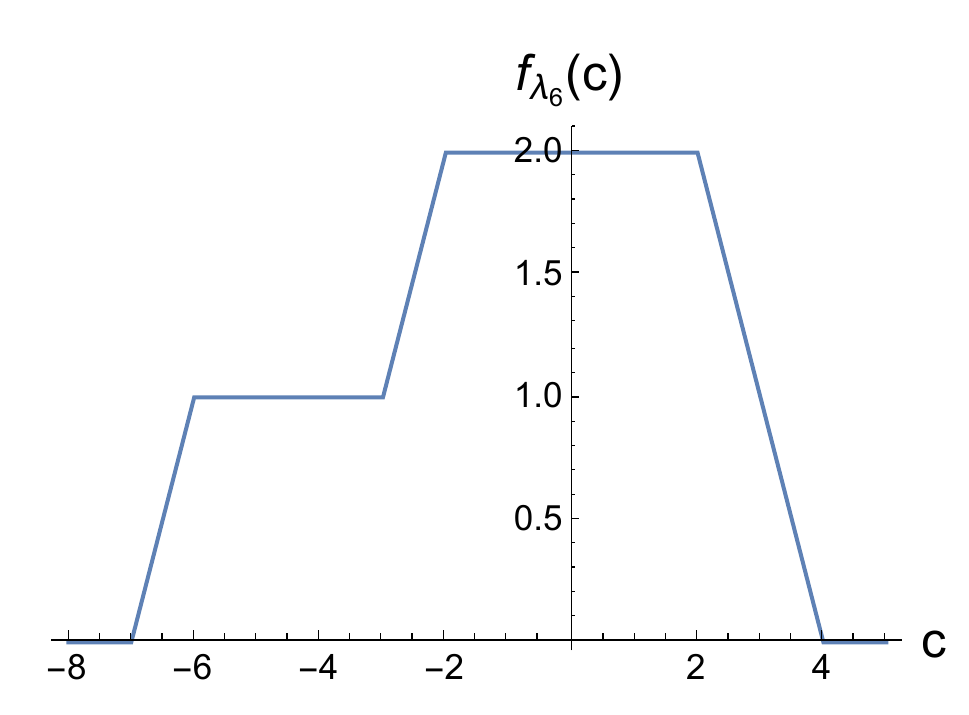} &
   \\
  \end{tabular}
  \caption{The CDF plots for the partitions degenerate in $\{\la c \ra,\la c^{2}\ra\}$ at $n=15$.}
  \label{n15CDFs}
\end{figure}

For $ \lambda_{1} = [ 6 , 3 , 2^3 ] ,  \lambda_{2} =[ 5^2 ,2, 1^3 ]  $, we have
\bea
	f_{\lambda_{1}} &=& \{ 1,2,2,2,2,2,1,1,1,1 \}, \hspace{10pt} \mathrm{over\;range} \hspace{10pt} [ -4,5 ] \\
	f_{\lambda_{2}} &=& \{1,1,1,1, 2,2,2,2,2, 1\}, \hspace{10pt} \mathrm{over\;range} \hspace{10pt} [ -5,4 ],\\
	\Delta_{\left(\lambda_{1},\lambda_{2}\right)} &=& \{1, 0, -1, -1, 0, 0, 0, 1, 1, 0, -1\}.
\eea
For $ \lambda_{3} = [ 7 , 4 , 2^2 ] ,  \lambda_{4} =[ 6^2 , 1^3 ]  $, we have
\bea
	f_{\lambda_{3}} &=& \{1, 2, 2, 2, 2, 2, 1, 1, 1,1\}, \hspace{10pt} \mathrm{over\;range} \hspace{10pt} [ -3,6 ] \\
	f_{\lambda_{4}} &=& \{1, 1, 1, 1, 2, 2, 2, 2, 2, 1\}, \hspace{10pt} \mathrm{over\;range} \hspace{10pt} [ -4,5 ],\\
	\Delta_{\left(\lambda_{3},\lambda_{4}\right)} &=& \{1, 0, -1, -1, 0, 0, 0, 1, 1, 0, -1\}.
\eea
For $ \lambda_{5} = [ 5 , 2^5 ] ,  \lambda_{6} = [ 4^2 , 2^2 , 1^3 ] $, we have
\bea
	f_{\lambda_{5}} &=& \{ 1,2,2,2,2,2,1,1,1,1 \}, \hspace{10pt} \mathrm{over\;range} \hspace{10pt} [ -5,4 ] \\
	f_{\lambda_{6}} &=& \{1,1,1,1, 2,2,2,2,2, 1\}, \hspace{10pt} \mathrm{over\;range} \hspace{10pt} [ -6,3 ],\\
	\Delta_{\left(\lambda_{5},\lambda_{6}\right)} &=& \{1, 0, -1, -1, 0, 0, 0, 1, 1, 0, -1\}.
\eea
The CDF plots for the three degenerate pairs are displayed in figure \ref{n15CDFs}.

\begin{figure}[h!]
\centering
  \begin{tabular}{@{}cccc@{}}
    \includegraphics[width=.48\textwidth]{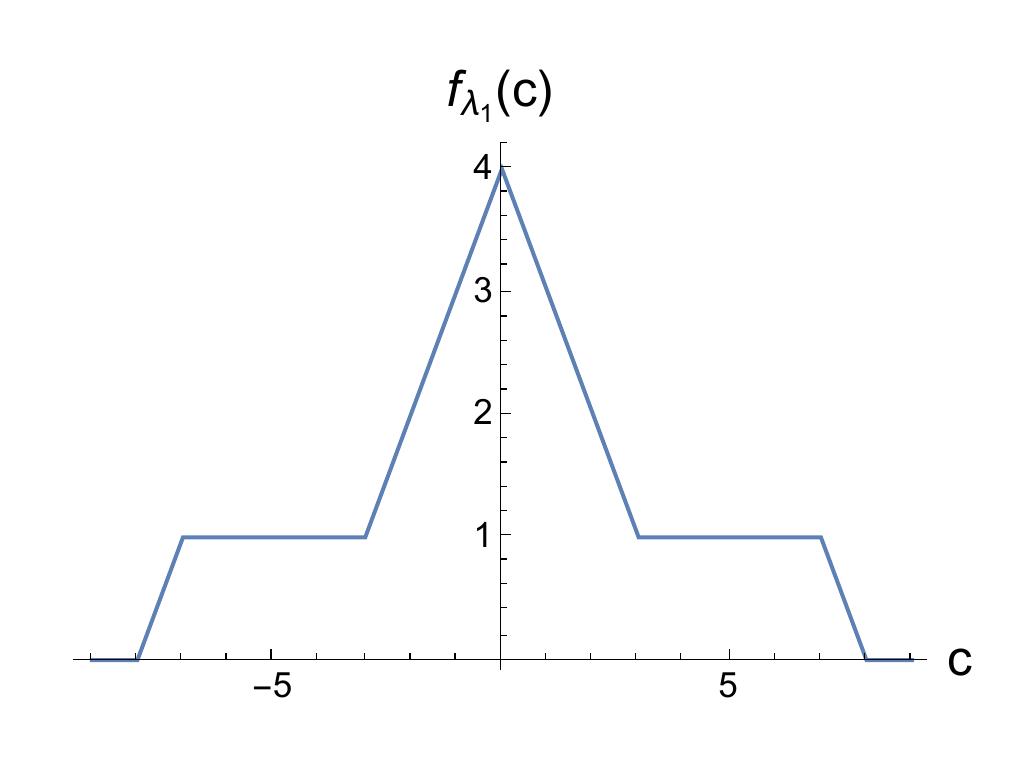} &
    \includegraphics[width=.48\textwidth]{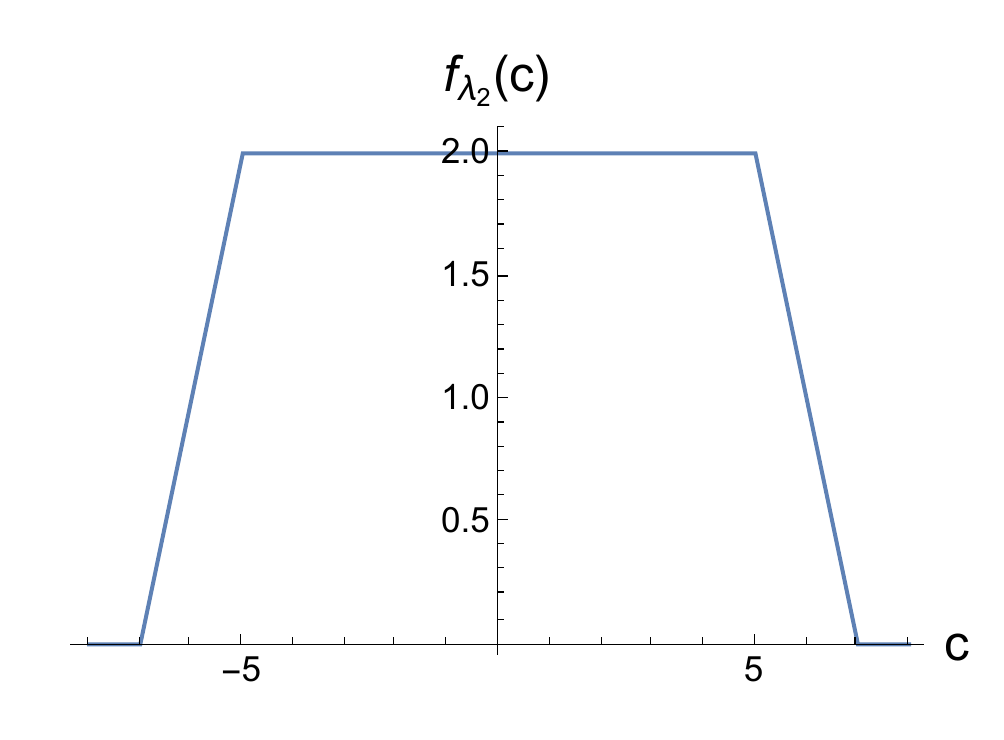} &
        \\
  \end{tabular}
  \caption{{\small{The CDF plots for the partitions degenerate in $\{\la c \ra,\la c^{2} \ra, \la c^{3} \ra \}$ at $n=24$.}}}
  \label{n24CDFs}
\end{figure}

As a final example, at $n = 24$, has a single degenerate pair for moments $\{ \langle  c \rangle, \langle  c^{2} \rangle, \langle  c^{3} \rangle \} $. See equation (\ref{eq:keq3deg}). The partitions are $\{ [8, 4^{3}, 1^{4}], [7^{2}, 2^{5}] \}$ and the values for $\{ \langle  c \rangle, \langle  c^{2} \rangle, \langle  c^{3} \rangle \} = \{ 0,292, 0 \}$, and the partitions are $\{ [8, 4^{3}, 1^{4}], [7^{2}, 2^{5}] \}$. The CDFs and $\Delta$ are
\bea
	f_{\lambda_{1}} &=& \{ 1,1,1,1,1,2,3,4,3,2,1,1,1,1,1\}, \hspace{10pt} \mathrm{over\;range} \hspace{10pt} [ -7,7 ] \\
	f_{\lambda_{2}} &=& \{1, 2, 2, 2, 2, 2, 2, 2, 2, 2, 2, 2, 1\}, \hspace{10pt} \mathrm{over\;range} \hspace{10pt} [ -6,6 ],\\
	\Delta_{\left(\lambda_{1},\lambda_{2}\right)} &=& \{-1, 0, 1, 1, 1, 0, -1, -2, -1, 0, 1, 1, 1, 0, -1\}.
\eea
The CDF plots for these two partitions are shown in figure \ref{n24CDFs}.

\section{ Summary and Outlook  } 

We have proved that central elements $T_2 , T_3 , \cdots , T_n $ associated with reduced cycle structures  $ \cC_{ (1) } , \cC_{ (2) } , \cdots , \cC_{ (n-1) } $ generate the centre $ \cZ ( \mC ( S_n) ) $. 
We then showed that restricting to a subset $ \{ T_2  , \cdots , T_k \} $ generates the centre for 
all $n$ up to $n_* (k)$. We used computational methods to determine $n_{*}(k)$ for $ k $ up to $6$. 
For the classes $T_2 , T_3 $ and the collection $ \{ T_2 , T_3 \}$, we computed the dimensions of the subspaces generated. We showed that these dimensions are directly encoded in the normalized characters of 
these conjugacy classes in Young diagrams $R$ with $n$ boxes, which are in turn related to Casimir eigenvalues for the $U(N)$ representation associated with $R$.  The multiplicities of the normalized characters can be used to quantify the amount of information available with specified sets such as $\{ T_2 \} $ , $ \{ T_3 \} $ or $ \{ T_2 , T_3 \}$, using Shannon entropy functions. These entropies were calculated and led to the conclusion that the dimensions as well as entropies give sensible measures of 
the amount of information available  with the normalized characters, equivalently specified Casimir charges. We presented some conjectures on the large $n$ behaviours of relative dimensions and entropies for 
$T_2$ and $ T_3$, based on the plausible expectation that at large $n$ the different measures should give the same ranking.

We have observed that the  power-sums of contents can be viewed as moments of a content distribution function. This is simply a discrete function on a finite set of points in the range $ [ - n+ 1, n -1] $, which was shown to uniquely determine a Young diagram.  Some initial steps in the 
direction of using CDFs in order to understand the degeneracies between normalized characters that occur 
for $n$ just above $ n_{*}(k)$ have been made. We have observed that differences of Young diagram CDFs
obey some diophantine vanishing moment conditions above $n_{*}(k)$. Using computational work, we obtained some instances of these solutions to diophantine equations. It is reasonable to expect that a combination of techniques from combinatorics and number theory will, in the future,  allow a general analytic treatment of these diophantine equations for general $k$ and provide further information on $n_* (k)$ as $k$ increases. 

The centre $ \cZ ( \mC ( S_n ) $ is one of a class of interesting permutation centralizer algebras which are relevant to multi-matrix and tensor invariants \cite{EHS,PCA}. There is a 2-parameter algebra $ \cA ( m  , n ) $ 
relevant to the 2-matrix system with $U(N)$ gauge symmetry, with structure closely related to Littlewood-Richardson coefficients. 
There is a $1$-parameter algebra $ \cK ( n )$ relevant to 3-index tensor systems, which is closely related to Kronecker coefficients. The algebras $ \cA ( m  , n ) $  were recently used to derive identities involving contents of Young diagrams, which have applications in quantum  information processing tasks
\cite{QIPCQF}: the present paper is another link between permutation algebras and information theoretic perspectives directly motivated, in the present case, by information theoretic questions in AdS/CFT.
Another connection between the 2-matrix system and small black holes in AdS/CFT is proposed in \cite{Bersmall}. It is evident that we are only beginning to scratch the surface of the story linking AdS/CFT, information and permutation algebras.  Analogous algebras play a role in matrix/tensor systems with $O(N)/Sp(N)$ symmetry  \cite{AABF0211,CDD1303,CDD1301,GKemp1,GKemp2,LBSR1804}. The results of this paper, developing the connection between Casimirs and the structure of permutation algebras should admit a generalization to these cases. It will be fascinating to explore these systems using the combination of analytic and computational techniques we have used here, to generate 
sequences analogous to $n_{*}(k)$ for these cases.

 \begin{center} 
 { \bf Acknowledgements} 
 \end{center} 
S.R. is supported by the STFC consolidated grant  ST/P000754/1 ``String Theory, Gauge Theory \& Duality'' and  a Visiting Professorship at the University of the Witwatersrand, funded by a Simons Foundation grant to the Mandelstam Institute for Theoretical Physics. We are grateful  to Robert de Mello Koch, Stephen Doty, Arun Ram, Joan Simon for very useful  conversations on the subject of this paper. 

\section{Appendix A: Mathematica Code}\label{ref:MTCA}

We begin by writing code to calculate the content polynomial or the content power sum for a partition specified by $P$.  
{\small{\begin{lstlisting}
ContentPowerSum[k_,P_]:= 
		Sum[Sum[(a-m)^k,{a,1,P[[m]]}],{m,1,Length[P]}]
ListContentPowerSums[k_,n_]:=
		Table[ContentPowerSum[k,P],{P,IntegerPartitions[n]}] 
\end{lstlisting}}}
The integer $k$ specifies the power of the terms in the sum, while $P$ specifies the actual partition. For example, taking partitions of 3, and $k=2$, the above code computes, for the three partitions $(3), (2,1)$ and $(1,1,1)$:
{\small{\begin{lstlisting}
In[1]:= ContentPowerSum[2,{3}]
        ContentPowerSum[2,{2, 1}]
        ContentPowerSum[2,{1, 1, 1}]

Out[1]= 5

Out[2]= 2

Out[3]= 5.
\end{lstlisting}}}
The definition in the second line of code ``ListContentPowerSums" gives a list of these power sums that runs over all the partitions on $n$. After running the above code for $k=1$ and $n=6$, we find:
{\small{\begin{lstlisting}
In[3]:= ListContentPowerSums [1,6]

Out[3]= {15, 9, 5, 3, 3, 0, -3, -3, -5, -9, -15}
\end{lstlisting}}}
Now we wish to compare the lists of the content power sums for different values of $k$. Below $S$ will be a set of positive integers specifying the powers of contents to be summed over; this function will produce the list of vectors of content power sums for the partitions of n, with the powers specified by $S$:
{\small{ \begin{lstlisting}
ListSetContentPowerSums[S_ ,n_ ]:= 
  Table[Table[ContentPowerSum[i,P],{i,S}],{P,IntegerPartitions[n]}] 
\end{lstlisting}}}
This code will allow us to compare the set $\{ \la c(R) \ra, \la c(R)^{2}\ra, \cdots \}$. For example
{\small\begin{lstlisting}
In[1]:= ListSetContentPowerSums[{1,2},6]

Out[1]= {{15,55}, {9,31}, {5,15}, {3,19}, {3,7}, {0,10}, 
        {-3,19}, {-3,7}, {-5,15}, {-9,31}, {-15,55}}.
 \end{lstlisting}}
We can see that the list $\{ \la c(R) \ra, \la c(R)^{2}\ra \}$ contain no degeneracies at $n=6$. The code to compute codimension data for $T_{2}$, then $T_{3}$ and then for $\{T_{2}, T_{3}\}$ respectively is found below.
{\small{\begin{lstlisting}
Table[Length[ListSetContentPowerSums[{1},n]]-
Length[DeleteDuplicates[ListSetContentPowerSums[{1},n]]],{n,Range[2,70]}]
  
 Table[Length[ListSetContentPowerSums[{2},n]]-
Length[DeleteDuplicates[ListSetContentPowerSums[{2},n]]],{n,Range[2,70]}]
  
Table[Length[ListSetContentPowerSums[{1,2},n]]-
Length[DeleteDuplicates[ListSetContentPowerSums[{1,2},n]]],{n,Range[2,70]}].
\end{lstlisting}}}
The idea here is simply to generate the list of content polynomials, and count the number of uniques elements by subtracting the length of the list when duplicates have been deleted. The above code generates codimension data for $n$ from $n=2$ to $n=70$.

\end{document}